\documentclass[manuscript,screen]{acmart}

\usepackage{threeparttable}
\usepackage[utf8]{inputenc}

\usepackage[ruled,linesnumbered,noend]{algorithm2e}
\SetAlgoLined
\SetKwInOut{Input}{Input}\SetKwInOut{Output}{Output}

\usepackage{xcolor}
\usepackage{setspace}
\usepackage{amsmath,amsfonts}
\usepackage{url}
\usepackage{subfigure}
\usepackage{colortbl}
\usepackage{multirow}
\usepackage{color}
\usepackage{graphicx}
\usepackage{amsmath}
\usepackage{amsfonts}
\usepackage{bbm}
\usepackage{pifont}
\usepackage{makecell}
\usepackage{cleveref}
\crefname{section}{§}{§§}
\Crefname{section}{§}{§§}
\usepackage{enumitem}
\usepackage{float}
\usepackage[most]{tcolorbox}
\usepackage{colortbl}
\AtBeginDocument{%
	\providecommand\BibTeX{{%
			\normalfont B\kern-0.5em{\scshape i\kern-0.25em b}\kern-0.8em\TeX}}}

\setcopyright{acmlicensed}
\copyrightyear{2018}
\acmYear{2018}
\acmDOI{XXXXXXX.XXXXXXX}

\acmISBN{978-1-4503-XXXX-X/18/06}

\begin{document}
	
	\title{TVDiag: A Task-oriented and View-invariant Failure Diagnosis Framework for Microservice-based Systems with Multimodal Data}
	
	\author{Shuaiyu Xie}
	\email{theory@whu.edu.cn}
	\orcid{0000-0001-7925-3788}
	\affiliation{%
		\institution{Wuhan University}
		\country{China}
	}
	
	\author{Jian Wang$^\ast$}
	\email{jianwang@whu.edu.cn}
	\orcid{0000-0002-1559-9314}
	\affiliation{%
		\institution{Wuhan University}
		\country{China}
	}
	
	\author{Hanbin He}
	\email{hhbsgsh@whu.edu.cn}
	\orcid{0009-0004-0871-9041}
	\affiliation{%
		\institution{Wuhan University}
		\country{China}
	}
	
	\author{Zhihao Wang}
	\email{zhihao_wang@whu.edu.cn}
	\orcid{0009-0008-5966-0325}
	\affiliation{%
		\institution{Wuhan University}
		\country{China}
	}
	
	\author{Yuqi Zhao}
	\email{yuqizhao@ccnu.edu.cn}
	\orcid{0000-0002-2642-5109}
	\affiliation{%
		\institution{Central China Normal University}
		\country{China}
	}
	
	\author{Neng Zhang}
	\email{nengzhang@ccnu.edu.cn}
	\orcid{0000-0001-8662-5690}
	\affiliation{%
		\institution{Central China Normal University}
		\country{China}
	}
	
	\author{Bing Li$^\ast$}
	\email{bingli@whu.edu.cn}
	\orcid{0000-0002-2165-2636}
	\affiliation{%
		\institution{Wuhan University}
		\country{China}
	}
	
	\authorsaddresses{$^\ast$Jian Wang and Bing Li are the corresponding authors of this paper. Authors’ addresses: Shuaiyu Xie, theory@whu.edu.cn, School of Computer Science, Wuhan University, China; Jian Wang, jianwang@whu.edu.cn, School of Computer Science, Wuhan University, China; Hanbin He, hhbsgsh@whu.edu.cn, School of Computer Science, Wuhan University, China; Zhihao Wang, zhihao\_wang@whu.edu.cn, School of Computer Science, Wuhan University, China; Yuqi Zhao, yuqizhao@ccnu.edu.cn,
		School of Computer Science, Central China Normal University, China; Neng Zhang, nengzhang@ccnu.edu.cn, School of Computer Science \& Hubei Provincial Key Laboratory of Artificial Intelligence and Smart Learning, Central China Normal University, China; Bing Li, bingli@whu.edu.cn, School of Computer Science, Wuhan University, China}

	\renewcommand{\shortauthors}{Shuaiyu Xie and Jian Wang, et al.}
	
	\begin{abstract}
		Microservice-based systems often suffer from reliability issues due to their intricate interactions and expanding scale. With the rapid growth of observability techniques, various methods have been proposed to achieve failure diagnosis, including root cause localization and failure type identification, by leveraging diverse monitoring data such as logs, metrics, or traces. However, traditional failure diagnosis methods that use single-modal data can hardly cover all failure scenarios due to the restricted information. Several failure diagnosis methods have been recently proposed to integrate multimodal data based on deep learning. These methods, however,  tend to combine modalities indiscriminately and treat them equally in failure diagnosis, ignoring the relationship between specific modalities and different diagnostic tasks. This oversight hinders the effective utilization of the unique advantages offered by each modality.
		To address the limitation, we propose \textit{TVDiag}, a multimodal failure diagnosis framework for locating culprit microservice instances and identifying their failure types (e.g., Net-packets Corruption) in microservice-based systems. \textit{TVDiag} employs task-oriented learning to enhance the potential advantages of each modality and establishes cross-modal associations based on contrastive learning to extract view-invariant failure information. Furthermore, we develop a graph-level data augmentation strategy that randomly inactivates the observability of some normal microservice instances to mitigate the shortage of training data. 
		Experimental results on three datasets show that \textit{TVDiag} outperforms state-of-the-art methods in multimodal failure diagnosis by at least 32.46\% and 3.08\% in terms of $HR@1$ and F1-score, respectively.
	\end{abstract}

	
	
	
	
	\maketitle
	
	\section{Introduction}
	\label{sec:introduction}
	Microservice architecture is gaining popularity in developing online systems because of its advantages of lower coupling, better resilience, and faster delivery. However, microservice-based systems frequently encounter reliability issues arising from their intricate interactions and expanding scale. For example, a failure in one microservice instance can propagate to other instances through message interactions between microservices, resulting in multiple instances becoming abnormal simultaneously \cite{peng2022trace}, \cite{qiu2020firm}, \cite{wang2022operation}. The abundance of abnormal microservice instances has turned the root cause localization into a labor-intensive and error-prone endeavor. Operators may also struggle with identifying numerous suspected failure types (e.g., out-of-memory, network exception, and I/O pressure), leading to significant time consumption in selecting appropriate recovery strategies. Consequently, it is imperative to devise an automated diagnosis tool to enhance operators' efficiency and reliability in failure mitigation.

	\begin{figure}[t]
		\centering
		
		\includegraphics[width=\linewidth]{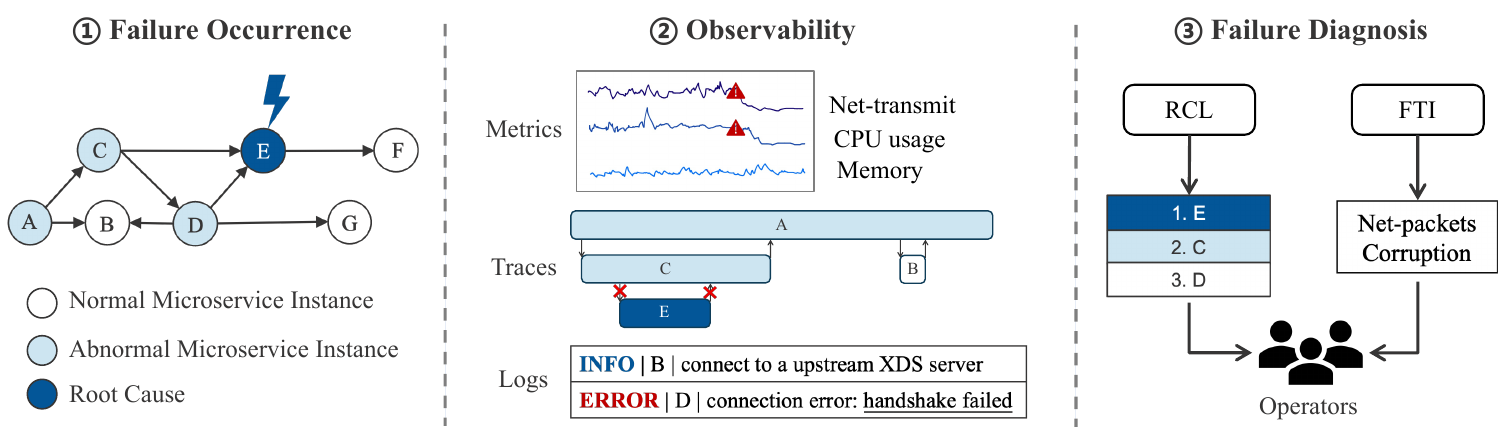}
		\caption{An example of failure diagnosis in a microservice-based system.  The network packets of microservice instance $E$ are damaged, causing a failure in the system. }
	\label{fig:process}
\end{figure}

\textbf{\underline{R}oot \underline{C}ause \underline{L}ocalization} (RCL) and \textbf{\underline{F}ailure \underline{T}ype \underline{I}dentification} (FTI) are two fundamental tasks in failure diagnosis for microservice-based systems.
RCL involves identifying the specific microservice instance responsible for the failure of a group of abnormal candidates. In contrast, FTI determines the type of the current failure, facilitating the selection of appropriate recovery measures. 
The growing observability of microservice-based systems allows for the utilization of multimodal monitoring data, including traces, logs, and metrics, to provide comprehensive insights into the status of microservice systems from various perspectives. More details of multimodal monitoring data can be found in Section \ref{sec:observability}.
Fig. \ref{fig:process} illustrates the two failure diagnosis tasks in a microservice-based system. Operators typically combine domain knowledge with the analysis of monitoring data, listing potential culprit microservices instances (i.e., RCL) and the failure type that may have triggered the incident (i.e., FTI).
Current approaches primarily concentrate on diagnosing failures using single-modal monitoring data, 
which have progressed in specific diagnosis tasks. For example, log-based methods focus on mining event patterns and transforming them into rules to classify failure types \cite{pan2021faster,fu2014digging, du2017deeplog,amar2019mining, he2018identifying, rosenberg2020spectrum, bansal2020decaf, li2020swisslog}. Trace-based methods, aided by rich dependency information, build system correlation graphs and simulate the failure propagation to infer root causes \cite{li2021practical,yu2021tracerank,li2022microsketch,yu2021microrank, rios2022localizing, mi2013toward,zhou2019latent,guo2020graph}. Metric-based methods extract features of historical metrics to diagnose failures \cite{li2022actionable,he2022graph, wu2021microdiag,meng2020localizing,zhang2021aamr,lin2018microscope,ma2020automap,chen2016causeinfer,liu2021microhecl,wu2020microrca,ma2020self,wu2021identifying,shan2019diagnosis}. 

Despite the success of single-modal failure diagnosis methods in many scenarios, some specific failures are difficult to identify using a single modality due to the limited information each modality provides \cite{yu2023nezha}. For instance, trace-based methods fall short in diagnosing hardware failures in the absence of fine-grained machine metrics \cite{lee2023eadro}. As an example, we can find the error call (i.e., $C \to E$) based on the traces in Fig. \ref{fig:process}, but we cannot ascertain the failure type (i.e., Net-packets Corruption) solely based on the provided traces. This is because many failure types can cause error calls, such as unavailable service, server-side exception, or Net-packets Corruption. Meanwhile, metrics and logs can furnish macroscopic clues (i.e., sudden drop in net transmission and handshake failed) about this failure type. Although metrics and logs encapsulate more detailed information, they are distributed on different microservice instances and provide only local information about a microservice or a host, necessitating additional expertise to establish causal relationships for the entire system.


Given the limitations of single-modal failure diagnosis, several multimodal failure diagnosis methods based on deep learning have been proposed to integrate diverse data sources for different diagnosis tasks. DiagFusion \cite{zhang2023robust} adopts an early fusion approach to convert multimodal data into event features uniformly during data processing. 
Eadro \cite{lee2023eadro} intermediately fuses the high-dimensional features of each modality during model learning.
Compared to single-modal failure diagnosis methods, these multimodal methods achieve significant performance improvements by combining complementary information from multiple views. 
Nevertheless, direct fusion is insufficient for extracting shared information among modalities, such as the system status and the abnormal microservice set.
Furthermore, these methods often ignore the relationship between diagnosis tasks and modalities. In reality, the extent of contribution from various modalities tends to differ across specific diagnosis tasks. 
For instance, in RCL, operators typically need to construct the propagation path of a failure in a microservice-based system and then trace it back to locate root causes. This process involves building a correlation graph between microservices based on traces. In contrast, for FTI, log messages contain extensive descriptions of failure behaviors, such as \textit{handshake failed} in Fig. \ref{fig:process}, which effectively contribute to FTI.
Therefore, tailoring diagnosis to task preferences can maximize the contribution of each modality.

This paper presents \textit{TVDiag}, a multimodal failure diagnosis framework designed to locate root causes and identify failure types in microservice-based systems. For each failure, we apply alert detection for multimodal data of each microservice instance and then aggregate this alert information using a graph neural network. We then perform multi-task learning for RCL and FTI, leveraging the common knowledge to enhance the learning of each task. In this phase, we integrate three components with prior knowledge to guide \textit{TVDiag} in leveraging the advantages of each modality and capturing shared information between them. The three components are described as follows:
\begin{enumerate}
	\item \textbf{Task-oriented learning}: Considering the preference of a task for specific modalities, we introduce a novel task-oriented feature learning method to enhance the potential contribution of each modality to corresponding tasks. 
	\item \textbf{Cross-modal association}: As the multimodal data of a failure can be considered representations from distinct perspectives, some view-invariant information for the failure is shared across all modalities. We design a cross-modal association method based on contrastive learning \cite{chen2020simple} to capture this view-invariance information.
	\item \textbf{Graph augmentation}: To mitigate the issue of insufficient labeled data and simulate failures with imperfect observability, we develop a graph-based data augmentation strategy by randomly inactivating non-root cause instances.
\end{enumerate}

We rigorously evaluated \textit{TVDiag} through comprehensive experiments utilizing three multimodal monitoring datasets, comprising two open-source datasets and one from an online benchmark. Remarkably, \textit{TVDiag} outperforms the state-of-the-art methods by 32.46$\%$ in $HR$@1 accuracy for root cause localization and improves F1-score by 3.08$\%$ for failure type identification. The main contributions are summarized as follows:

\begin{itemize}
	
	\item We propose \textit{TVDiag}, a multimodal failure diagnosis framework that locates root causes and identifies failure types in microservice-based systems. \textit{TVDiag} integrates a task-oriented learning method to amplify the potential advantages of specific modalities in failure diagnosis. Furthermore, \textit{TVDiag} extracts view-invariant failure information shared across multimodal data based on cross-modal associations.
	
	
	
	\item We introduce a graph-based data augmentation strategy that involves random inactivation of non-root cause instances, thereby alleviating the problem of insufficient training data. This strategy can be easily integrated into existing frameworks without altering the primary network structure.
	
	\item We evaluate our framework using three multimodal datasets, achieving substantial improvement across various metrics. The source code and experimental data are publicly available \cite{TVDiag}.

\end{itemize}

The rest of the paper is organized as follows. In Section \ref{sec:background}, we provide an overview of the basic preliminaries in microservice-based systems and define the problem to be addressed in the paper. Section \ref{sec:motivation} delineates the motivation for using multimodal data. Furthermore, we offer insights into the preferences and view-invariance characteristics exhibited within multimodal data. In Section \ref{sec:approach}, we present a comprehensive description of our system. Section \ref{sec:evaluation} details the evaluation metrics and presents the experimental results obtained. Section \ref{sec:related-work} offers the related work about failure diagnosis. The conclusion and future work are discussed in Section \ref{sec:conclusion}.


\begin{figure}[t]
	\begin{minipage}
		{0.52\linewidth}
		\flushleft
		\includegraphics[width = 0.9\linewidth]{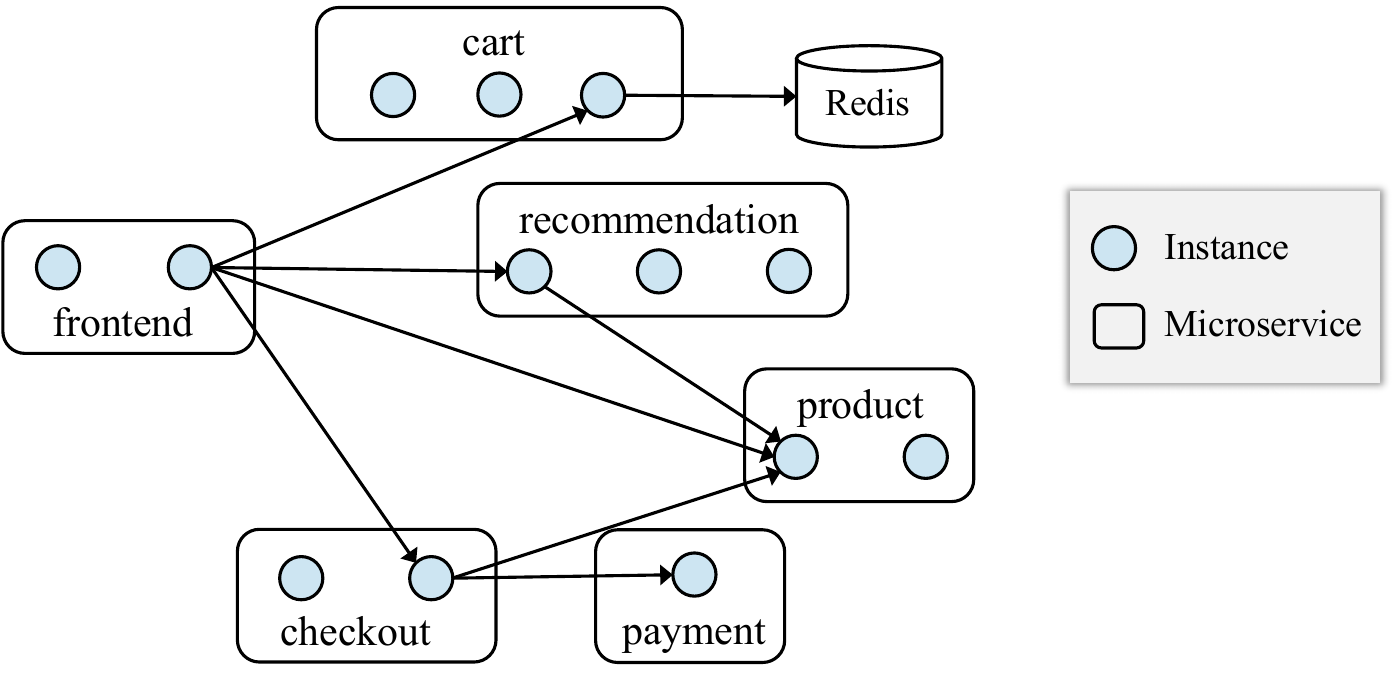}
		\caption{Part of correlation graph in Online Boutique \cite{OnlineBoutique}.}
		\label{fig:hipster-example}
	\end{minipage}
	\begin{minipage}{0.4\linewidth}
		\flushright
		\includegraphics[width = \linewidth]{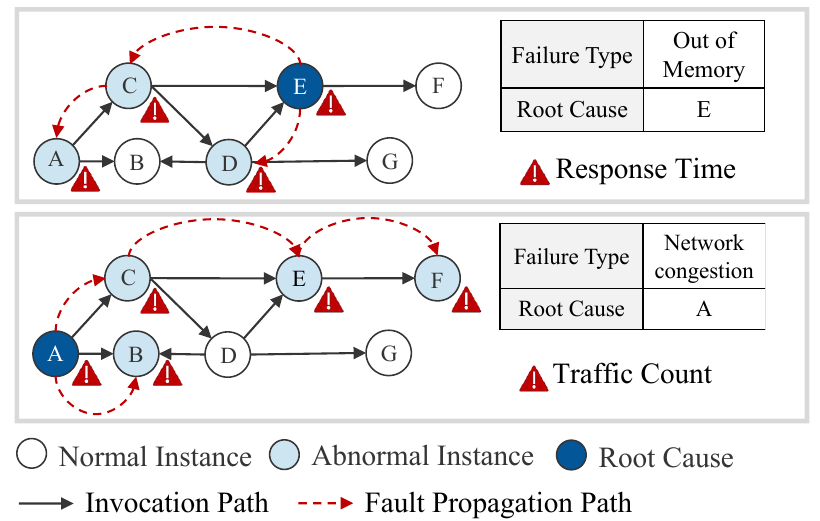}
		\caption{Two examples of failure propagation.}
		\label{fig:failure-propagation}
	\end{minipage}
	\vspace{-4mm}
\end{figure}

\section{Background}
\label{sec:background}

\subsection{Microservice Architecture and Failure}
\label{sec:MI-failure}

\textbf{Microservice and Instance}. 
Microservice architecture aims to achieve lower coupling and faster delivery by breaking down a monolithic application into several independent microservices, each handling specific business tasks autonomously and communicating with one another \cite{xie2024pbscaler, li2022enjoy, peng2022trace}. To handle high volumes of incoming requests, each microservice typically runs a group of functionally identical instances, which act as workers processing specific business tasks. Fig. \ref{fig:hipster-example} illustrates a partial internal architecture of Online Boutique \cite{OnlineBoutique}, a microservice-based e-commerce system developed by Google, highlighting the relationship between microservices and instance affiliation. When a user initiates a request to purchase a product in this system, the request will be received by a specific instance of the \textit{frontend} microservice and then forwarded to instances belonging to other microservices. Although instances affiliated with a microservice undertake the same business task, their performance can vary significantly due to discrepancies in deployment environments and received traffic. Our focus is on locating the instance-level root cause, as identifying the microservice-level root cause can confuse operators due to the presence of numerous candidate instances.

\textbf{Failure}. A failure is characterized as the accumulation of multiple anomalies, which can lead to degraded system performance or even system unresponsiveness. 
According to MEPFL \cite{zhou2019latent}, failures are categorized into four broad categories: monolithic, multi-instance, configuration, and asynchronous interaction.
Typical examples of such failures include out-of-memory, network congestion, and process exits, etc.

In a microservice-based system, a single failure in an instance can propagate to other instances due to frequent and intricate communication between microservices. Unfortunately, the trajectories and directions of failure propagation are often diverse and unpredictable \cite{xie2023impacttracer}. Fig. \ref{fig:failure-propagation} depicts two failure propagation scenarios in microservice-based systems. In the upper part, an out-of-memory malfunction causes a suboptimal performance for $E$, subsequently affecting its upstream instances waiting for its response. In the bottom part of Fig. \ref{fig:failure-propagation}, network congestion significantly reduces the transmission of network packages from instance $A$, leading to anomalous traffic counts in downstream instances. Both situations give rise to numerous anomalous instances and suspicious propagation trajectories. Although experienced operators can diagnose noticeable failures in one instance, failure propagation between instances complicates this process and impedes quick system recovery. 
Li et al. \cite{li2022actionable} also observed that most failures in online systems are recurring, meaning they repeatedly occur in different locations within microservice-based systems. Therefore, summarizing historical experiences and extracting failure fingerprints is essential for accurately diagnosing failures and implementing appropriate recovery measures. 

\subsection{Observability}
\label{sec:observability}
\begin{figure}
	\centering
	\includegraphics[width=0.8\linewidth]{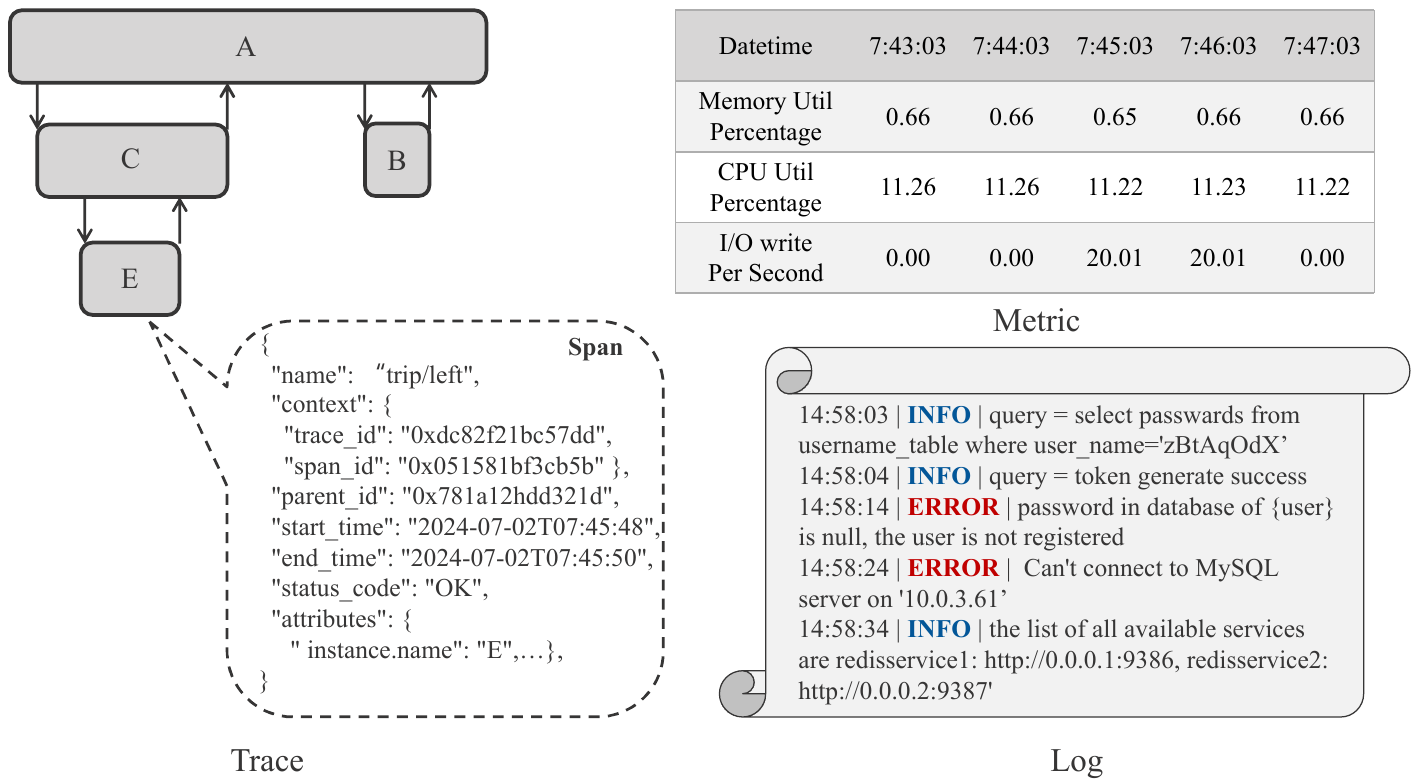}
	\caption{Examples of data structures for three types of monitoring data.}
	\label{fig:background-observarbility}
	\vspace{-4mm}
\end{figure}

Observability refers to the ability to measure the internal state of a system through monitoring data \cite{Jay2023What}. As illustrated in Section \ref{sec:introduction}, the monitoring data is typically composed of three types: metrics, logs, and traces. These multimodal data have specific data structures and provide insights into microservice-based systems from different perspectives.
\begin{itemize}[left=0pt]
	\item[---] \textbf{Traces}. Traces capture the execution path of each user request and record the information of all instances on that pathway. As shown in Fig. \ref{fig:background-observarbility}, a trace can be regarded as a tree, where each node records information related to an instance. A node is also called a span. Each span contains context (e.g., trace ID and span ID) and attributes (e.g., service name and instance name). Moreover, spans record the status of instances (e.g., start time, end time, and status code). For the sake of simplicity, we use the instance name to represent each span in all figures. 
	\item[---] \textbf{Metrics}. Metrics record the trends of system status and business performance in the form of time series. Generally, metrics are collected at fixed intervals and stored in numerical form. Fig. \ref{fig:background-observarbility} lists three metrics related to system resource usage,  recorded at one-minute intervals. 
	\item[---] \textbf{Logs}. Logs document events at different levels for each instance in a semi-structured text format. Logs typically consist of three elements: timestamps, log levels (e.g., INFO, WARN, and ERROR), and messages. Fig. \ref{fig:background-observarbility} depicts three types of INFO-level logs and two types of ERROR-level logs. We can initially judge the severity of an event based on its log level and analyze the message to further understand the event's context and consequences. For example, the message of the first ERROR-level log indicates that a nonexistent user was queried in the database, while the second ERROR-level log relates to a MySQL connection failure. As such, logs, especially ERROR-level logs, have superiority in determining failure types.
\end{itemize}
Despite the usefulness of these modalities, operators face considerable challenges in diagnosing failures because of the heterogeneity of these data sources and the interference of large volumes of irrelevant data \cite{Stephen2022What}.


\subsection{Problem Statement}
Our work aims to pinpoint the root cause and classify the failure type based on multimodal monitoring data once a failure is detected. Hence, we must address two interconnected problems: \textbf{Root Cause Localization} (RCL) and \textbf{Failure Type Identification} (FTI).
Failure diagnosis can assist operators in maintaining the reliability of the online system. As depicted in Fig. \ref{fig:process}, RCL results in a ranking list $\{E,C,D,\cdots\}$ in descending order, enabling operators to identify the root cause instance $E$ within a narrowed scope. Regarding FTI, we treat this task as a classification problem and categorize the current failure into a specific type, such as out-of-memory, facilitating the development of appropriate recovery measures (e.g., allocating more memory to instance $E$).

Given the captured traces $\mathcal{T}$, system metrics $\mathcal{M}$, and collected logs $\mathcal{L}$ of all instances over a period of time, we are committed to designing an automated diagnostic tool $\mathcal{D}$ that identifies the underlying failure type $t$ and ranks the culprit instances $\mathcal{R}$. We formulate our diagnosis process as:
\begin{equation}
	\mathcal{D}\left(\mathcal{M}, \mathcal{T}, \mathcal{L},  \Theta \right) \to t, \mathcal{R},
\end{equation}
where $\Theta$ represents the learnable parameters of $\mathcal{D}$. 

\section{Motivation}
\label{sec:motivation}
To gain a deeper understanding of the impact of various monitoring data on failure diagnosis, we deployed Online Boutique \cite{OnlineBoutique} and deliberately injected multiple typical failures into its environment. Traces were collected by the Jaeger platform\footnote{https://www.jaegertracing.io/}, metrics were gathered using Prometheus\footnote{https://prometheus.io/}, and logs were sourced from the corresponding file system.

\subsection{The Necessity of Using Multimodal Data}
\begin{figure}[t]
	\centering
	\includegraphics[width=\linewidth]{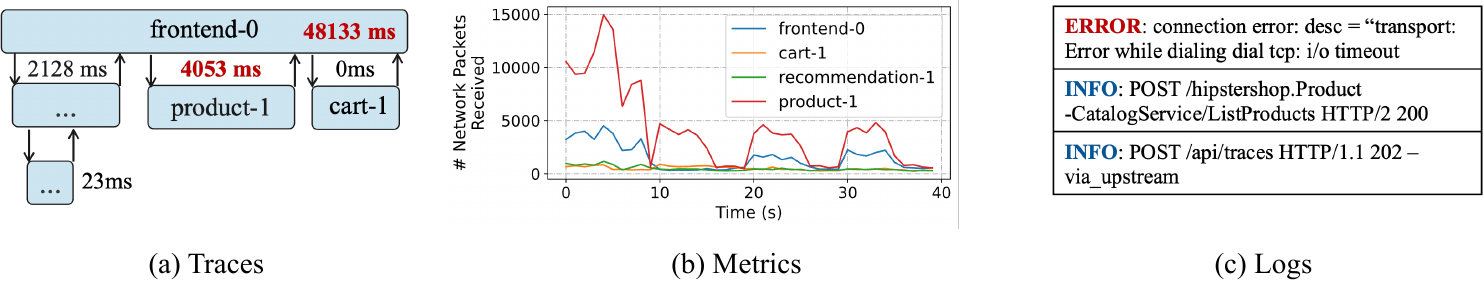}
	\vspace{-4mm}
	\caption{A real case of multimodal monitoring data.}
	\label{fig:Observability}
	\vspace{-4mm}
\end{figure}

Failure types in microservice-based systems are diverse and complex, ranging from insufficient external hardware resources to defects in the execution logic of a microservice. As mentioned in Section \ref{sec:introduction}, any single-modal diagnosis method can hardly cover all failure scenarios due to the limited information offered by a single modality. To illustrate this point, we intentionally dropped some network packets received by one instance (named \textit{product-}1) of the \textit{product} microservice in Fig. \ref{fig:hipster-example}. In this scenario, \textit{product-}1 is identified as the root cause, while Net-packets-loss represents the corresponding failure type.
Fig. \ref{fig:Observability} shows real samples of multimodal monitoring data captured within a specific time window. 

An operator might pinpoint the culprit instance \textit{product-}1 by analyzing captured traces. However, determining the failure type (Net-packets-loss) would be challenging due to insufficient details on network packets in traces. To further identify the failure type, the operator would need to check for keywords such as "tcp: i/o timeout" in the logs. If the logs lack relevant information,  the operator would have to observe and analyze various metrics. Finally, the operator would confirm the failure type as Net-packets-loss due to a significant decrease in the number of network packets received by \textit{product-}1. To sum up, it is plausible to integrate the superiority of three modal information for failure diagnosis.

\subsection{Preferences and View-invariance in Multimodal Data}
\label{sec:motivation-2}

During our empirical diagnosis of failures using multimodal monitoring data, we observed intriguing phenomena and patterns that provide valuable insights for our design.

Firstly, we noticed a significant disparity in the contribution of each modality to different diagnosis tasks. Fig. \ref{fig:Observability} shows a motivating example concerning Net-packets-loss failure in a microservice-based system. 
As mentioned in Section \ref{sec:background}, messages in ERROR-level logs often record information about failure types, indicating a preference for the log modality in the FTI task. In this example, the "tcp:i/o timeout" in the ERROR-level log suggests a network-related failure type.
In addition, by analyzing the execution path and instance status displayed in the trace, we can deduce an abnormal invocation pair (i.e., \textit{frontend}-0$\to$\textit{product-}1), which indicates that the root cause lies in the downstream instance \textit{product-}1. Metrics show a significant downward trend in the network packets received by \textit{product-}1. Coupled with the "tcp:i/o timeout" provided by the logs, we can infer that the failure type is Net-packets-loss rather than reduced workload. This indicates that metrics and traces are crucial for the RCL task, while logs and metrics are important for identifying failure types. These observations suggest that different tasks exhibit biases towards certain modalities, necessitating unequal attention to each modality in specific diagnostic tasks.

Secondly, multimodal monitoring data share some view-invariant information when a failure occurs. Although each modality offers distinct perspectives on system status, certain shared information is shared across multiple modalities. We refer to this shared information as view-invariant information. As shown in Fig. \ref{fig:Observability}, we can independently infer the system status (i.e., "abnormal") through abnormal latency in traces, irregular metric fluctuations, or ERROR-level logs. We can also deduce the abnormal phenomenon (i.e., "performance degradation") based on the high latency of traces or "timeout" keywords in logs. Besides, normal instances (e.g., "\textit{cart-1"}) are identifiable in traces and metrics. The exploration and amplification of this view-invariant information facilitate the construction of cross-modal associations, contributing to the reduction of the scope of failure diagnosis.

\begin{figure*}[t]
	\centering
	\includegraphics[width=\textwidth]{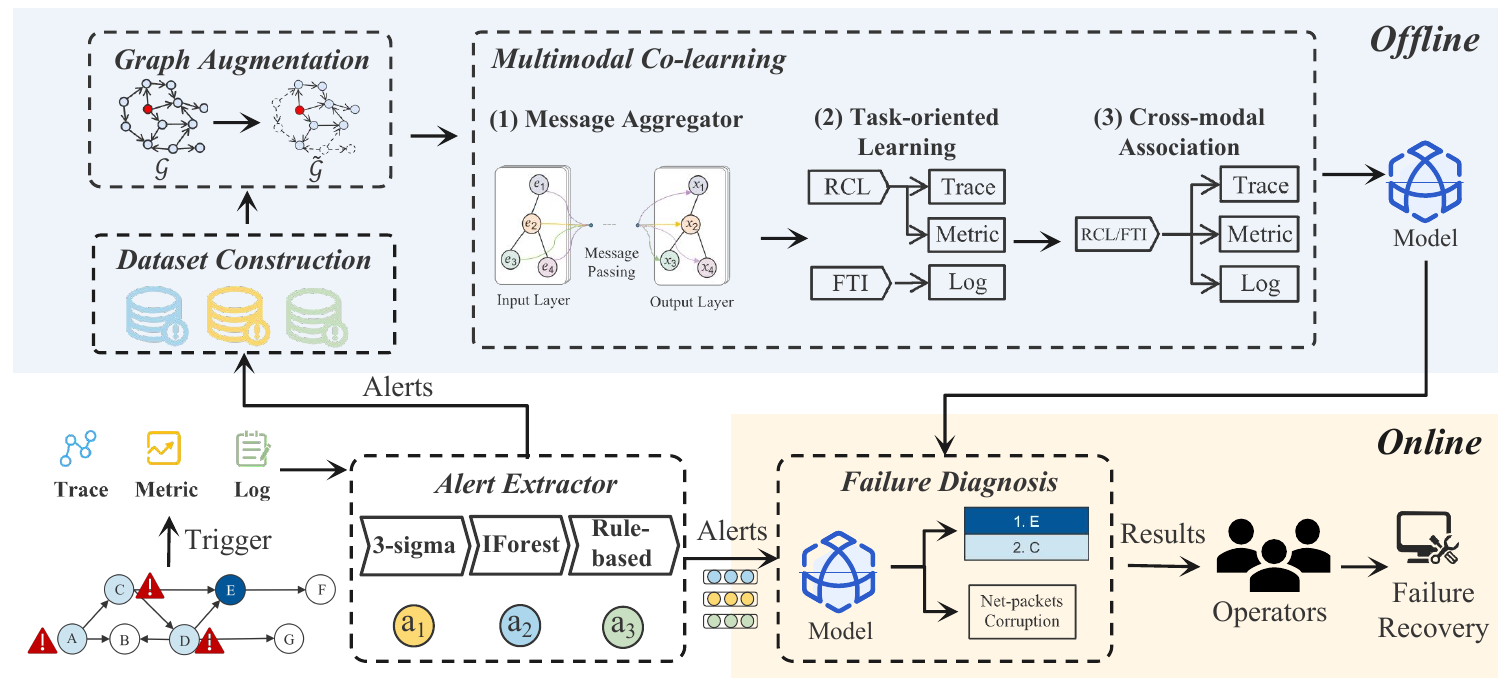}
	\vspace{-4mm}
	\caption{Overview of \textit{TVDiag}.}
	\label{fig:structure}
	\vspace{-4mm}
\end{figure*}
\vspace{-2mm}

\section{Approach}
\label{sec:approach}
Fig. \ref{fig:structure} depicts the overall structure of our proposed multimodal failure diagnosis framework, \textit{TVDiag}, encompassing an alert extractor, an offline training phase, and an online diagnosis phase. 
In the \textbf{Alert Extractor} component, we extract alerts from the historical multimodal data for offline training and detect anomalies for online
diagnosis (\cref{sec:AE}).
During the offline training phase, we transform the alerts into learnable features and build an alert dataset.  \textit{TVDiag} then trains a diagnosis model based on this pre-constructed multimodal dataset. In the online diagnosis phase, alerts are extracted in real time from multimodal data, and the diagnosis model is utilized to predict root causes and failure types.

The offline training process is composed of three components: (1) \textbf{Dataset Construction}: Firstly, \textit{TVDiag} builds a correlation graph between microservice instances based on traces. We then represent extracted alerts by high-dimensional features and bind them to affiliated instances, creating a unified alert dataset prepared for training (\cref{sec:DP}). (2) \textbf{Graph Augmentation}: To enrich the training alert dataset and enhance the generalizability of the diagnosis model, we design a graph-based data augmentation strategy by randomly inactivating non-root cause microservice instances (\cref{sec:AUG}). (3) \textbf{Multimodal Co-learning}: Considering the preference of diagnosis tasks for specific modalities, we propose a task-oriented learning method to amplify modality-specific advantages. Furthermore, we establish cross-modal associations using contrastive learning to extract view-invariant failure information (\cref{sec:MC}). 

The online diagnosis part reuses the alert extractor to collect alerts from multimodal monitoring data. Next, \textit{TVDiag} transforms these alerts into learnable features which the diagnosis model can understand. In the \textbf{Failure Diagnosis} module, our objective is to fuse multimodal alert features and then employ the diagnosis model to localize root causes and determine failure types (\cref{sec:failure-diagnosis}).


\begin{figure}[t]
	\begin{minipage}
		{0.48\linewidth}
		\flushleft
		\small
		\captionof{table}{Templates and examples of extracted alerts.}
		\label{tab:alert-template}
		\begin{tabular}{l|l|l}
			
			\toprule
			\textbf{Modality} & \textbf{Alert Template} & \textbf{Example} \\ \midrule
			Metric & \begin{tabular}[c]{@{}l@{}}( \textit{reporterId}, \\ \textit{metricName},\\ \textit{abnormalDirection})\end{tabular} & \begin{tabular}[c]{@{}l@{}}("\textit{product-}1",  \\ "networkReceiveMB", \\ "down")\end{tabular} \\ \hline
			Trace & \begin{tabular}[c]{@{}l@{}}( \textit{reporterId}, \\ \textit{parentId}, \\ \textit{operationName}, \\ \textit{abnormalType})\end{tabular} & \begin{tabular}[c]{@{}l@{}}("\textit{product-}1", \\ "\textit{frontend}-0", \\ "GetProduct", \\ "500")\end{tabular} \\ \hline
			Log & \begin{tabular}[c]{@{}l@{}}(  \textit{reporterId}, \\ \textit{logKey})\end{tabular} & \begin{tabular}[c]{@{}l@{}}( "\textit{product-}1", \\ "13" )\end{tabular} \\ \bottomrule
		\end{tabular}
	\end{minipage}
	\begin{minipage}{0.51\linewidth}
		\flushright
		\includegraphics[width = \linewidth]{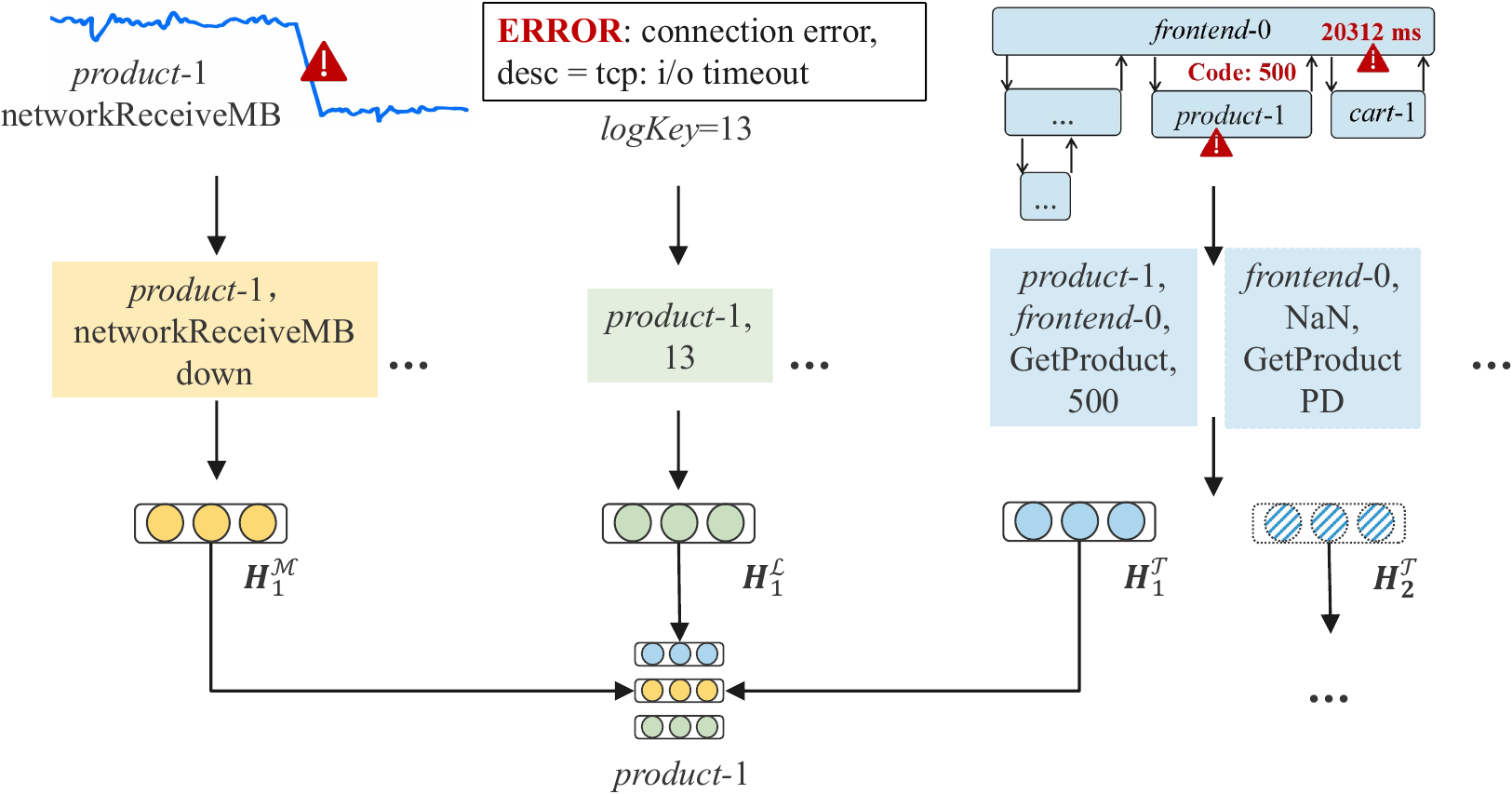}
		\vspace{-4mm}
		\caption{Illustration of the alert extractor and dataset construction. Three alert features $(\textbf{H}_1^\mathcal{M},\textbf{H}_1^\mathcal{T},\textbf{H}_1^\mathcal{L})$ are grouped by their \textit{reporterId} (i.e., \textit{product}-1).}
		\label{fig:alert-process}
	\end{minipage}
\end{figure}

\subsection{Alert Extractor}
\label{sec:AE}
This module abstracts valuable alert information from multimodal monitoring data, formulating them with custom templates. The variation in the collection interval and the heterogeneity in data structure pose challenges to the alignment of multimodal data. Inspired by DiagFusion \cite{zhang2023robust}, \textit{TVDiag} separately converts multimodal heterogeneous monitoring data into unified alerts. These alerts document anomalies from multimodal data using fixed text templates, facilitating straightforward encoding into learnable semantic features. More specifically, \textit{TVDiag} detects alerts from metrics, logs, and traces across all microservice instances. Table \ref{tab:alert-template} presents the templates and examples of alerts for each modality. The common element shared among all templates is defined as (\textit{reporterId}, $\ast$), where \textit{reporterId} represents the unique ID of the microservice instance reporting the alert, and $\ast$ records the modality-specific information. 

(1) \textbf{Trace Alert}. We adopt the Isolation Forest (IForest) \cite{liu2008isolation} to detect anomalies and generate alerts based on the response time and status code of each invocation pair from traces. As a widely used anomaly detection method, IForest constructs decision trees by randomly isolating data points and identifies anomalies by observing the average path length required to isolate them. The extraction performance of IForest will be further discussed in section \ref{rq:2}.
Operators typically detect system failures through interactions, response times, and status codes in traces \cite{guo2020graph}. High response time often implies performance degradation, while abnormal status codes indicate business errors.
Here, $\ast$ is elaborated as (\textit{parentId}, \textit{operationName}, \textit{abnormalType}), capturing the caller of the reporter instance, the operation, and the current abnormal type. Note that we define the abnormal type here as either abnormal status codes or performance degradation (PD). For example, the alert ("\textit{product-}1", "\textit{frontend}-0", "GetProduct", "500") indicates that the status code for the operation (i.e., GetProduct) of the invocation pair (\textit{frontend}-0 $\to$ \textit{product-}1) is abnormal. Meanwhile, the alert ("\textit{product-}1", "\textit{frontend}-0", "GetProduct", "PD") signifies that the invocation pair has a notably high latency.

(2) \textbf{Metric Alert}. Due to the potential generation of thousands of metrics in a microservice system, training a separate IForest model for each metric incurs significant time overhead. Consequently, we opt for the 3-sigma rule \cite{pukelsheim1994three}, which, while slightly less performant, offers rapid processing speed as an alternative. For a given metric (e.g., networkReceiveMB) under inspection, we collect the numerical fluctuations $\left[m_1, m_2,\cdots,m_n\right]$ over a time period. Subsequently, we compute the mean $\mu$ and standard variance $\sigma$ for these fluctuations. If $m_n$ exceeds $\mu+3\sigma$ (or is less than $\mu-3\sigma$), it is deemed an alert with the "up" (or 'down') direction in this metric. The metric name and the abnormal direction are then recorded as an alert.

\begin{figure}[t]
	\begin{minipage}{0.46\linewidth}
		\flushleft
		\includegraphics[width = 0.9\linewidth]{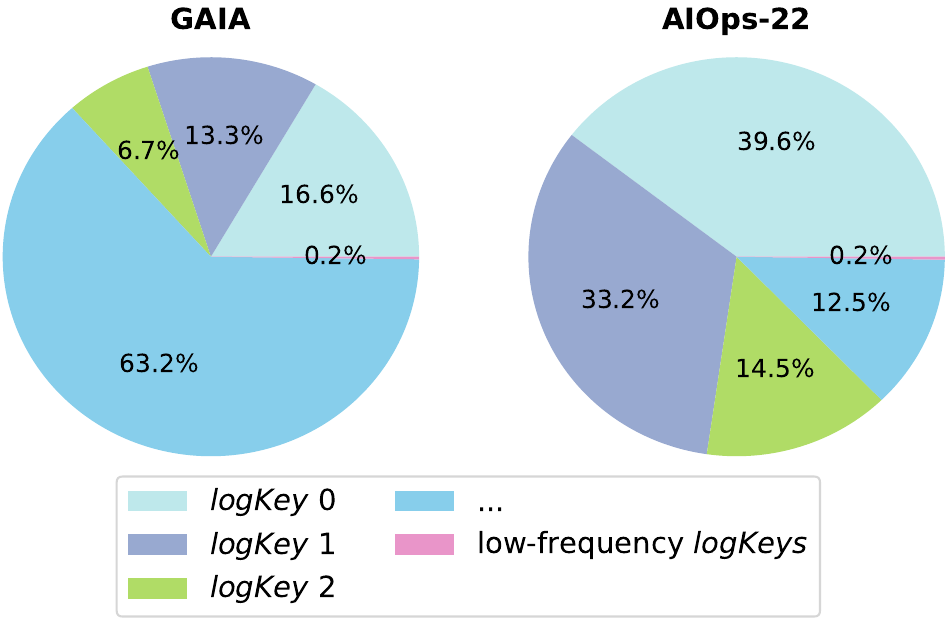}
		\caption{Log volumn distribution of \textit{logKeys}. The top 3 high-frequency \textit{logKeys} (i.e., \textit{logKeys} 0, 1, 2) account for 36.6\% to 87.3\% logs on two datasets, while low-frequency \textit{logKeys} occupy only 0.2\% logs.}
		\label{fig:logkey-long-tail}
	\end{minipage}
	\begin{minipage}
		{0.53\linewidth}
		\flushleft
		\captionof{table}{Distribution of valuable and ERROR-level \textit{logKeys} across two datasets. S1, S2, and S3 are experienced students responsible for identifying and labeling valuable \textit{logKeys} in failure diagnosis.}
		\label{tab:user-study}
		\resizebox{\linewidth}{!}{
			\begin{tabular}{lllccccc}
				
				\toprule
				\multirow{2}{*}{Dataset} & \multirow{2}{*}{LogKey} & \multirow{2}{*}{Category} & \multicolumn{5}{c}{Distribution (\%)} \\
				\cline{4-8}
				&  &  &  & S1 & S2 & S3 & P-value\\
				\midrule
				\multirow{4}{*}{$\mathcal{A}$} & \multirow{3}{*}{Valuable} & Low-frequency & - & 71.43 & 55.56 & 77.78 &8.7e-3\\
				&  & High-frequency & - & 28.57 & 44.44 & 22.22 & -\\
				&  & ERROR-level & - & 100.00 & 55.56 & 100.00 & 2.6e-2\\
				
				\cline{2-3}
				& \multirow{2}{*}{ERROR-level} & Low-frequency & 77.78 & - & - & - &-\\
				&  & High-frequency & 22.22 & - & - & - &-\\
				\midrule
				\multirow{4}{*}{$\mathcal{B}$} & \multirow{3}{*}{Valuable} & Low-frequency & - & 100.00 & 66.67 & 100.00 &3.9e-3\\
				&  & High-frequency & - & 0.00 & 33.33 & 0.00&- \\
				&  & ERROR-level & - & 100.00 & 66.67 & 100.00&3.9e-3 \\
				
				\cline{2-3}
				& \multirow{2}{*}{ERROR-level} & Low-frequency & 100.00 & - & - & - &-\\
				&  & High-frequency & 0.00 & - & - & -&-\\
				\bottomrule
		\end{tabular}}
	\end{minipage}
	\vspace{-4mm}
\end{figure}

(3) \textbf{Log Alert}. As depicted in \cite{du2017deeplog, du2018spell, he2017drain}, a semi-structured log message can be parsed into a fixed part (i.e., \textit{logKey}) and a variable part. For example, the \textit{logKey} for the log "upload business logs on \underline{2021-08-15} failed" is "upload business logs on $\#$ failed." \textit{TVDiag} parses logs using Drain \cite{he2017drain} and matches each log with a \textit{logKey}. Because some \textit{logKeys} are typically associated with specific failures, we detect these valuable \textit{logKeys} as log alerts. However, manually assessing the value of all \textit{logKeys} across different systems is labor-intensive and time-consuming. Furthermore, failure-unrelated and common \textit{logKeys} should be avoided, as they can introduce unnecessary noise or even mislead the model. Therefore, an automated detector is needed to identify more valuable \textit{logKeys} while filtering out noise that is irrelevant to failures.

We argue that \textit{ERROR-level and low-frequency \textit{logKeys} are more helpful in failure diagnosis}.
Fig. \ref{fig:logkey-long-tail} shows the distribution of the top 3 high-frequency \textit{logKeys} and low-frequency \textit{logKeys} on two open-source datasets, that is, GAIA \cite{GAIA} and AIOps-22 \cite{AIOps}, which will be detailed in \cref{sec:exp-design}. We sort all \textit{logKeys} in descending order by occurrence frequency and select the latter half as low-frequency \textit{logKeys}. We observe that low-frequency \textit{logKeys} only occupy 0.2\% of total logs on two datasets. To further verify our assumption, we invited three Ph.D. students with over five years of experience in software development and maintenance to label the valuable \textit{logKeys} that are helpful for failure diagnosis. Note that they make judgments based solely on the value of the information within the \textit{logKeys}, without perceiving their frequencies. Based on the annotations provided by the three Ph.D. students, as shown in Table \ref{tab:user-study}, we conducted T-tests to determine whether the proportion of low-frequency or ERROR-level \textit{logKeys} among valuable entries is significantly higher than that of high-frequency \textit{logKeys}. For low-frequency \textit{logKeys} and ERROR-level \textit{logKeys}, the P-values obtained from the two datasets are below the significance threshold of 0.05, indicating that the majority of valuable \textit{logKeys} are low-frequency or ERROR-level, thereby confirming our hypothesis.
Besides, we observe that ERROR-level \textit{logKeys} are typically low-frequency because failures are not common in online systems.

In terms of log alert production,
DiagFusion \cite{zhang2023robust} randomly samples a batch of \textit{logKeys} as alerts. However, several high-frequency \textit{logKeys} occupy the majority of total logs \cite{yu2023logreducer}. Besides, these high-frequency \textit{logKeys} contain numerous redundant information that is useless for failure diagnosis, such as records of common queries. Random sampling risks preserving this redundancy and overlooking low-frequency \textit{logKeys}, including many valuable \textit{logKeys} which are generally rare and helpful for failure diagnosis. To alleviate this problem, \textit{TVDiag} employs two rules for alert generation:
\begin{itemize}[left=0pt]
	\item \textbf{Rule 1}: All ERROR-level \textit{logKeys} are treated as alerts. ERROR-level \textit{logKeys} can be identified by searching the predefined keywords.
	\item \textbf{Rule 2}: The top-$k$ \textit{logKeys} with the lowest occurrence frequency in history are treated as alerts. Note that $k$ (with a default value of 0.5) is a hyper-parameter that can be adjusted by operators according to their preferences.
\end{itemize}
Using these two rules, \textit{TVDiag} retains ERROR-level and low-frequency \textit{logKeys} as alerts to efficiently preserve more valuable \textit{logKeys}, while eliminating the redundancy of
high-frequency \textit{logKeys}.

\subsection{Dataset Construction}
\label{sec:DP}
For each historical failure, we extract alerts from three modalities and analysis traces to build a correlation graph $\mathcal{G}=\left \langle \mathcal{V},\mathcal{E} \right \rangle$, where $\mathcal{V}$ is a set of microservice instances, and $\mathcal{E}$ is a set of invocation pairs from traces. Note that the edges in $\mathcal{E}$ are bidirectional, covering all possible directions of failure propagation mentioned in Section \ref{sec:MI-failure}. Each alert is assigned to the corresponding microservice instance $v$ ($v \in \mathcal{V}$) based on the \textit{reporterId}. For each microservice instance $v$, DiagFusion regards alerts of three modalities as a sentence, with each alert considered a word, and utilizes fastText \cite{bojanowski2017enriching}, a word embedding technology widely used in natural language processing, to encode the alerts into a learnable feature. However, prematurely merging the alerts of three modalities into one feature is not conducive to extracting the high-dimensional information unique to each modality. In contrast, for each instance $v$, \textit{TVDiag} transforms alerts of different modalities into a tuple of features, denoted as $\textbf{H}_v = \left(\textbf{H}_v^\mathcal{M},\textbf{H}_v^\mathcal{T},\textbf{H}_v^\mathcal{L}\right)$, where $\textbf{H}_v^\mathcal{M}$, $\textbf{H}_v^\mathcal{T}$, and $\textbf{H}_v^\mathcal{L}$ denote the features of metrics, traces, and logs in microservice instance $v$, respectively. As shown in Fig. \ref{fig:alert-process}, \textit{TVDiag} transforms multimodal alerts into learnable features, and then group them by their \textit{reporterId}.  We combine the correlation graph $\mathcal{G}$ and the corresponding alert features of all historical failures into the alert dataset.


\begin{figure}[t]
	\centering
	\includegraphics[width=0.8\linewidth]{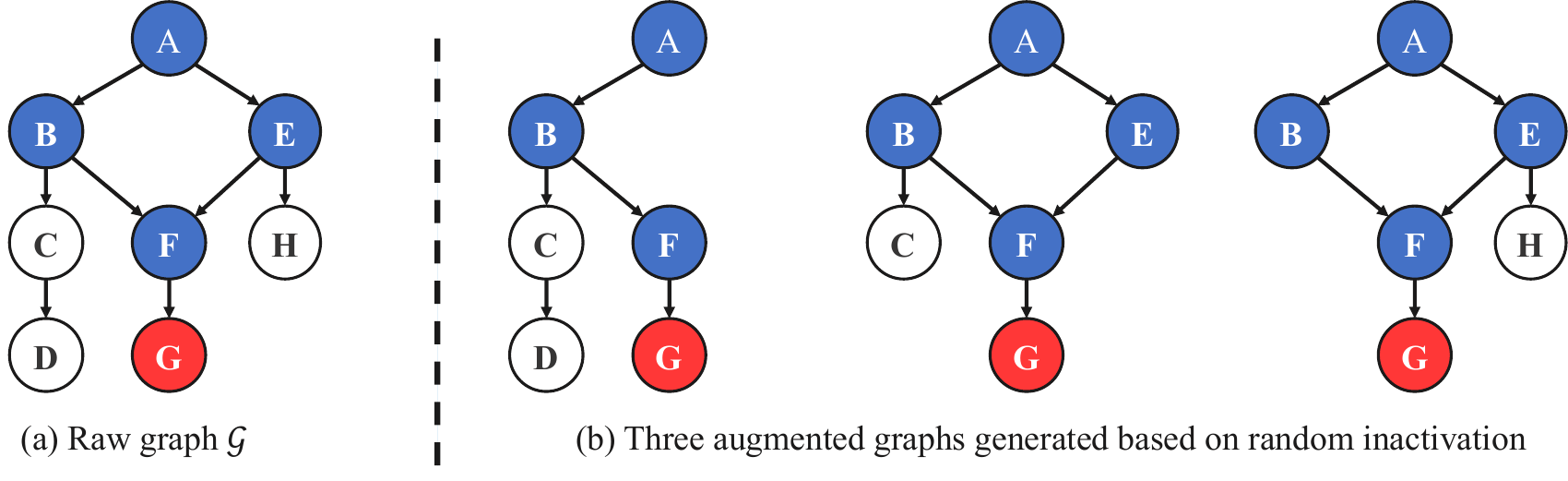}
	\caption{Example of graph augmentation. The red node is the root cause of the current failure, the blue nodes are abnormal instances, and the blank nodes denote normal instances. All augmented individuals are essentially subgraphs derived from the raw graph.}
	\label{fig:aug}
	\vspace{-4mm}
\end{figure}

\subsection{Graph Augmentation}
\label{sec:AUG}
Manually scrutinizing a large volume of monitoring data and subsequently labeling failures is extremely arduous.
To mitigate the problem of inadequate labeled data and enhance the generalizability of our framework, we implement a graph-based data augmentation strategy that involves random inactivation for non-root cause microservice instances. 
During the failure period, the microservice-based system consists of two types of microservice instances: root cause instances and non-root cause instances. While non-root cause instances can be further divided into abnormal instances and normal instances.
The rationale of our graph-based data augmentation strategy is that the lack of observability for some non-root cause microservice instances has little impact on the judgment of the root cause and failure type. 
Disabling the observability in normal microservice instances will not significantly affect failure diagnosis, as confirmed by existing anomaly-based failure diagnosis methods \cite{zhang2021aamr, lin2018microscope, wu2020microrca}. Meanwhile, the absence of observability in some abnormal non-root cause microservice instances does simulate pod-killing scenarios, enhancing \textit{TVDiag} to adapt to situations with incomplete observability.

Based on the above motivations, we can reasonably hypothesize that inactivating a small proportion of non-root cause nodes will not significantly alter the semantic information of the graph $\mathcal{G}$. Inspired by the node-dropping strategy proposed by GraphCL \cite{you2020graph}, we design a graph-based data augmentation method. As depicted in Fig. \ref{fig:aug}, for the instance correlation graph $\mathcal{G}$, \textit{TVDiag} randomly discards $m$ nodes along with their related edges to generate an augmented graph  $\mathcal{\tilde{G}}$. Here we calculate $m$ as:
\vspace{-2mm}
\begin{equation}
	m=\lfloor p \cdot \left| \mathcal{V} \right| \rfloor, 
\end{equation}
where $p$ denotes the inactivation probability and $\left| \mathcal{V} \right|$ is the number of microservice instances. We apply the graph augmentation strategy to each sample in the training alert dataset, generating a mutated instance correlation graph $\mathcal{\tilde{G}}$ corresponding to the original graph $\mathcal{G}$. Notably,  $\mathcal{\tilde{G}}$ and $\mathcal{G}$ share the same root cause and failure type. By incorporating $\mathcal{\tilde{G}}$ into the training dataset, we alleviate data scarcity issues and enhance the model performance.

\subsection{Multimodal Co-learning}
\label{sec:MC}
\begin{figure}
	\centering
	\includegraphics[width=0.9\linewidth]{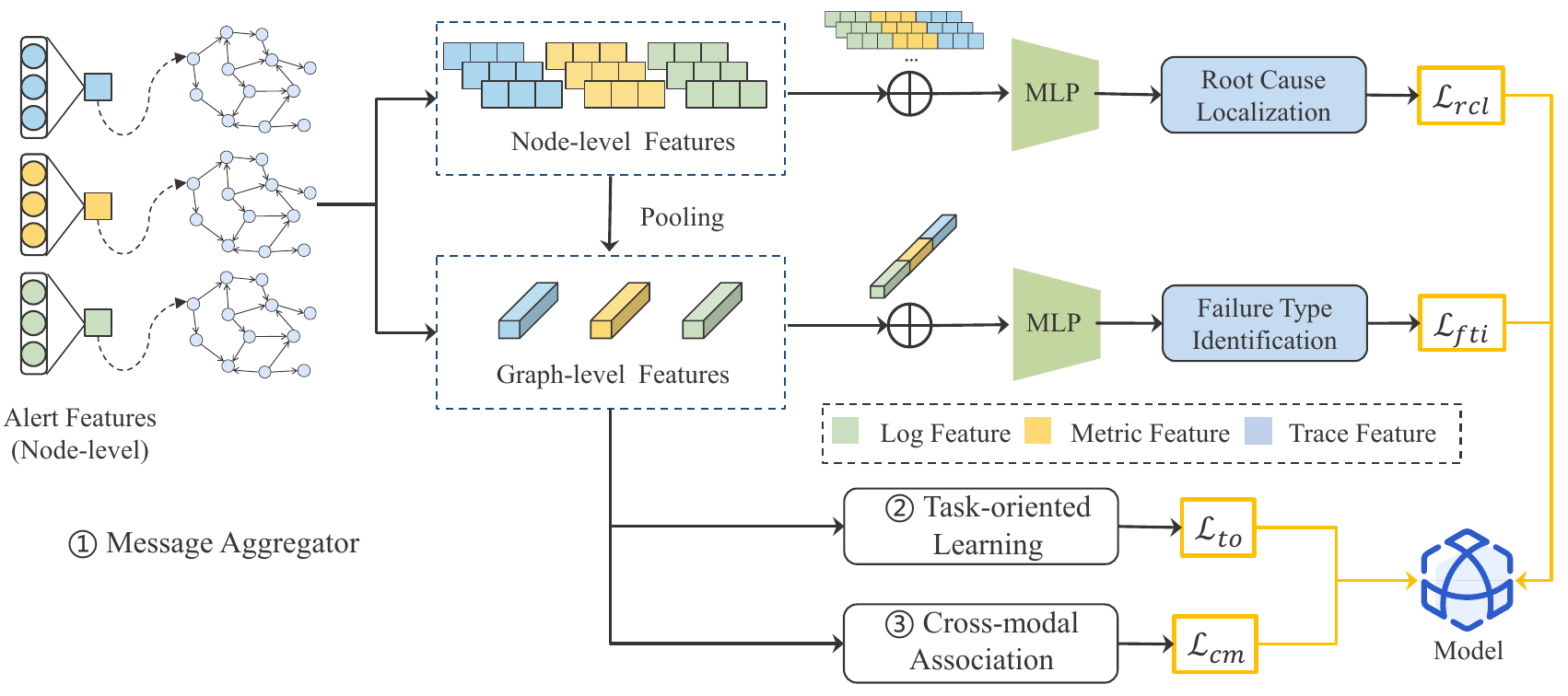}
	\caption{Training process of our proposed method. Specifically, the multimodal co-learning module (\cref{sec:MC}) first integrates the alert features for three modalities in the message aggregator stage, thereby acquiring multimodal node-level and graph-level features that describe the failure from different perspectives. Then, we conduct task-oriented learning and cross-modal association to refine the graph-level features, resulting in two losses $\mathcal{L}_{to}$ and $\mathcal{L}_{cm}$. In the RCL module, we concatenate the three multimodal features (node-level) for each node and score them individually. Regarding the FTI module, we concatenate the three graph-level features and use a multilayer perceptron (MLP) to infer the failure type. (\cref{sec:failure-diagnosis}).}
	\label{fig:multi-co-learning}
\end{figure}

\textit{TVDiag} shares and integrates knowledge among different modalities by co-learning multimodal features. Fig. \ref{fig:multi-co-learning} shows the overall training process of \textit{TVDiag}. Considering the failure propagation between microservices \cite{wu2020microrca, yu2021microrank}, \textit{TVDiag} performs message passing and feature aggregation to obtain graph-level features for each modality. In this process, we conduct task-oriented learning to enhance the advantages of specific modalities under corresponding tasks. Moreover, \textit{TVDiag} builds cross-modal associations based on contrastive learning for three modalities, enabling the proposed method to extract the invariant information for failures from different views.

\begin{algorithm}[t]
	\setstretch{1.25}
	\caption{Message Aggregator}
	\label{al:ma}
	\Input{Instance correlation graph $\mathcal{G}=\left \langle \mathcal{V},\mathcal{E} \right \rangle$; alert features \{$(\textbf{H}_v^\mathcal{M},\textbf{H}_v^\mathcal{T},\textbf{H}_v^\mathcal{L}), \forall v \in \mathcal{V}$ \};  total layers $N_l$; learnable aggregator functions $(a^\mathcal{M}, a^\mathcal{T}, a^\mathcal{L})$; weight matrixs $(\textbf{W}^\mathcal{M}, \textbf{W}^\mathcal{T}, \textbf{W}^\mathcal{L})$;  nonlinear activation function $\sigma$; neigborhood function $N(\cdot)$ } 
	\Output{Graph-level features $(\textbf{F}^\mathcal{M}, \textbf{F}^\mathcal{T}, \textbf{F}^\mathcal{L})$, Node-level features $\{(\textbf{E}_v^\mathcal{M}, \textbf{E}_v^\mathcal{T}, \textbf{E}_v^\mathcal{L})|v\in \mathcal{V}\}$}
	\DontPrintSemicolon
	\SetNoFillComment
	\SetKwData{concat}{concat}\SetKwData{normalize}{normalize}
	$\textbf{E}_{v,0}^\mathcal{M}, \textbf{E}_{v,0}^\mathcal{T}, \textbf{E}_{v,0}^\mathcal{L} \leftarrow \textbf{H}_v^\mathcal{M},\textbf{H}_v^\mathcal{T},\textbf{H}_v^\mathcal{L}, \forall v \in \mathcal{V}$\\
	\For{$l \leftarrow 1$ \KwTo $N_l$}{
		\For{$v \in \mathcal{V}$}{
			\For{$D \in \{\mathcal{M}, \mathcal{T}, \mathcal{L}\}$}{
				$\textbf{E}_{N(v),l+1}^D \leftarrow a^D(\{\textbf{E}_{u,l}^D, \forall u \in N(v)\})$\\
				$\textbf{E}_{v,l+1}^D \leftarrow \sigma(\textbf{W}_l^D\cdot $\concat($\textbf{E}_{v,l}^D, \textbf{E}_{N(v),l+1}^D)$)\\
				$\textbf{E}_{v,l+1}^D \leftarrow$  
				\normalize($\textbf{E}_{v,l+1}^D$)
			}
		}   
	}
	\For{$D \in \{\mathcal{M}, \mathcal{T}, \mathcal{L}\}$}{
		$\textbf{F}^D \leftarrow \mathop{\max}_{v \in \mathcal{V}} \left( \textbf{E}_{v,N_l}^D \right)$
	}
	\Return{$(\textbf{F}^\mathcal{M}, \textbf{F}^\mathcal{T}, \textbf{F}^\mathcal{L})$, $\{(\textbf{E}_v^\mathcal{M}, \textbf{E}_v^\mathcal{T}, \textbf{E}_v^\mathcal{L})|v\in \mathcal{V}\}$}
\end{algorithm}

\textbf{Message Aggregator}: Given the instance correlation graph $\mathcal{G}$ (or the augmented graph $\mathcal{\tilde{G}}$) in the training dataset, \textit{TVDiag} performs message passing with three graph encoders on the features of each modality to simulate failure backtracking, yieding three graph-level features. The core of these graph encoders are multi-layer graph neural networks.
The whole process of message aggregator is depicted in Algorithm \ref{al:ma}. We use ${\mathcal{M}, \mathcal{T}, \mathcal{L}}$ to represent the three modalities: metric, trace, and log. Firstly, three alert features $\left(\textbf{H}_v^\mathcal{M},\textbf{H}_v^\mathcal{T},\textbf{H}_v^\mathcal{L} \right)$ serve as the initial embeddings of each node $v$ in $\mathcal{G}$ (Line 1). For each modality $D \in \{\mathcal{M}, \mathcal{T}, \mathcal{L}\}$, $\textbf{E}_{v,l}^D$ denote the $l$-th layer embedding of node $v$. 
By distilling high-dimensional representation of neighborhoods $N(v)$, each node $v$ assimilates alert information from others and calculate its own embedding $\textbf{E}_{v,l+1}^D$ in the new layer (Lines 5-7). $a^D$ and $\textbf{W}_l^D$ denote the learnable aggregator function and weight matrix of $l$-th layer for modality $D$. $\sigma$ represents the nonlinear activation function. Herein, we leverage GraphSAGE \cite{hamilton2017inductive} as the backbone of each layer in graph encoders. Unlike training distinct embeddings for each node, GraphSAGE learns aggregation from neighborhoods, which better aligns with the extraction of failure backtracking patterns. We set up multi-layer GraphSAGEs to enable nodes to perceive the topology and node feature information of a larger neighborhood. 

After traversing $N_l$ GraphSAGE layers, we obtain the graph-level feature $\textbf{F}^D$ for each modality $D$ by summarizing all node embeddings $\textbf{E}_{v,N_l}^D$ in $\mathcal{G}$ through a max pooling layer (Line 9). 
In other words, $\textbf{F}^D$ is the maximum value of all node embeddings in the $N_l$ layer. We apply the above operations to the node embeddings for all modalities, yielding three graph-level features (i.e., $\textbf{F}^\mathcal{M}$, $\textbf{F}^\mathcal{T}$, and $\textbf{F}^\mathcal{L}$) and all node-level features in the last layer (i.e., $\{(\textbf{E}_v^\mathcal{M}, \textbf{E}_v^\mathcal{T}, \textbf{E}_v^\mathcal{L})|v\in \mathcal{V}\}$).

\begin{figure}[t]
	\centering{
		\includegraphics[width=0.8\textwidth]{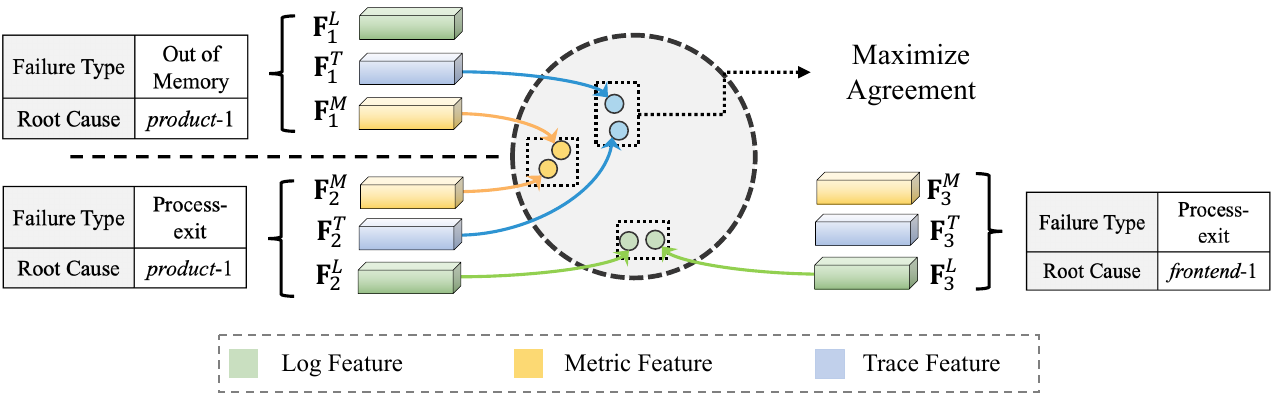}
	}
	\caption{Example of task-oriented learning.}
	\label{fig:TO}
\end{figure}

\textbf{Task-oriented Learning}: Rather than directly fusing the graph-level features of three modalities, \textit{TVDiag} amplifies the potential contribution of each modality to corresponding tasks. The fundamental principle is to maximize the commonality among modality-specific features that share the same task label. To illustrate the process of task-oriented learning, we take the root cause localization (RCL) task with the trace modality as an example, noting that this approach can be applied to other modalities and tasks as well. For the RCL task, we posit that two sets of traces with the same root cause should exhibit maximum coherence at the feature level. This is because the affected microservices and the failure paths caused by the same culprit are likely to be similar. Let $\mathbb{S}=\{s_1,s_2, \cdots,s_n \}$ represent a mini-batch with $n$ failure samples in the training dataset, where each sample $s_i=(\textbf{F}_i^\mathcal{M}, \textbf{F}_i^\mathcal{T}, \textbf{F}_i^\mathcal{L},r_i, t_i)$ contains three graph-level features, a root cause $r_i$, and a failure type $t_i$. As shown in Fig. \ref{fig:TO}, we define the trace features ($\textbf{F}_1^\mathcal{T}$ and $\textbf{F}_2^\mathcal{T}$) of two samples with the same root cause (\textit{product-}1) as a positive pair, striving to bring them closer in the feature space and extract their shared knowledge about the root cause. Naturally, trace features ($\textbf{F}_1^\mathcal{T}$ and $\textbf{F}_3^\mathcal{T}$) with different root causes are considered a negative pair.

\begin{algorithm}[t]
	\setstretch{1.15}
	\caption{Task-oriented Learning}
	\label{al:to}
	
	\Input{Mini-batch $\mathbb{S}=\{s_1,s_2, \cdots,s_n \}$} 
	\Output{Task-oriented loss $\mathcal{L}_{to}$}
	\DontPrintSemicolon
	\SetNoFillComment
	
	\SetKwData{TOLoss}{TOLoss}
	\SetKwInOut{Initialization}{Initialization}
	$\mathbb{P}, \mathbb{N}$: positive set and negative set\\
	\For{$s_i \in \mathbb{S}$}{ 
		$r_i, t_i \leftarrow $ Get the root cause and failure type from $s_i$\\
		\For{$s_j \in \mathbb{S}$}{
			\eIf{$j\neq i$ and $r_j=r_i$}{
				$\mathbb{P}_i^\mathcal{M}, \mathbb{P}_i^\mathcal{T} \leftarrow \mathbb{P}_i^\mathcal{M} \cup \{j\}, \mathbb{P}_i^\mathcal{T} \cup \{j\}$
			}{
				$\mathbb{N}_i^\mathcal{M}, \mathbb{N}_i^\mathcal{T} \leftarrow \mathbb{N}_i^\mathcal{M} \cup \{j\}, \mathbb{N}_i^\mathcal{T} \cup \{j\}$\\
			}
			
			\eIf{$t_j=t_i$}{
				$\mathbb{P}_i^\mathcal{L} \leftarrow \mathbb{P}_i^\mathcal{L} \cup \{j\}$\\
			}{
				$\mathbb{N}_i^\mathcal{L} \leftarrow \mathbb{N}_i^\mathcal{L} \cup \{j\}$\\
			}
		}
		
	}
	\For{$D \in \{\mathcal{M}, \mathcal{T}, \mathcal{L}\}$}{
		$\mathcal{L}_{to}^\mathcal{D} \leftarrow$ \TOLoss($\mathbb{P}^D$, $\mathbb{N}^D$) \tcp{Calculate task-oriented loss according to Eq. \ref{eq:to}.}
	}
	$\mathcal{L}_{to} \leftarrow \mathcal{L}_{to}^\mathcal{M} + \mathcal{L}_{to}^\mathcal{T} + \mathcal{L}_{to}^\mathcal{L}$ \\
	\Return{$\mathcal{L}_{to}$}
	
\end{algorithm}

We adopt a supervised contrastive learning approach \cite{chen2022representation, khosla2020supervised} to enhance the agreement between two features in a positive pair and distance the two features in a negative pair. The whole process is illustrated in Algorithm \ref{al:to}. 
$\mathbb{P}$ and $\mathbb{N}$ on Line 1 denote two sets that contain positive pairs and negative pairs, respectively. For each sample $s_i$, \textit{TVDiag} initializes its positive set and negative set of three modalities (Lines 4-12). 
In addition to traces, we also employ root cause $r_i$ as the guidance for the positive/negative set construction of metrics. More concretly, we regard other samples with the same root cause (i,e., $r_i$) as the positive samples of $s_i$ (Lines 5-8). This is because metrics have shown considerable potential in the RCL task \cite{li2022actionable,he2022graph}. Additionally, the abundant failure details present in logs play a crucial role in the FTI task \cite{yuan2019approach}. Therefore, we use failure type $f_i$ to guide the selection of postive/negative pairs of logs (Lines 9-12). In the following Lines 13-14, we iterate through the three modalities and calculate their task-oriented losses, guiding our model to extract commonalities within the positive pairs. For any modality $D \in \{\mathcal{M}, \mathcal{T}, \mathcal{L}\}$, the task-oriented loss of modality $D$ can be formulated as:
\begin{equation}
	\label{eq:to}
	\mathcal{L}_{to}^D=\sum_{i=1}^{n}{\frac{-1}{\left| \mathbb{P}_i^D\right|}}\sum_{j \in \mathbb{P}_i^D}{\ln\left[\frac{\phi(\textbf{F}_i^D,\textbf{F}_j^D)}{\phi(\textbf{F}_i^D,\textbf{F}_j^D)+\sum_{z\in \mathbb{N}_i^D}{\phi(\textbf{F}_i^D,\textbf{F}_z^D)}}\right]}.
\end{equation} $\phi\left(\textbf{F},\textbf{F}^{\prime}\right)$ represents the probability that the $\textbf{F}^{\prime}$ is the positive pair of \textbf{F}, which is calculated by:
\begin{equation}
	\label{eq:distance}
	\phi\left(\textbf{F},\textbf{F}^{\prime}\right)=\exp\left(sim\left( \textbf{F}, \textbf{F}^{\prime}\right)/\tau\right),
\end{equation}
where $\tau$ is the temperature parameter that adjusts the attention to difficult samples \cite{wang2021understanding}. $sim\left(\cdot\right)$ measures the similarity (e.g., the cosine similarity) between two features. 
Eq. \ref{eq:rcl_loss} enforces the model to optimize the commonality of a positive pair (i.e., reducing the $\phi(\textbf{F}_i^D,\textbf{F}_j^D)$) and amplify the difference of a negative pair (i.e., improving the $\phi(\textbf{F}_i^D,\textbf{F}_z^D)$).
In this way, \textit{TVDiag} can effectively identify the common and effective information of one modality by magnifying the similarity between the specific-modality samples under the same root cause or same failure type. Finally, we output the total task-oriented loss $\mathcal{L}_{to}$ by summing the corresponding losses of the three modalities (Line 15). 


\begin{figure}[t]
	\centering{
		\includegraphics[width=0.8\textwidth]{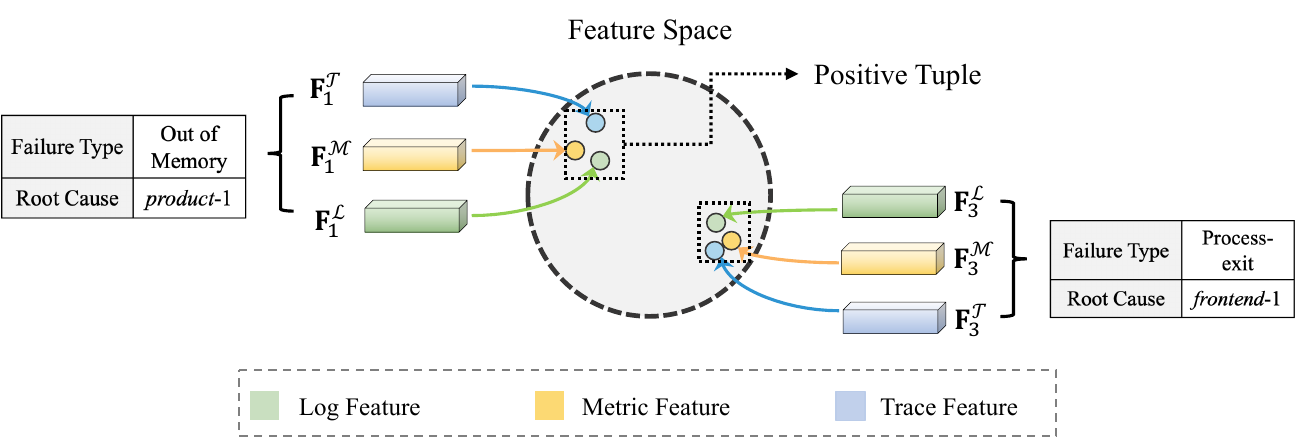}
	}
	\caption{Example of cross-modal association.}
	\vspace{-4mm}
	\label{fig:CM}
\end{figure}

\textbf{Cross-modal Association}: Different modalities provide varying views of the same failure.
These modalities share some hidden view-invariant information, such as abnormal microservice instances, abnormal periods, and failure degrees. Motivated by this, we preserve the view-invariant information by building cross-modal associations through contrastive learning. As shown in Fig. \ref{fig:CM}, \textit{TVDiag} treats $\left(\textbf{F}_i^\mathcal{M},\textbf{F}_i^\mathcal{T},\textbf{F}_i^\mathcal{L}\right)$, which belong to the same sample $s_i$, as a positive tuple that should be tightly clustered in the feature space, while features of other samples with different root causes or failure types are dispersed. 

Given a mini-batch of samples $\mathbb{S}=\left\{s_1,s_2,\cdots,s_n\right\}$, where each sample $s_i$ contains the features of three modalities (i.e., $\textbf{F}_i^\mathcal{M}, \textbf{F}_i^\mathcal{T},$ and $ \textbf{F}_i^\mathcal{L}$) for a failure, we set metrics as the core view and seek to bring other views (i.e., traces and logs) closer to metrics for each sample $s_i$. Metrics are suitable as the core view of cross-modal association because they share many commonalities with other modalities in various tasks. As mentioned in Section \ref{sec:motivation-2}, both metric types and log messages are usually utilized to assess the system status and failure phenomena. Besides, there is common information between metrics and traces in identifying normal or abnormal microservices. Therefore, we decide to establish the correlation between metrics and traces ($\mathcal{M}\sim\mathcal{T}$), and between metrics and logs ($\mathcal{M}\sim\mathcal{L}$). As traces and logs move closer to metrics, they also move closer to each other.

More specifically, we minimize the distances between $\textbf{F}_i^\mathcal{M}$ and $\textbf{F}_i^\mathcal{T}$, $\textbf{F}_i^\mathcal{M}$ and $\textbf{F}_i^\mathcal{L}$ separately, achieving closer proximity among the three modalities. Given two modalities $D_1$ and $D_2$, the learning objective is defined as:
\begin{equation}
	\mathcal{L}_{cm}^{D_1 \sim D_2}=\frac{1}{2n}\sum_{i=1}^{n}{\left(l\left(\textbf{F}_i^{D_1}, \textbf{F}_i^{D_2}\right) + l\left(\textbf{F}_i^{D_2}, \textbf{F}_i^{D_1}\right)\right)},
\end{equation}
where $l\left(\textbf{F}_i^{D_1},\textbf{F}_i^{D_2}\right)$ represents the loss between two features, and the loss is defined as follows:
\begin{equation}
	l\left(\textbf{F}_i^{D_1},\textbf{F}_i^{D_2}\right)=-\ln \\{\frac{\phi\left(\textbf{F}_i^{D_1},\textbf{F}_i^{D_2}\right)}{\sum_{a=1}^{n}{\mathbbm{l}_{a\neq i}\phi\left(\textbf{F}_i^{D_1},\textbf{F}_a^{D_1}\right)} + {\sum_{b=1}^{n}{\phi\left(\textbf{F}_i^{D_1},\textbf{F}_b^{D_2}\right)}}}},
\end{equation}
where $\mathbbm{l}_{a\neq i}$ takes on the value 1 if $a\neq i$, and 0 otherwise. The term $\mathcal{L}_{cm}^{D_1 \sim D_2}$ is intended to facilitate the features of $D_1$ and $D_2$ in learning a more common consensus between them.
We formulate the loss of cross-modal association as:
\begin{equation}
	\label{eq:cm}
	\mathcal{L}_{cm}=\mathcal{L}_{cm}^{\mathcal{M}\sim \mathcal{T}}+\mathcal{L}_{cm}^{\mathcal{M}\sim \mathcal{L}},
\end{equation}
where $\mathcal{L}_{cm}^{\mathcal{M}\sim \mathcal{T}}$ represents the contrastive loss between metric features and trace features, and $\mathcal{L}_{cm}^{\mathcal{M}\sim \mathcal{L}}$ signifies the contrastive loss between the metric features and log features. Guided by the loss of cross-modal association $\mathcal{L}_{cm}$, the view-invariant information between different modalities is effectively amplified. 

\begin{figure}[t]
	\centering{
		\includegraphics[width=\textwidth]{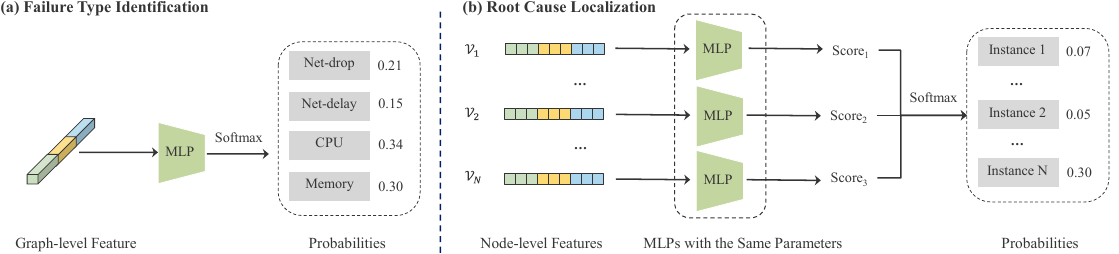}
	}
	\caption{Illustration of failure diagnosis based on features from different levels.}
	\vspace{-4mm}
	\label{fig:downstream}
\end{figure}

\subsection{Failure Diagnosis}
\label{sec:failure-diagnosis}
To comprehensively consider the information from all three modalities and perform failure diagnosis, \textit{TVDiag} separately concatenates the features of node-level ($\{(\textbf{E}_v^\mathcal{M}, \textbf{E}_v^\mathcal{T}, \textbf{E}_v^\mathcal{L})|v\in \mathcal{V}\}$) and graph-level ($\textbf{F}^\mathcal{M}$, $\textbf{F}^\mathcal{T}$, and $\textbf{F}^\mathcal{L}$), which are then used as inputs for downstream failure diagnosis tasks:
\begin{equation}
	\textbf{F}=\textbf{F}^\mathcal{M}\oplus \textbf{F}^\mathcal{T} \oplus \textbf{F}^\mathcal{L},
\end{equation}
\begin{equation}
	\textbf{E}_v=\textbf{E}_v^\mathcal{M}\oplus \textbf{E}_v^\mathcal{T} \oplus \textbf{E}_v^\mathcal{L}.
\end{equation}
The fused feature $\textbf{F}$ aggregates multimodal graph-level information, encompassing the global semantic context of the microservice-based system, whereas $\textbf{E}_v$ focuses on the local information specific to instance $v$. As shown in Fig. \ref{fig:downstream}, we leverage a multilayer perceptron (MLP) to identify the failure type (FTI) of the microservice-based system based on the graph-level fused feature $\textbf{F}$. The output size of this MLP equals the number of failure types that have occurred in the past. \textit{TVDiag} chooses the cross-entropy loss to quantify the discrepancy between predicted failure types and actual ones:
\begin{equation}
		\label{eq:fti_loss}
		\mathcal{L}_{fti}=-\frac{1}{T}\sum_{i=1}^{T}\sum_{k=1}^{C}{y_{i,k}\log(p_{i,k})},
	\end{equation}
where $T$ denotes the number of samples and $C$ indicates the number of failure types. $y_{i,k}=1$ if the failure type of the $i$-th sample is $k$, and 0 otherwise. The value $p_{i,k}$ represents the probability of determining the failure type of $s_i$ as $k$.
	
Regarding the root cause localization (RCL), we do not fix the output size to match the number of instances, as we do for FTI, because the number of microservice instances may vary due to their inherent elasticity. Instead, we set the MLP output size to 1, representing a score indicating the likelihood of each node being the root cause. The MLP processes the feature from each instance $v$ and generates the score based on the corresponding node-level feature $E_v$. Finally, the scores of all nodes are transformed into probabilities through a softmax function. Similarly, \textit{TVDiag} defines the loss as: 
\begin{equation}
	\label{eq:rcl_loss}
	\mathcal{L}_{rcl}=-\frac{1}{T}\sum_{i=1}^{T}\sum_{j=1}^{N}{y_{i,j}\log(p_{i,j})},
\end{equation}
where $N$ is the total number of microservice instances. Here, $y_{i,j}=1$ if the root cause of the $i$-th sample ($s_i$) is $j$, and 0 otherwise.

Intuitively, RCL and FTI can benefit from shared knowledge, such as the category of abnormal metrics, unusual log templates, and abnormal microservices, among others. Therefore, RCL and FTI can be regarded as two complementary multi-classification tasks that share knowledge. Accordingly, we train these two diagnosis tasks jointly, leveraging the shared knowledge to enhance the learning of each other. To achieve this, we train these two MLPs jointly to increase the performance of both subtasks while saving training and inference time.

Because $\mathcal{L}_{to}$ and $\mathcal{L}_{cm}$ are essentially contrastive losses, we scale them by multiplying a hyper-parameter $\omega$, preventing the model from neglecting the optimization of the most critical diagnostic tasks (i.e., $\mathcal{L}_{rcl}$ and $\mathcal{L}_{fti}$). The overall loss is a combination of four components:
\begin{equation}
	\label{eq:rawLoss}
	\mathcal{L}=\alpha\cdot\mathcal{L}_{rcl}+\beta\cdot\mathcal{L}_{fti} + \gamma\cdot \omega\cdot\mathcal{L}_{to} + \delta\cdot \omega\cdot\mathcal{L}_{cm},
\end{equation}
where $\alpha$, $\beta$, $\gamma$, and $\delta$ are weighting factors for each component. $\mathcal{L}_{rcl}$ and $\mathcal{L}_{fti}$ denote the losses mentioned in Equations. \ref{eq:fti_loss}-\ref{eq:rcl_loss}, which aim to optimize two diagnosis tasks (i.e., RCL and FTI). The loss of task-oriented learning $\mathcal{L}_{to}$ and the loss of cross-modal association $\mathcal{L}_{cm}$ serve as auxiliary guides to refine our model in addition to optimizing two diagnosis tasks.

To avoid assigning static weights that might be suboptimal, we follow the previous research \cite{liebel2018auxiliary} and convert the static weights ($\alpha$, $\beta$, $\gamma$, and $\delta$) into learnable parameters that can be dynamically tuned during model training. Equation \ref{eq:rawLoss} is then reformulated as:
\begin{equation}
	\mathcal{L}=\sum_{1\leq z \leq4}\frac{1}{2\cdot \theta_z^2}\cdot\mathcal{L}\left(z\right)+\ln{\left(1+\theta_z^2\right)},
\end{equation}
where $\theta_z$ is the learnable parameters and $\mathcal{L}\left(z\right)$ represents the loss of each component, i.e., $\mathcal{L}_{to}$, $\mathcal{L}_{cm}$, $\mathcal{L}_{rcl}$ or $\mathcal{L}_{fti}$. 

Finally, \textit{TVDiag} generates a ranking list $\mathcal{R}$ for RCL and identifies a failure type $t$ for FTI. These results narrow down the diagnosis scope and guide operators in making recovery decisions.

\section{EVALUATION}
\label{sec:evaluation}
To examine the effectiveness of the proposed \textit{TVDiag}, we conducted comparisons with state-of-the-art failure diagnosis methods on three datasets. This section will answer the following research questions:
\begin{itemize}[left=0pt]
	\item[---] \textbf{RQ1: How does \textit{TVDiag} perform compared to existing methods?} 
	
	The failure diagnosis encompasses two tasks: root cause localization (RCL) and failure type identification (FTI). For the RCL task, we assessed and compared our method with four single-modal baselines and three multimodal methods. For the FTI task, we compared \textit{TVDiag} with a log-based method and a multimodal approach. Additionally, we derived three traditional machine learning methods from \textit{TVDiag} as supplementary baselines.
	
	\item[---] \textbf{RQ2: What is the effectiveness of each component in \textit{TVDiag}?} 
	
	\textit{TVDiag} consists of five main modules: alert extractor, graph augmentation (AUG), task-oriented learning (TO), cross-modal association (CM), and dynamic weights (DW). To investigate the contribution of these modules, we conducted an ablation study by individually removing each one.
	
	\item[---] \textbf{RQ3: How do hyper-parameters influence \textit{TVDiag}?} 
	
	\textit{TVDiag} includes several hyper-parameters, such as the inactivation probability $p$, the contrastive loss scale $\omega$, the number of graph layers $l$, and the temperature parameter $\tau$. We conducted comprehensive comparison experiments on each hyper-parameter to verify their effect. 
	
	
	
	\item[---] \textbf{RQ4: How does each modality impact the two diagnosis tasks in \textit{TVDiag}?}
	
	Integrating multimodal monitoring data is crucial for failure diagnosis. We performed comparison experiments by ablating one or two modalities to investigate their contribution to failure diagnosis. Furthermore, we quantified the contribution ratio of each  modality in each inference, providing operators with actionable insights to prioritize modalities in subsequent analysis. To ascertain the mutual effect between two fundamental failure diagnosis tasks, we calculated their inter-task affinity during the training process.
	
	\item[---] \textbf{RQ5: How does \textit{TVDiag} perform in terms of efficiency?}
	
	To meet the demands of fast online diagnosis, the online inference phase of \textit{TVDiag} is constrained within a limited time frame. We also need to control the offline training overhead to accommodate scenario changes. We analyzed the training and inference costs of \textit{TVDiag} and compared them with existing baselines to assess its efficiency.
	
\end{itemize}

\begin{table*}[t]
	\small
	\caption{Dataset Statistics}
	\label{tab:dataset-setting}
	\vspace{-2mm}
	\begin{threeparttable}
		\begin{tabular}{c|c|c|c|ll|ll}
			\Xhline{0.8pt}
			\multicolumn{1}{c|}{\cellcolor[HTML]{FFFFFF}Dataset} & 
			\multicolumn{1}{c|}{\cellcolor[HTML]{FFFFFF}\# of Microservices} &
			\multicolumn{1}{c|}{\cellcolor[HTML]{FFFFFF}\# of Instances} &\multicolumn{1}{c|}{\cellcolor[HTML]{FFFFFF}\# of Failure Types}& \multicolumn{2}{c|}{\cellcolor[HTML]{FFFFFF}\# of Records} &\multicolumn{2}{c}{\cellcolor[HTML]{FFFFFF}\# of Types\tnote{1}}\\ \hline
			& &  &  & Metric & 217,461,390 & Metric types & 3320 \\
			& &  &  & Trace & 3,084,066 & Trace types & 122 \\
			\multirow{-3}{*}{$\mathcal{A}$} &  \multirow{-3}{*}{5} &\multirow{-3}{*}{10} & \multirow{-3}{*}{5} & Log & 87,974,577 & LogKeys & 49 \\ \hline
			& &  &  & Metric & 25,914,595 &Metric types& 404 \\
			& &  &  &Trace & 63,283,650 &Trace types & 63 \\
			\multirow{-3}{*}{$\mathcal{B}$}  &\multirow{-3}{*}{10} & \multirow{-3}{*}{40} &\multirow{-3}{*}{9} & Log & 95,003,038 &LogKeys & 119 \\ \hline
			& &  &  & Metric & 77,376,704 &Metric types& 293 \\
			& &  &  &Trace & 7,297,688 &Trace types & 143 \\
			\multirow{-3}{*}{$\mathcal{C}$}  &\multirow{-3}{*}{14} & \multirow{-3}{*}{21} &\multirow{-3}{*}{7} & Log & 20,365,334 &LogKeys & 422 \\
			\Xhline{0.8pt}
		\end{tabular}
		
		\begin{tablenotes}
			\footnotesize
			\item[1] A metric type can be represented by (\textit{instance}, \textit{metric name}). We define traces of the same type as those that execute the same set of operations in the same order. LogKeys are the predefined templates of logs.
		\end{tablenotes}
	\end{threeparttable}
\end{table*}

\subsection{Experimental Design}
\label{sec:exp-design}
\textbf{Datasets}: Experiments were conducted on two open-source microservice datasets ($\mathcal{A}$, $\mathcal{B}$) with multimodal monitoring data, where $\mathcal{A}$ is derived from the GAIA dataset\cite{GAIA}, and $\mathcal{B}$ is based on the AIOps-22 dataset\cite{AIOps}. Besides, we collected an additional dataset $\mathcal{C}$ with diverse failure data from a widely-used application, further validating the effectiveness of \textit{TVDiag}. Table \ref{tab:dataset-setting} shows the statistics of three datasets.

\begin{itemize}
	\item[---] \textbf{Dataset} $\mathcal{A}$. The GAIA dataset has been widely employed as an evaluation dataset in various papers \cite{zhang2023robust, zhang2023efficient, sui2023logkg}. It records metrics, traces, and logs from the MicroSS simulation system in July 2021. MicroSS is a business simulation system that consists of ten microservices and several middleware components like Redis, MySQL, and Zookeeper. To simulate real-world failures, various types of instance-level failures, such as memory anomalies and login failures, were intentionally injected into MicroSS.
	\item[---] \textbf{Dataset} $\mathcal{B}$. The AIOps-22 dataset originates from the training data released by the AIOps 2022 Challenge, where failures across three levels (node, service, and instance) were injected into a web-based e-commerce platform. To replicate a scenario involving a microservice-based online system with large-scale multimodal monitoring data, AIOps-22 collected four business metrics, 400 performance metrics, traces, and logs to describe the status of running microservices. To align with the failure level of the GAIA dataset, we retained 113 instance-level failure records, which were manually injected by the sponsor. 
	\item[---] \textbf{Dataset} $\mathcal{C}$. To further verify the effectiveness of \textit{TVDiag} in diagnosing various online failures, we deployed a well-known microservice-based system: Sockshop \cite{Sockshop} in a Kubernetes cluster. We induced diverse application failures using Chaos Mesh \cite{chaosmesh} and gathered approximately 70GB of system telemetry data for subsequent testing. To mimic resource issues, we injected common failures into this application, including CPU-hog and memory-stress. For network issues, we simulated various failures resulted by poor network conditions, such as packet-corruption, packet-loss, and network-delay. Furthermore, we simulated code defects by either killing pods or setting their status to failure. Specifically, we injected these failures into all stateless microservices and repeated the process five times.
\end{itemize}

\textbf{Competing Approaches}: 
We utilized state-of-the-art multimodal and single-modal methods in RCL and FTI, respectively, as baselines. Notably, we selected baseline methods that can be directly applied to most generic multimodal monitoring data. These baselines do not require invasive code modifications in microservices to introduce auxiliary information. For the RCL task, we considered the following methods:

\begin{itemize}
	\item[---] \textbf{DiagFusion} \cite{zhang2023robust} is a multimodal method that fuses all modalities during data preparation by consolidating them into events. Subsequently, these events are encoded with two topology-adaptive graphs for downstream diagnosis tasks. For the RCL task, DiagFusion utilizes a topology adaptive graph convolution network to learn failure patterns and generates a ranking list of candidate root causes.
	\item[---] \textbf{Eadro} \cite{lee2023eadro} presents a multimodal failure diagnosis framework that integrates anomaly detection and RCL. It designs separate feature extractors for three modalities and then fuses the multimodal features. Using the graph attention networks, Eadro learns the dependency-aware status of microservices and optimizes the fused feature in the downstream diagnosis task.
	
	
	\item[---] \textbf{RF-RCL} is a variant of \textit{TVDiag} that replaces the failure diagnosis model with a typical ensemble model, Random Forest (RF).
	
	\item[---] \textbf{MicroRank} \cite{yu2021microrank} is a trace-based diagnosis method for the RCL task, synthesizing the Pagerank algorithm and spectrum analysis to infer root causes.
	
	\item[---] \textbf{TraceRCA} \cite{li2021practical} proposes a trace-based root cause localization framework. It first mines several suspicious service sets using the FP-Growth algorithm. For each microservice within these sets, TraceRCA scores it according to the discrepancies between the number of abnormal and normal traces.
	
	\item[---] \textbf{MicroRCA} \cite{wu2020microrca} introduces a metric-based diagnosis method that utilizes the Pagerank algorithm to pinpoint the root cause within the anomalous subgraph.
	
	\item[---] \textbf{NicroHECL} \cite{liu2021microhecl} is a metric-based root cause localization approach that dissects the golden metrics to analyze failure propagation and rank candidate microservices.
\end{itemize}

For the FTI task, we chose a log-based method and DiagFusion as the baseline. Besides, we derived three variants of \textit{TVDiag} by replacing the multimodal co-learning module with three traditional machine-learning methods:

\begin{itemize}
	\item[---] \textbf{LogCluster} \cite{lin2016log} is a log-based problem identification method that categorizes failures by clustering logs and matching them with a knowledge base.
	\item[---] \textbf{DiagFusion} \cite{zhang2023robust}: For the FTI task, DiagFusion jointly trains a topology adaptive graph convolution network with the network used for RCL, sharing the failure knowledge with each other and identifying the failure types. Note DiagFusion statically assigns equal weights to the two tasks.
	\item[---] \textbf{Medicine} \cite{tao2024giving} is a modality-independent failure diagnosis framework that identifies failure types based on multimodal monitoring data. To preserve the unique characteristics of each modality, it encodes them separately and further adopts multimodal adaptive optimization to balance the learning progress.
	\item[---] \textbf{DT-FTI} is a variant of \textit{TVDiag} that sets the failure diagnosis model as a decision tree (DT). It concatenates the multimodal alert features in \cref{sec:DP} and uses them as the input to this model, which outputs the classification results of failure types. 
	\item[---] \textbf{SVM-FTI} is a variant of \textit{TVDiag} that replaces the failure diagnosis model with the support vector machine (SVM).
	\item[---] \textbf{LGB-FTI} is a variant of \textit{TVDiag} that replaces the failure diagnosis model with the typical ensemble model, i.e., LightGBM (LGB).
\end{itemize}

\textbf{Evaluation Criteria}: For RCL, a ranking list $\mathcal{R}$ comprising root cause candidates is produced. We employed widely used $HR@k$ (hit ratio of top-$k$), $Avg@k$ (average hit ratio of top-$k$) \cite{wu2020microrca, yu2021microrank, li2022actionable}, and $NDCG@k$ (normalized discounted cumulative gain of top-$k$) \cite{lee2023eadro, liu2024integrating} to measure the performance of RCL. $HR@k$ is formulated as:
\begin{equation}
	HR@k=\frac{1}{N}\sum_{i=1}^{N}{r_i \in \mathcal{R}_i{[1:k]}},
\end{equation}
where $N$ is the number of all failures, $\mathcal{R}_i{[1:k]}$ denotes the top-$k$ results of the $i$-th ranking list $\mathcal{R}_i$, and $r_i$ is the ground truth for the $i$-th failure. The average hit ratio for top-$k$ results is defined as:
\begin{equation}
	Avg@k=\frac{1}{k}\sum_{i=1}^{k}{HR@i}.
\end{equation}
In general, the instances with higher ranks should inherently receive higher weights. We introduced $MRR@k$ to further assess the relevance of the top-$k$ results with the root cause:
\begin{equation}
	MRR@k=\frac{1}{N}\sum_{i=1}^{N}{{\frac{{rel}_i}{rank_i}}},
\end{equation}
where $rank_i$ represents the position of the root cause within the ranking list $\mathcal{R}$ for the $i$-th sample. ${rel}_i$ is 1 if the $rank_i \leq k$, and 0 otherwise. Higher rankings of root causes imply better $MRR@k$ scores.

For the FTI task, we used typical classification indicators to measure the performance of baselines: $Precision=\frac{TP}{TP+FP}$, $Recall=\frac{TP}{TP+FN}$, and F1-score=$2\cdot\frac{Precision\cdot Recall}{Precision+Recall}$, where $TP$ and $FP$ denote the number of true-positive results and the number of false-positive results, respectively. $FN$ represents the number of samples whose failure types are not identified. Higher values of these metrics indicate better performance.

\textbf{Implementation Details}: For the division of the training and test sets in dataset $\mathcal{A}$, 
we followed the division method proposed by DiagFusion \cite{zhang2023robust} for fair comparison.
We directly split training samples and test samples based on the sample collection time, using samples from earlier time for training and later ones for testing to avoid data leakage. 
For dataset $\mathcal{B}$, we randomly shuffled the dataset and used 80$\%$ of the data for training and the remaining 20$\%$ for testing. 
The alert embedding dimension was set to 128. 
Regarding alert extraction in dataset $\mathcal{A}$, we selected the 40 minutes prior to failure injection as the training window for alert extractors, and the 10 minutes following the failure injection as the alert extraction window. For dataset $\mathcal{B}$, both the training window and the alert extraction window were set to five minutes. We configured the alert extraction window of dataset $\mathcal{C}$ to 10 minutes and collected additional normal telemetry data for training of alert detectors.
We set the dimension of the hidden layer in GraphSAGE to 64 and the output dimension to 32. Herein, we designated the aggregator function of GraphSAGE as LSTM to achieve better performance \cite{hamilton2017inductive}.

The implementation of \textit{TVDiag} was based on PyTorch 1.12.0+cu116 and DGL 0.9.1+cu116. All experiments were conducted on a workstation with an NVIDIA GeForce GTX 1080 GPU and 64 GB of RAM. We publicized our source code and experimental data in \cite{TVDiag}. During model training, we used the Adaptive Moment Estimation (Adam) optimizer with an initial learning rate of $1\times10^{-3}$ and a weight decay of $1\times10^{-4}$. To prevent overfitting and improve efficiency, we implemented an early stopping strategy based on changes in loss, halting the training process when the loss has not decreased for an extended period. The temperature parameter $\tau$ was set to 0.3. 

\subsection{Performance Comparison (RQ1)}

\textbf{Objectives:} We investigate the effectiveness of \textit{TVDiag} in RCL and FTI tasks by comparing it with existing single-modal and multimodal approaches. We further explore the performance of multimodal methods in root cause localization across different failure types.

\textbf{Experimental design:} Regarding the RCL task, we conducted a comparison experiment with four single-modal methods (i.e., MicroRCA, MicroHECL, TraceRCA, and MicroRank) and three multimodal methods, namely Eadro, DiagFusion, and RF-RCL. We adopted four metrics as the measurements: $HR@1$, $HR@3$, $Avg@3$, and $MRR@3$. We further utilized the T-test to determine whether \textit{TVDiag}'s performance on various metrics is significantly greater than that of the baseline methods. For the FTI task, we selected a typical log-based method (LogCluster) and a multimodal method (DiagFusion) as the baselines. More specifically, we replaced the backbone of \textit{TVDiag} with three traditional machine learning models (SVM, DT, and LightGBM), concatenating alert features from the three modalities as input.

\begin{table}[t]
	\caption{Root cause localization comparison.}
	\vspace{-2mm}
	\label{tab:RCL-performance}
	\begin{threeparttable}
		\resizebox{\linewidth}{!}{
			\begin{tabular}{@{}cccccccccccc@{}}
				\toprule
				Dataset            & Modality      & Method     & P-value     & $HR@1$           & \multicolumn{1}{c}{$\Uparrow HR@1$}    & $HR@3$          & \multicolumn{1}{c}{$\Uparrow HR@3$} & $Avg@3$          & \multicolumn{1}{c}{$\Uparrow Avg@3$} & $MRR@3$         & $\Uparrow MRR@3$  \\ \midrule
				\multirow{8}{*}{$\mathcal{A}$} & Metric        & MicroRCA     &3.4e-05   & 0.181          & 319.337\%                     & 0.385          & 136.104\%                  & 0.287          & 194.077\%                   &0.268                &208.396\%         \\
				& Metric         & MicroHECL    &2.4e-06   &0.118           &  543.220\%                   &0.233          &   290.129\%               &0.177           &  376.836\%                 &0.167              &395.900\%      \\
				& Trace         & TraceRCA    &1.3e-05   & 0.192          & 295.313\%                     & 0.364          & 149.725\%                 & 0.291          &    190.034\%               & 0.270               & 205.922\%       \\
				& Trace         & MicroRank    &8.6e-05  & 0.192          &     295.313\%                 & 0.419          &   116.945\%                & 0.308          &       174.026\%            & 0.288               &   187.145\%      \\
				
				& Multimodality & Eadro        &1.6e-03   & 0.304          &149.671\%  & 0.592          &  53.547\%                  & 0.473          &  78.436\%                   &   0.437           & 89.347\%       \\
				& Multimodality & DiagFusion   &3.1e-02   & 0.404          & \multicolumn{1}{c}{87.871\%}  & 0.801          & 13.483\%                    & 0.636          & 32.704\%                    &0.586                &41.001\%         \\
				& Multimodality & RF-RCL   &3.4e-02   &0.573           & \multicolumn{1}{c}{32.461\%}  & 0.815          &11.534\%                 &0.719        &17.385\%                   & 0.686          & 20.423\%   \\
				& Multimodality & \textit{\textbf{TVDiag}} & - & \textbf{0.759} & \multicolumn{1}{c}{-}        & \textbf{0.909} & \multicolumn{1}{c}{-}     & \textbf{0.844} & \multicolumn{1}{c}{-}      & \textbf{0.827} & -       \\ \midrule
				\multirow{8}{*}{$\mathcal{B}$} & Metric        & MicroRCA     &6.7e-07   & 0.058          & 1441.379\%                    & 0.155          & 507.097\%                  & 0.107          & 763.551\%                   & 0.099               &829.223\%
				\\
				& Metric         & MicroHECL    &6.8e-07   & 0.044          &1931.818\%                     &0.142           &  562.676\%                & 0.094          & 882.979\%                  & 0.085             &974.414\%      \\
				& Trace         & TraceRCA    &2.8e-09   & 0.150          &  496.000\%                    & 0.204          & 361.275\%
				& 0.183          &  404.918\%
				& 0.176               & 422.412\%
				\\
				& Trace         & MicroRank    &4.4e-09  & 0.171         & 422.807\%                     & 0.227          & 314.537\%
				& 0.205          & 350.732\%
				& 0.197               &    364.611\%
				\\
				& Multimodality & Eadro       &6.3e-04    & 0.226          & 295.575\%                     & 0.478          & 96.862\%
				& 0.367          & 151.771\%
				&   0.339             & 170.852\%
				\\
				& Multimodality & DiagFusion   &4.3e-08   & 0.205          & 336.098\%                     & 0.273          & 244.689\%
				& 0.239          & 301.739\%
				& 0.229               &300.655\%
				\\
				& Multimodality & RF-RCL   & 7.5e-03  &    0.635      & \multicolumn{1}{c}{40.787\%}  &  0.829         & 13.510\%               &   0.739      &    25.034\%               &0.719           & 27.456\%   \\
				& Multimodality & \textit{\textbf{TVDiag}} &-& \textbf{0.894} & \multicolumn{1}{c}{-}         & \textbf{0.941} & \multicolumn{1}{c}{-}      & \textbf{0.924} & \multicolumn{1}{c}{-}       & \textbf{0.917} &  -       \\ \midrule
				\multirow{8}{*}{$\mathcal{C}$}& Metric & MicroRCA & \multicolumn{1}{r}{6.4E-06} & 0.063  & \multicolumn{1}{r}{1319.048\%} & 0.231  & \multicolumn{1}{r}{307.359\%} & 0.145  & \multicolumn{1}{r}{537.241\%} & 0.132  & \multicolumn{1}{r}{594.571\%} \\
				& Metric & MicroHECL & \multicolumn{1}{r}{1.8E-05} & 0.054  & \multicolumn{1}{r}{1555.556\%} & 0.245  & \multicolumn{1}{r}{284.082\%} & 0.147  & \multicolumn{1}{r}{530.000\%} & 0.132  & \multicolumn{1}{r}{593.695\%} \\
				& Trace & TraceRCA & \multicolumn{1}{r}{1.3E-07} & 0.116  & \multicolumn{1}{r}{670.690\%} & 0.220  & \multicolumn{1}{r}{327.727\%} & 0.171  & \multicolumn{1}{r}{439.300\%} & 0.161  & \multicolumn{1}{r}{469.462\%} \\
				& Trace & MicroRank & \multicolumn{1}{r}{1.7E-06} & 0.054  & \multicolumn{1}{r}{1555.556\%} & 0.200  & \multicolumn{1}{r}{370.500\%} & 0.140  & \multicolumn{1}{r}{560.000\%} & 0.121  & \multicolumn{1}{r}{655.632\%} \\
				& Multimodality & Eadro & \multicolumn{1}{r}{1.4E-04} & 0.191  & \multicolumn{1}{r}{368.063\%} & 0.415  & \multicolumn{1}{r}{126.747\%} & 0.317  & \multicolumn{1}{r}{191.789\%} & 0.291  & \multicolumn{1}{r}{214.883\%} \\
				& Multimodality & DiagFusion & \multicolumn{1}{r}{9.0E-04} & 0.519  & \multicolumn{1}{r}{72.254\%} & 0.710  & \multicolumn{1}{r}{32.535\%} & 0.612  & \multicolumn{1}{r}{51.063\%} & 0.597  & \multicolumn{1}{r}{53.531\%} \\
				& Multimodality & RF-RCL& \multicolumn{1}{r}{5.1E-03} & 0.622  & \multicolumn{1}{r}{43.730\%} & 0.817  & \multicolumn{1}{r}{15.177\%} & 0.715  & \multicolumn{1}{r}{29.291\%} & 0.701  & \multicolumn{1}{r}{30.820\%} \\
				& Multimodality & \textbf{\textit{TVDiag}} & -     & \textbf{0.863}  & -     & \textbf{0.954}  & -     & \textbf{0.913}  & -     & \textbf{0.903}  & - \\
				\bottomrule

		\end{tabular}}
		
		\begin{tablenotes}
			\footnotesize
			\item $\Uparrow$ indicates the improvement rate of \textit{TVDiag} compared to other methods.
		\end{tablenotes}
	\end{threeparttable}
	\vspace{-4mm}
\end{table}

\begin{table}[t]
	\caption{Failure type identification comparison.}
	\vspace{-2mm}
	\label{tab:FTI-performance}
	\small
	\begin{tabular}{@{}cccccccccc@{}}
		\toprule
		Dataset            & Modality      & Method          & P-value & $Precision$      & $\Uparrow Precision$ & $Recall$         & $\Uparrow Recall$  & F1-score       & $\Uparrow$ F1-score \\ \midrule
		            \multirow{7}[1]{*}{$\mathcal{A}$} & Log   & LogCluster & 1.4E-02 & 0.888 & 4.167\% & 0.915 & 3.607\% & 0.901 & 3.885\% \\
		        & Multimodality & DiagFusion & 3.3E-04 & 0.852 & 8.568\% & 0.838 & 13.126\% & 0.84  & 11.429\% \\
		        & Multimodality & Medicine & 4.1E-06 & 0.669 & 38.266\% & 0.667 & 42.129\% & 0.648 & 44.444\% \\
		        & Multimodality & DT-FTI & 9.4E-03 & 0.889 & 4.049\% & 0.892 & 6.278\% & 0.89  & 5.169\% \\
		        & Multimodality & SVM-FTI & 9.9E-02 & 0.883 & 4.757\% & 0.935 & 1.390\% & 0.908 & 3.084\% \\
		        & Multimodality & LGB-FTI & 5.2E-02 & 0.892 & 3.700\% & 0.928 & 2.155\% & 0.907 & 3.197\% \\
		        & Multimodality & \textit{\textbf{TVDiag}} & -     & \textbf{0.925} & -     & \textbf{0.948} & -     & \textbf{0.936} & - \\
		        \midrule
		        \multirow{7}[2]{*}{$\mathcal{B}$} & Log   & LogCluster & 1.1E-02 & 0.434 & 103.226\% & 0.159 & 447.799\% & 0.147 & 495.918\% \\
		        & Multimodality & DiagFusion & 3.4E-04 & 0.533 & 65.478\% & 0.489 & 78.119\% & 0.505 & 73.465\% \\
		        & Multimodality & Medicine & 4.7E-04 & 0.307 & 187.296\% & 0.244 & 256.967\% & 0.244 & 259.016\% \\
		        & Multimodality & DT-FTI & 5.0E-05 & 0.448 & 96.875\% & 0.42  & 107.381\% & 0.422 & 107.583\% \\
		        & Multimodality & SVM-FTI & 3.6E-03 & 0.807 & 9.294\% & 0.784 & 11.097\% & 0.774 & 13.178\% \\
		        & Multimodality & LGB-FTI & 7.0E-04 & 0.604 & 46.026\% & 0.58  & 50.172\% & 0.556 & 57.554\% \\
		        & Multimodality & \textit{\textbf{TVDiag}} & -     & \textbf{0.882} & -     & \textbf{0.871} & -     & \textbf{0.876} & - \\
		        \midrule
		        \multirow{7}[2]{*}{$\mathcal{C}$} & Log   & LogCluster & 1.9E-06 & 0.233 & 278.541\% & 0.228 & 282.018\% & 0.213 & 311.268\% \\
		        & Multimodality & DiagFusion & 7.9E-05 & 0.338 & 160.947\% & 0.332 & 162.349\% & 0.328 & 167.073\% \\
		        & Multimodality & Medicine & 5.0E-04 & 0.497 & 77.465\% & 0.469 & 85.714\% & 0.454 & 92.952\% \\
		        & Multimodality & DT-FTI & 6.4E-05 & 0.376 & 134.574\% & 0.365 & 138.630\% & 0.366 & 139.344\% \\
		        & Multimodality & SVM-FTI & 3.2E-03 & 0.384 & 129.688\% & 0.315 & 176.508\% & 0.282 & 210.638\% \\
		        & Multimodality & LGB-FTI & 9.7E-04 & 0.442 & 99.548\% & 0.39  & 123.333\% & 0.388 & 125.773\% \\
		        & Multimodality & \textit{\textbf{TVDiag}} & -     & \textbf{0.649} & -     & \textbf{0.622} & -     & \textbf{0.636} & - \\
		        \bottomrule
	\end{tabular}
	\vspace{-2mm}
\end{table}

\begin{figure*}[t]
	\centering{
		\subfigure[ $HR@k$ of dataset $\mathcal{A}$]{
			\includegraphics[width=0.8\textwidth]{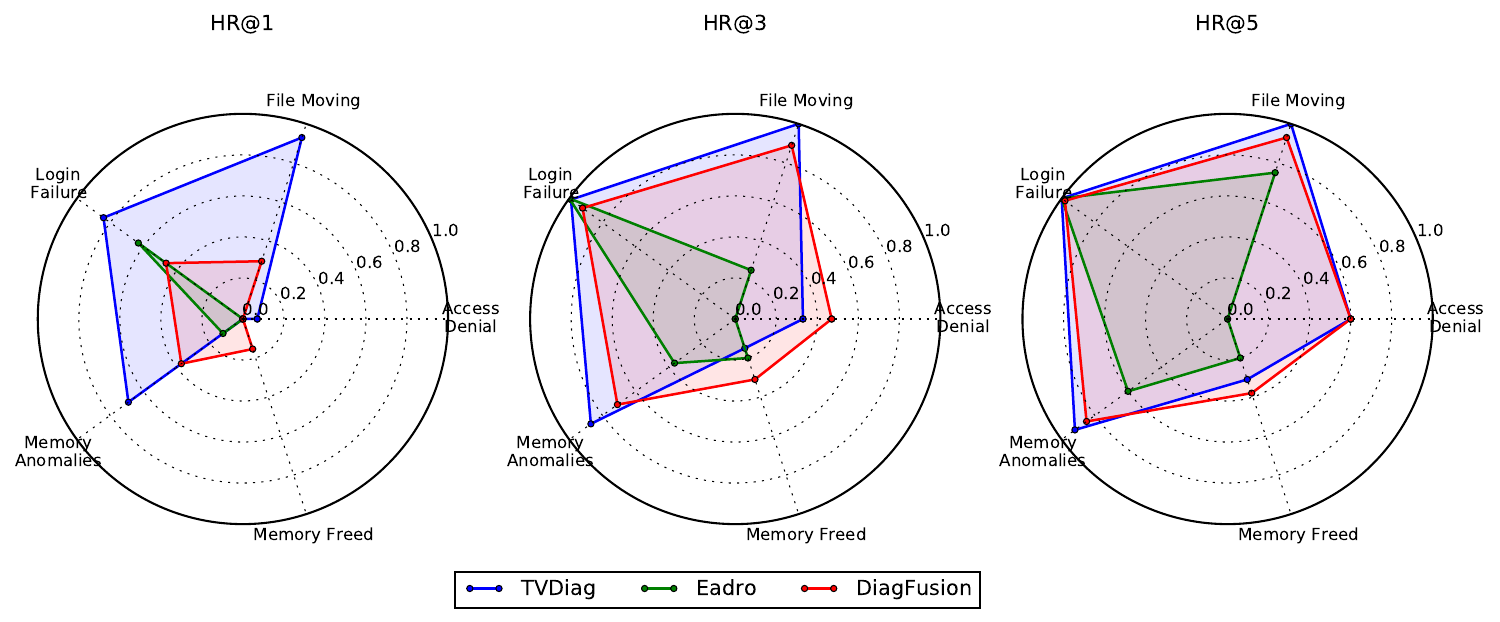} } 
		\subfigure[ $HR@k$ of dataset $\mathcal{B}$]{  \includegraphics[width=0.8\textwidth]{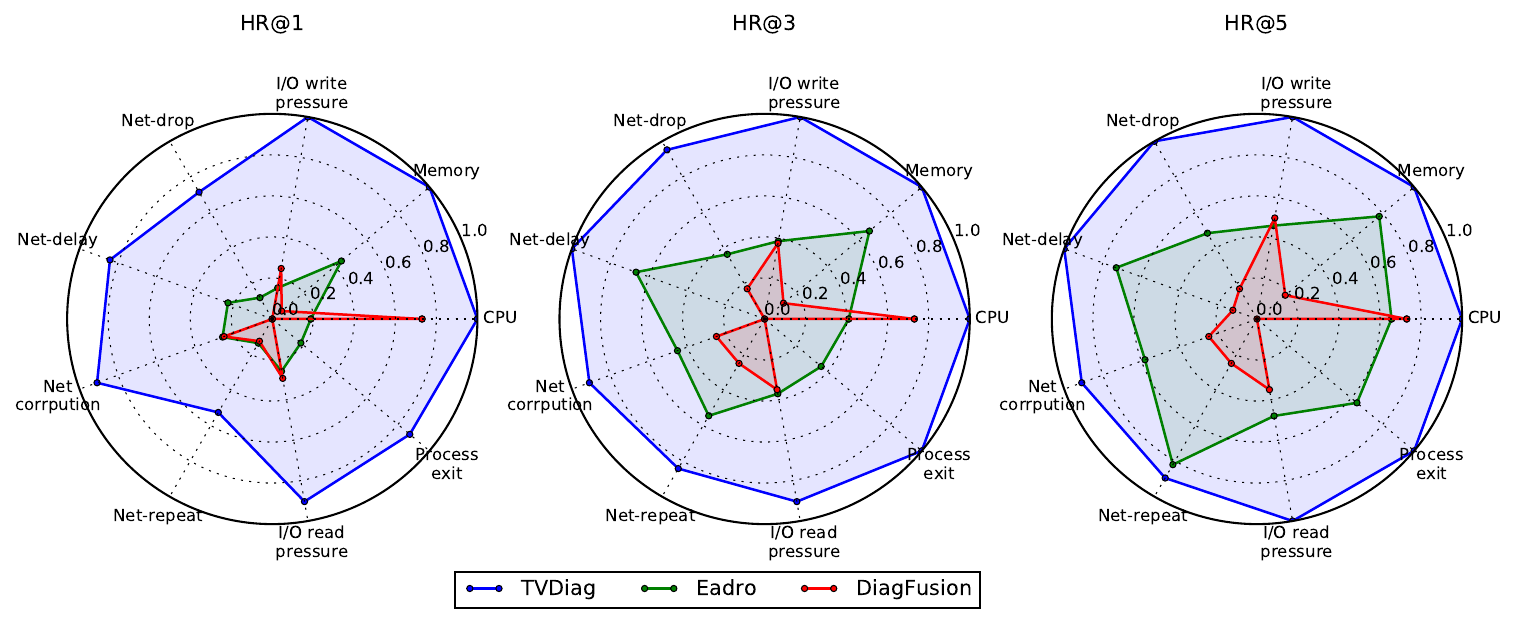}}
		\subfigure[ $HR@k$ of dataset $\mathcal{C}$]{  \includegraphics[width=0.8\textwidth]{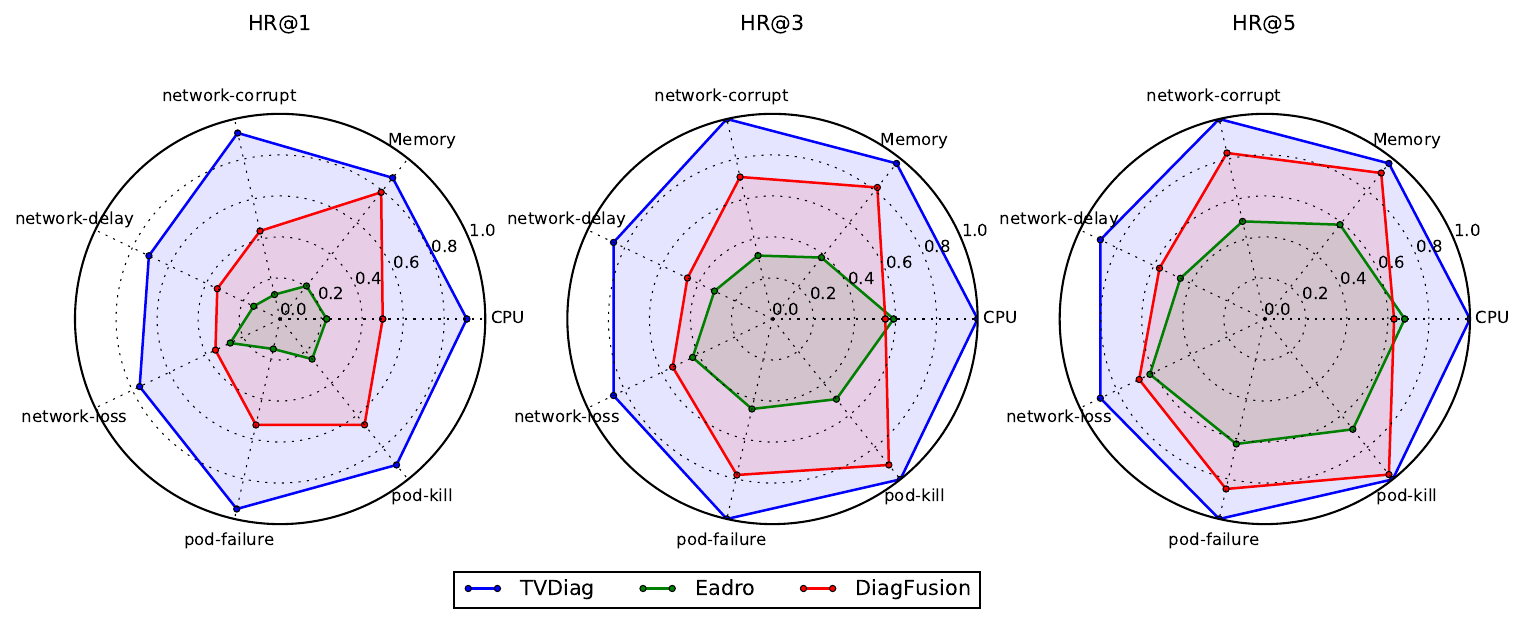}}
	}
	\vspace{-4mm}
	\caption{Performance comparison of $HR@k$ for different failure types.}
	\label{fig:leida}
\end{figure*}

\textbf{Results:}
Tables \ref{tab:RCL-performance} and \ref{tab:FTI-performance} present the evaluation results of different methods on the three datasets. \textit{TVDiag} outperforms all state-of-the-art multimodal-based methods in both the RCL task and the FTI task. As shown in Table \ref{tab:RCL-performance}, \textit{TVDiag} significantly surpasses all baseline methods for the RCL task, achieving at least a 32.46$\%$ improvement in $HR@1$ accuracy and a 17.39$\%$ improvement in $Avg@3$ compared to other methods. Compared to most baseline methods, the improvement of \textit{TVDiag} in all metrics is significant, with a P-value less than 0.05. Notably, for dataset $\mathcal{B}$ with more than 40 instances, \textit{TVDiag} achieves a high $HR@1$ accuracy of 0.89, indicating its robust ability to locate the root cause of failures despite the increased number of microservice instances. In contrast, DiagFusion underperforms in dataset $\mathcal{B}$ because it first locates the microservice and then infers the failure instance based on the anomaly degree, which may mistakenly select those abnormal and non-root cause instances. It is worth noting that multimodal methods are consistently more effective than single-modal ones with regard to the RCL task, indicating that integration of multi-perspective data contributes to the performance of root cause localization.

For multimodal diagnosis methods, we further provided an intuitive representation of the performance across different failure types. As shown in Fig. \ref{fig:leida}, \textit{TVDiag} yields a better performance than DiagFusion and Eadro across all datasets. This superiority is more pronounced in $HR@1$, manifesting as a larger coverage area. A higher $HR@1$ indicates that operators incur fewer trial-and-error costs in root cause identification, enhancing diagnostic accuracy and failure recovery efficiency.

Table \ref{tab:FTI-performance} depicts the performance comparison for the FTI task, measured by F1-score, $Precision$, and $Recall$. \textit{TVDiag} can effectively identify the failure types, achieving the best F1-score, $Precision$, and $Recall$ among all datasets. Compared to DiagFusion, which integrates all modalities in the alert extraction process, \textit{TVDiag} demonstrates a substantial improvement in the F1-score, with enhancements ranging from 11.43\% to 167.07\% on three datasets. This underscores the efficacy of separately learning specific features for different modalities. LogCluster exhibits suboptimal performance on dataset $\mathcal{B}$ due to the scarcity of informative log templates within this dataset. This phenomenon further emphasizes the necessity of incorporating data from multiple perspectives (e.g., metrics) for the FTI task. Compared to the other two datasets, Dataset $\mathcal{C}$ exhibits a reduced log volume and limited metric diversity, which constrains human differentiation of certain failure patterns under restricted observability. For instance, both packet-loss and packet-corruption failures manifest nearly identical declines in packet transfer sizes. Consequently, all methods exhibit performance degradation of F1-score on this dataset. Nevertheless, \textit{TVDiag} still achieves nearly 30\% F1-score superiority, demonstrating its capability to extract discriminative latent features from challenging samples.

\begin{tcolorbox}[
	colframe=gray,
	width=\linewidth,boxrule=0.1mm,
	arc=1.0mm,left=0.2mm, auto outer arc,
	breakable]		
	\textbf{Answer to RQ1:} For the RCL task, \textit{TVDiag} significantly surpasses all baseline methods by 32.46\% in $HR@1$ and 17.39\% in $Avg@3$ across three datasets. In terms of the FTI task, \textit{TVDiag} demonstrates the most outstanding effectiveness, with an improvement rate of at least 3.08\% compared to state-of-the-art baselines.
\end{tcolorbox}

\subsection{Ablation Study (RQ2)}
\label{rq:2}
\textbf{Objectives: } We aim to explore the contribution of different modules within \textit{TVDiag}, including one module in the online diagnosis phase and four modules in the offline training phase.


\textbf{Experimental design: } As mentioned in Section \ref{sec:approach}, the training of \textit{TVDiag} comprises four main components: graph augmentation (AUG), task-oriented learning (TO), cross-modal association (CM), and dynamic weights (DW). AUG denotes the graph augmentation that relies on random inactivation for non-root cause instances. TO and CM represent task-oriented learning and cross-modal association, respectively. DW dynamically assigns weights
to each component in the overall loss $\mathcal{L}$. These derived methods can be summarized as follows:

\begin{itemize}
	\item[---] \textit{TVDiag} $w/o$ AUG: \textit{TVDiag} ablates the graph augmentation module described in \cref{sec:AUG}.
	\item[---] \textit{TVDiag} $w/o$ TO: \textit{TVDiag} ablates the task-oriented learning module described in \cref{sec:MC}.
	\item[---] \textit{TVDiag} $w/o$ CM: \textit{TVDiag} ablates the cross-modal association module described in \cref{sec:MC}.
	\item[---] \textit{TVDiag} $w/o$ DW: \textit{TVDiag} statically assigning equal weights to all components of Equation \ref{eq:rawLoss}.
\end{itemize}
In this section, we conducted a comprehensive ablation study across two open-source datasets, denoted as $\mathcal{A}$ and $\mathcal{B}$, comparing our proposed \textit{TVDiag}, which includes all components, with the remaining four variants. 

To further demonstrate that our enhanced alert extractor can capture more valuable information, we also conducted a performance comparison on all datasets by substituting it with the alert extractor of DiagFusion \cite{zhang2023robust}. Since both methods share the same extraction strategy for metrics, no comparison is necessary in that aspect. We utilized these two alert extractors to separately generate alert datasets for both trace and log modalities. Subsequently, we conducted failure diagnosis on the two extracted alert datasets, eliminating interference from other modalities.

Finally, we ablated specific alert fields to explore their contribution to failure diagnosis. Specifically, we selected alert fields that, when ablated, will not compromise the key information related to the modality. Therefore, we choose one field (\textit{abnormalDirection}) from the metric modality and two fields (\textit{operationName} and \textit{abnormalType}) from the trace modality. To avoid interference from other modalities, the experiments are conducted on a single modality only. The variants are listed as follows:
\begin{itemize}
	\item[---] $w/ $ Metric: \textit{TVDiag} only retains the metric modality.
	\item[---] $w/ $ Trace: \textit{TVDiag} only retains the trace modality.
	\item[---] $w/o$ abDirection: $w/ $ Metric ablates the \textit{abnormalDirection} field in Table \ref{tab:alert-template}.
	\item[---] $w/o$ opName: $w/ $ Trace ablates the \textit{operationName} field in Table \ref{tab:alert-template}.
	\item[---] $w/o$ abType: $w/ $ Trace ablates the \textit{abnormalType} field in Table \ref{tab:alert-template}.
\end{itemize}

\begin{table}[t]
	\caption{Experimental results of the ablation study for graph augmentation (AUG), task-oriented learning (TO), cross-modal association (CM), and dynamic weights (DW). Note that the results for RCL are marked in blue, while the results for FTI are marked in gray.}
	\label{tab:ablation}
	\begin{tabular}{@{}c >{\columncolor{cyan!10}}c >{\columncolor{cyan!10}}c >{\columncolor{cyan!10}}c >{\columncolor{gray!20}}cc>{\columncolor{cyan!10}}c>{\columncolor{cyan!10}}c>{\columncolor{cyan!10}}c>{\columncolor{gray!20}}c@{}}
		\toprule
		\multirow{2}{*}{Method} & \multicolumn{4}{c}{Dataset $\mathcal{A}$} &  & \multicolumn{4}{c}{Dataset $\mathcal{B}$} \\ \cmidrule(lr){2-5} \cmidrule(l){7-10} 
		&$HR@1$ &$HR@3$ & $Avg@3$ &F1-score &  & $HR@1$ & $HR@3$ & $Avg@3$ &F1-score \\ \midrule
		\textit{TVDiag} $w/o$ AUG & 0.675  & 0.879  & 0.794  & 0.905  &       & 0.882  & 0.959  & 0.922  & 0.838  \\
		\textit{TVDiag} $w/o$ TO & 0.754  & \textbf{0.911}  & 0.843  & 0.920  &       & 0.835  & 0.953  & 0.902  & 0.864  \\
		\textit{TVDiag} $w/o$ CM & 0.757  & 0.908  & \textbf{0.845}  & 0.931  &       & 0.888  & \textbf{0.959}  & \textbf{0.929}  & 0.868  \\
		\textit{TVDiag} $w/o$ DW & 0.757  & 0.908  & 0.843  & 0.935  &       & 0.847  & 0.941  & 0.900  & 0.858  \\
		\textit{TVDiag} & \textbf{0.759}  & 0.909  & 0.844  & \textbf{0.936}  &       & \textbf{0.894}  & 0.941  & 0.924  & \textbf{0.876}  \\ \bottomrule
	\end{tabular}
\end{table}

\begin{table}[t]
	\caption{Performance comparison of alert extractors.}
	
	\label{tab:extractor}
	\begin{tabular}{c|c|c|>{\columncolor{cyan!10}}c>{\columncolor{cyan!10}}c>{\columncolor{cyan!10}}c>{\columncolor{gray!10}}c}
		\toprule
		Dataset&Extractor & Modality & $HR@1$ &$HR@3$ & $Avg@3$ & F1-score \\
		\midrule
		\multirow{5}{*}{$\mathcal{A}$} &IForest & Trace & \textbf{0.410} & \textbf{0.583} & \textbf{0.506} & \textbf{0.732} \\
		&3-sigma & Trace & 0.398 &0.604 & 0.504 & 0.731 \\
		&Rule-based & Log & \textbf{0.337} &\textbf{0.639} & \textbf{0.518} & \textbf{0.945} \\
		&Sampling & Log & 0.301 & 0.644& 0.508 & 0.877 \\ \midrule
		\multirow{5}{*}{$\mathcal{B}$} &IForest & Trace & \textbf{0.435} & \textbf{0.647} & \textbf{0.537} & \textbf{0.528} \\
		&3-sigma & Trace & 0.102 &0.205 & 0.155 & 0.221 \\
		&Rule-based &Log & \textbf{0.235} &\textbf{0.300} & \textbf{0.274} & \textbf{0.567} \\
		&Sampling & Log & 0.080 & 0.182& 0.125 & 0.194\\
		\bottomrule
	\end{tabular}
\end{table}

\textbf{Results: } 
The results of the ablation study are listed in Table \ref{tab:ablation}. We elaborate on four key results as follows: (1) Our task-oriented learning module makes a macroscopic contribution, as \textit{TVDiag} achieves superior performance on both datasets when compared to \textit{TVDiag} $w/o$ TO. This is mainly because the TO component maximizes the agreement among samples of a specific modality that share the same task label, facilitating the extraction of useful potential information for that task. (2) Indispensable view-invariant information exists across different modalities. \textit{TVDiag} outperforms \textit{TVDiag} $w/o$ CM on $HR@1$ and F1-score across two datasets, owing to the preservation of view-invariant information (e.g., system status and normal microservices). (3) The integration of graph augmentation reinforces the overall failure diagnosis. The graph augmentation (AUG) method involves a node-dropping strategy simulating the missing data phenomenon in data collection, enhancing the generalization of \textit{TVDiag}.  This results in an average increase of 4.82\% in $HR@1$ and 3.45\% in F1-score. 
Moreover, the TO and CM modules usually require sufficient data in a batch to support contrastive learning. The inclusion of AUG here alleviates the problem of insufficient data and reduces the cost of manual labeling, leading to better performance in variants that incorporate AUG compared to those without it. (4) The DW module adaptively balances the training process of the RCL and FTI tasks, thereby promoting the overall effectiveness of \textit{TVDiag}. The variant that ablates dynamic weights exhibits a decline across almost all evaluation metrics on both datasets compared to \textit{TVDiag}. In conclusion, using all components of \textit{TVDiag} together achieves the best performance across most evaluation criteria. All components of \textit{TVDiag} contribute to the overall performance improvement.

As mentioned in Section \ref{sec:AE}, \textit{TVDiag} employs the IForest method and a rule-based approach to extract alerts from traces and logs, respectively. In contrast, DiagFusion achieves the same functionality using the 3-sigma and sampling methods. As summarized in Table \ref{tab:extractor}, the performance of the alert extractor in \textit{TVDiag} surpasses that of DiagFusion across two modalities and two diagnosis tasks. The IForest extractor exhibits superior performance in the trace modality due to the robustness of the IForest model, which is capable of handling noise in a small number of traces. Concerning logs, the sampling method of DiagFusion may overlook low-frequency abnormal logs with significant value, leading to the loss of crucial information. In contrast, \textit{TVDiag}'s two rules ensure that low-frequency abnormal logs are retained as alerts, providing robust criteria for failure diagnosis.

\begin{figure}[t]
	\centering{
		\includegraphics[width=\textwidth]{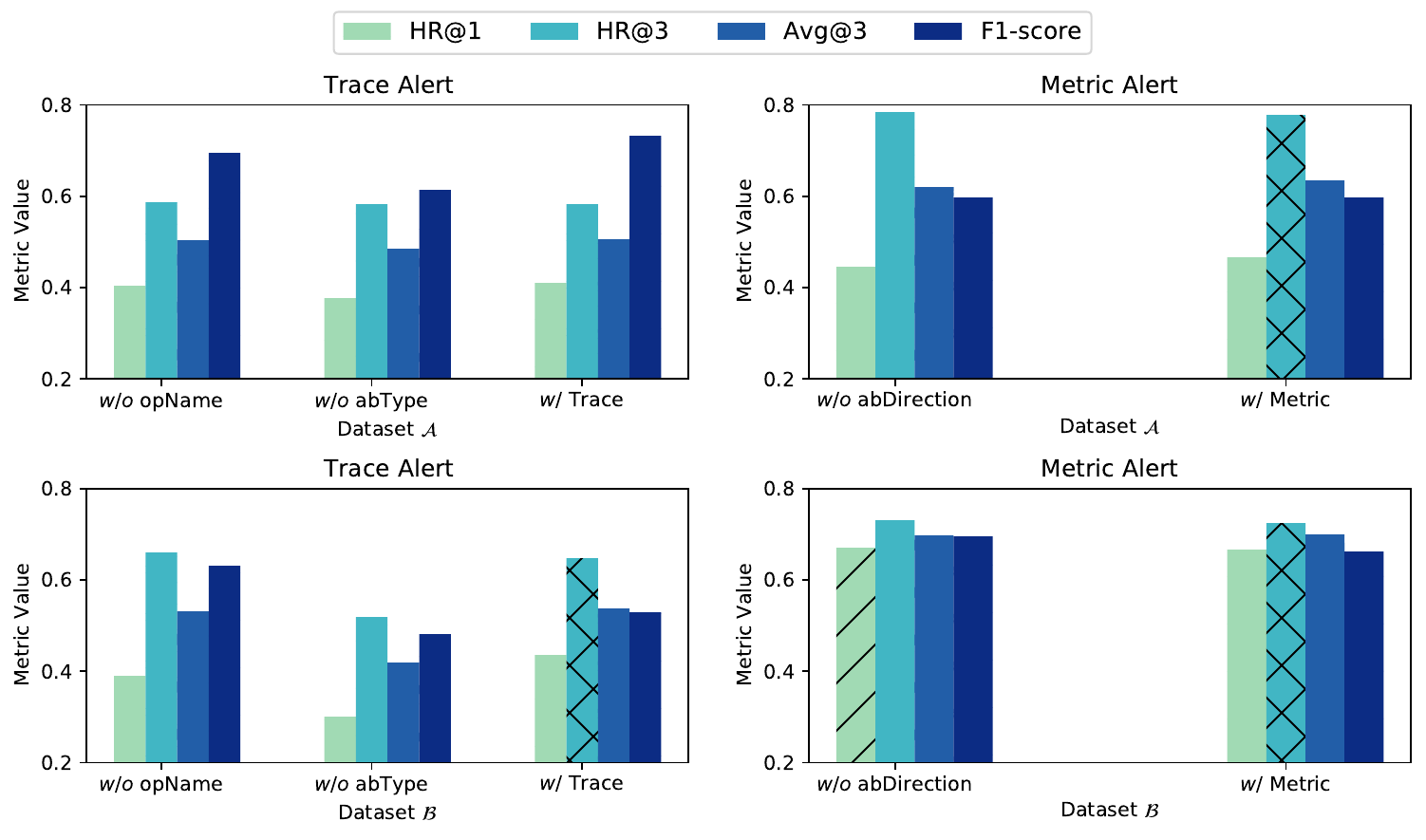} } 
	\vspace{-2mm}
	\caption{Ablation analysis of different alert fields.}
	\vspace{-4mm}
	\label{fig:attribute-impact}
	
\end{figure}

Fig. \ref{fig:attribute-impact} depicts the impact of different alert fields in failure diagnosis. Obviously, the \textit{operationName} and \textit{abnormalType} in trace alerts contribute to failure diagnosis, as the two related variants significantly underperform the $w/$ Trace variant across two datasets. The \textit{operationName} field defines the trace alerts at the interface level, eliminating false alerts caused by significant delay differences between different operations. Besides, \textit{abnormalType} contributes an average increase of 7.90\% to the F1-score, as it helps with the initial classification of certain failure types. For example, net-delay typically sets this field to 'PD', while process-exits often result in an abnormal status code for this field. Similarly, $w/$ Metric slightly outperforms the variant of $w/o$ \textit{abDirection}. Note that operators can also introduce additional fields based on their experience to provide the model with more fault-related cues.

\begin{tcolorbox}[
	colframe=gray,
	width=\linewidth,boxrule=0.1mm,
	arc=1.0mm,left=0.2mm, auto outer arc,
	breakable]		
	\textbf{Answer to RQ2:} The combination of four components (i.e., AUG, TO, CM, and DW) contributes to the performance of \textit{TVDiag}, resulting in the highest $HR@1$ for RCL and F1-score for FTI. Our proposed alert extractor, which incorporates the IForest method and a rule-based approach, exhibits a noticeable improvement across all metrics to the baseline method. 
\end{tcolorbox}




\subsection{Sensitivity Analysis of Hyper-parameters (RQ3)}

\textbf{Objectives:} We investigate the effect of four main hyper-parameters of \textit{TVDiag}: the inactivation probability ($p$), the scale factor ($\omega$), the number of graph layers ($l$), and the temprature parameter ($\tau$).

\textbf{Experimental design:} The inactivation probability $p$ controls the number of dropped nodes in the instance correlation graph $\mathcal{G}$ in \textit{TVDiag}. To measure the effect of different inactivation probabilities on the diagnosis task, we selected values for $p$ within the range of 0.1 to 0.9. 
The overall loss in Eq. \ref{eq:rawLoss} consists of four components: the task-oriented loss $\mathcal{L}_{to}$, the cross-modal association loss $\mathcal{L}_{cm}$, RCL loss $\mathcal{L}_{rcl}$, and FTI loss $\mathcal{L}_{fti}$.
Our primary goal is to optimize the two diagnosis tasks (i.e., reducing $\mathcal{L}_{rcl}$ and $\mathcal{L}_{fti}$), while the role of the contrastive losses ($\mathcal{L}_{cm}$ and $\mathcal{L}_{to}$) are mainly to enhance the quality of the features. We scale down $\mathcal{L}_{cm}$ and $\mathcal{L}_{to}$ to prevent the model from neglecting the optimization of two diagnosis tasks. We varied the contrastive loss scale $\omega$ in the range of 0 to 0.9 and evaluated its impact. Additionally, we regulated the number of graph layers $l$ from 1 to 4 to strike a balance between model performance and training time. The temperature parameter $\tau$ used in Eq. \ref{eq:distance} adjusts the contrast intensity between samples, thereby affecting the attention to difficult samples. We conducted experiments on $\tau$ in the range of 0.1 to 0.9 and evaluated its impact. All parameter experiments were conducted on two open-source datasets, denoted as $\mathcal{A}$ and $\mathcal{B}$.

\begin{figure}[t]
	\centering{
		\subfigure[ Inactivation probability ($p$) ]{
			\includegraphics[width=0.48\textwidth]{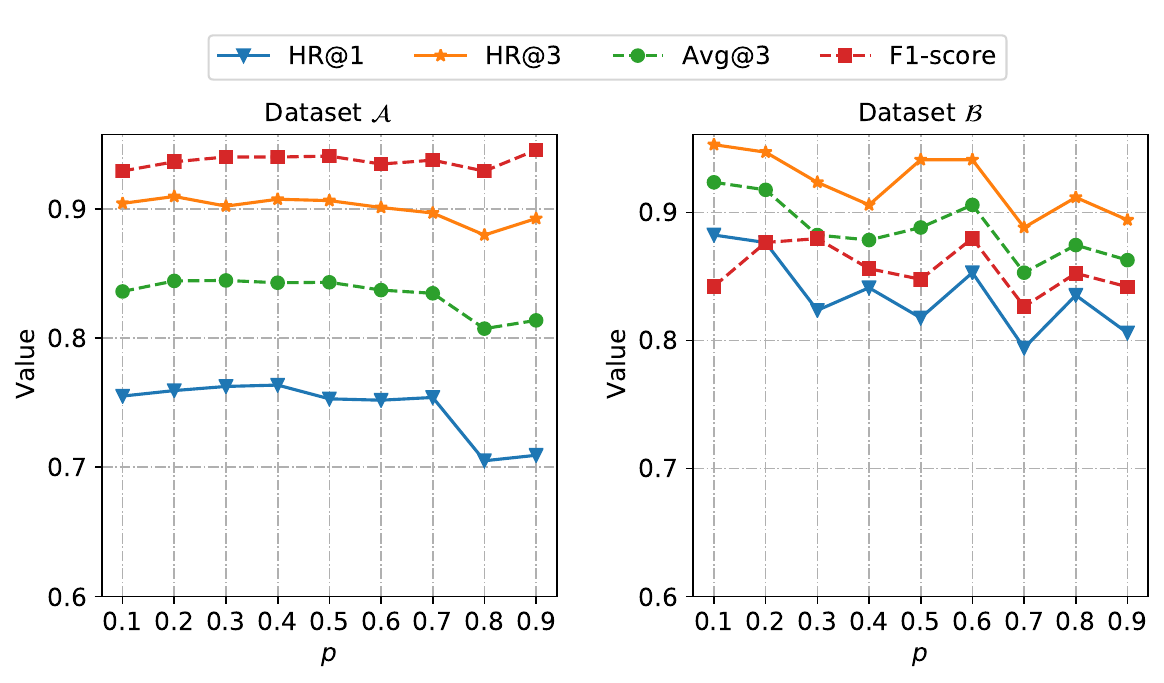} } \hspace{0.5mm}
		\subfigure[ Contrastive loss scale ($\omega$) ]{  \includegraphics[width=0.48\textwidth]{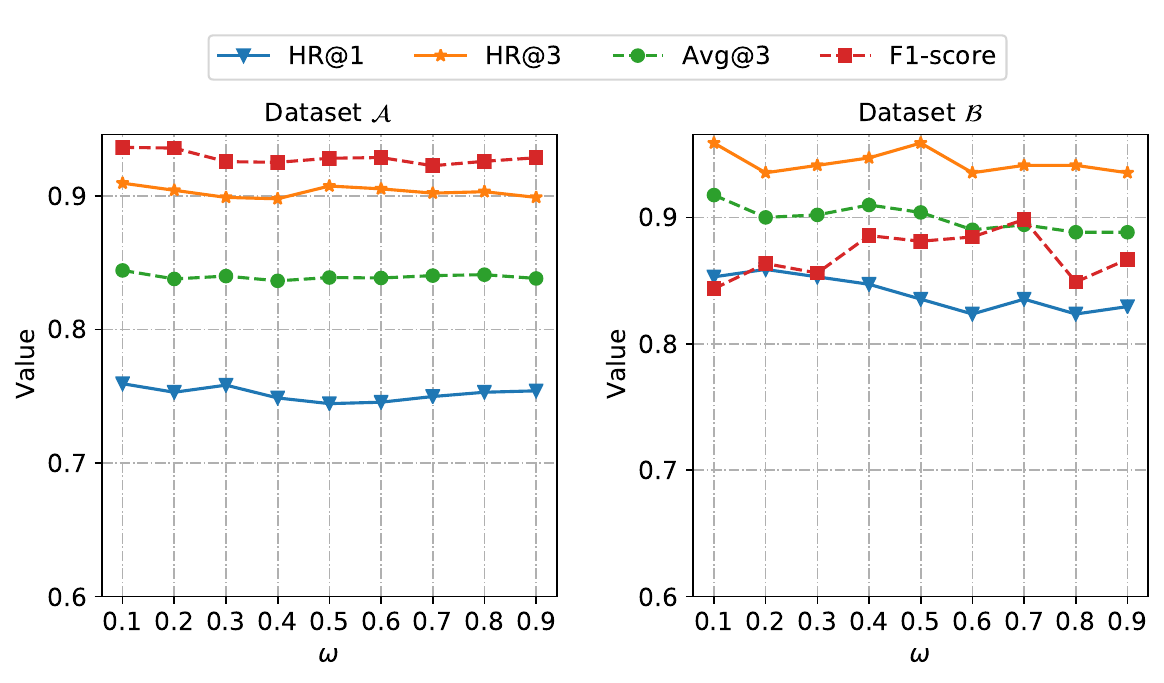}}
		\subfigure[ Temperature parameter ($\tau$) ]{  \includegraphics[width=0.48\textwidth]{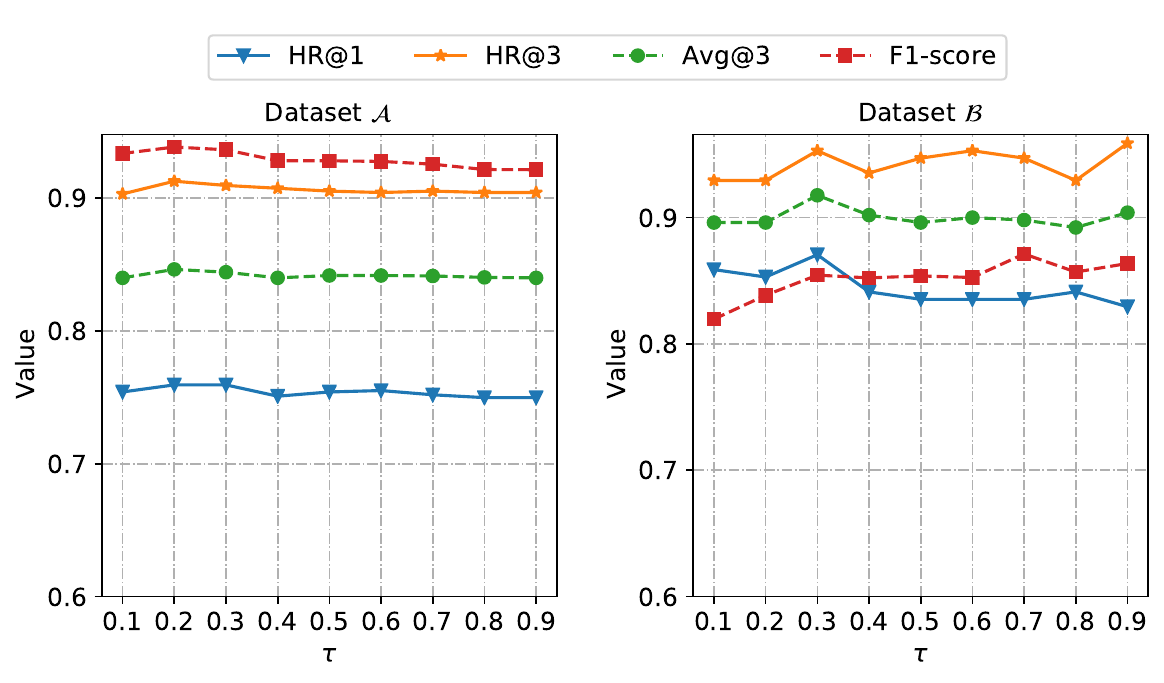}}\hspace{0.5mm}
		\subfigure[ Number of graph layers ($l$) ]{  \includegraphics[width=0.48\textwidth]{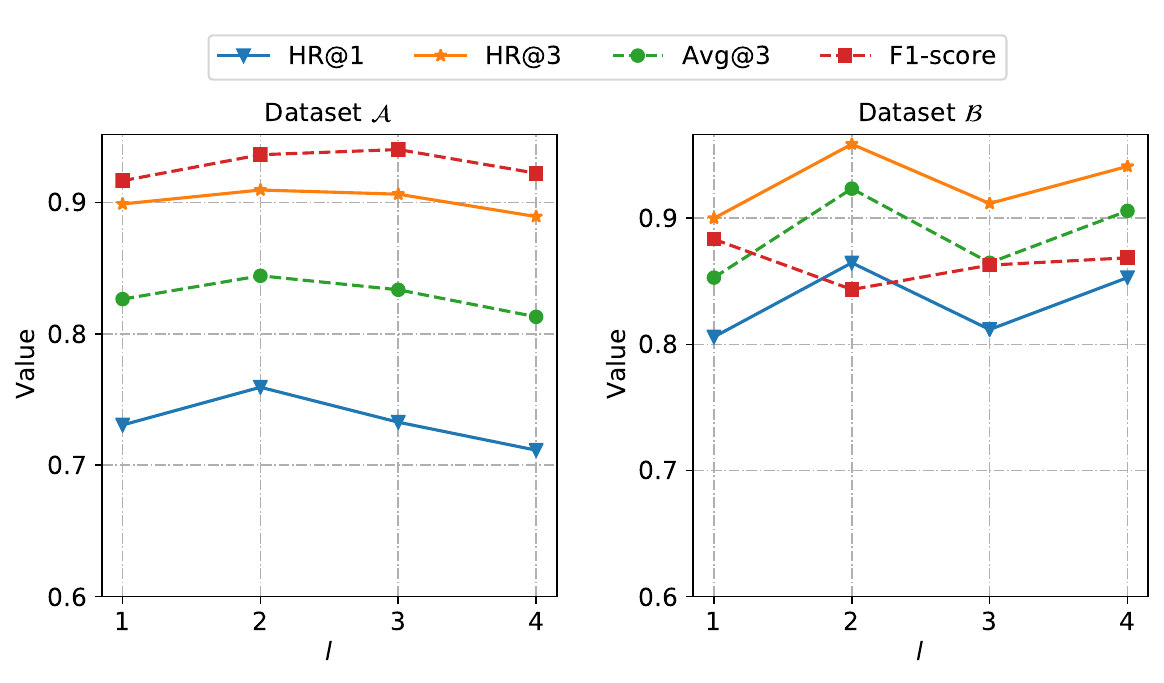}}
	}
	\vspace{-2mm}
	\caption{Effect of the inactivation probability ($p$), the contrastive loss scale ($\omega$), the number of graph layers ($l$), and the temperature parameter ($\tau$).}
	
	\label{fig:hyper}
\end{figure}

\textbf{Results:} Fig. \ref{fig:hyper} presents the effectiveness ($HR@1$, $HR@3$, $Avg@3$, and F1-score) of these hyper-parameters across a range of values. Since the number of graph layers 
$l$ significantly impacts efficiency, we also include the training time as a measurement for it.

\vspace{-2mm}

\begin{itemize}
	\item[---] \textbf{Inactivation probability}: The principle of graph augmentation is equivalent to adding a batch of samples, which share the root cause labels of the original samples. Therefore, we can randomly deactivate a proportion $p$ of non-root cause (non-label) pods to get an augmented sample. As shown in Fig. \ref{fig:hyper} (a), \textit{TVDiag} achieves outstanding performance on both $Avg@3$ and F1-score across two datasets when $p$=0.2. When the $p$ is relatively small, the augmented graph differs slightly from the original graph, resulting in modest performance across two datasets. As the number of abandoned nodes increases, the remaining nodes receive information from only a limited number of neighborhoods, leading to the final graph-level feature focusing solely on the local information. 
	
	\item[---] \textbf{Contrastive loss scale}: The training objective of \textit{TVDiag} is to optimize the performance of two diagnosis tasks and reduce the contrastive loss, which is the sum of losses from task-oriented learning and cross-modal association, namely $\mathcal{L}_{con}$. We adopt contrastive loss scale $\delta$ to adjust the weight of $\mathcal{L}_{con}$. When $\mathcal{L}_{con}$ is set to 0, it is equivalent to removing the TO and CM modules from \textit{TVDiag}. Fig. \ref{fig:hyper} (b) depicts the performance of \textit{TVDiag} for five scale factor settings on the two diagnosis tasks. When $\delta=0.1$, \textit{TVDiag} achieves the best performance in RCL and FTI tasks, indicating that the low $\mathcal{L}_{con}$ indeed serves as a small portion of prior knowledge to guide \textit{TVDiag} in learning view-invariant information and task-oriented features, rather than dominating the entire network's training. 
	
	\item[---] \textbf{Temprature parameter}: When calculating the distance between two samples, $\tau$ is used to smooth the similarity between two samples. A low $\tau$ encourages the model to spotlight the disparity between the most similar and the most dissimilar sample pairs. Conversely, a high $\tau$ indicates that the model is not sensitive to the difficult negative samples. Fig. \ref{fig:hyper} (c) illustrates the performance of failure diagnosis across various values of the temperature parameter $\tau$. \textit{TVDiag} yields the best performance when $\tau$ is set to 0.2.
	
	\item[---] \textbf{Number of graph layers}: The number of graph layers, $l$, represents the depth of the model in propagating and aggregating information from neighborhoods. In theory, as $l$ increases, the graph neural network can assimilate higher-order neighborhood information from more distant nodes. However, deeper layers also entail higher computational complexity and an increased risk of over-fitting. As depicted in Fig. \ref{fig:hyper}, \textit{TVDiag} achieves the best performance when $l$ is set to 2 across two datasets. Although there is a marginal decrease in F1-score on dataset $\mathcal{B}$ when $l$ is set to 2, \textit{TVDiag} considerably outperforms other variants on root cause localization considerably, showing an improvement of nearly 5\%. Therefore, we set $l$ as 2 for both two datasets.
\end{itemize}

\begin{tcolorbox}[
	colframe=gray,
	width=\linewidth,boxrule=0.1mm,
	arc=1.0mm,left=0.2mm, auto outer arc,
	breakable]		
	\textbf{Answer to RQ3:} \textit{TVDiag} performs best with a suitable inactivation probability of 0.2 in augmented graphs. Regarding the training process, it can be observed that low-temperature parameters and low contrastive loss scales could have a positive effect on the final diagnosis performance.  We set the number of graph layers as two in \textit{TVDiag}, balancing the failure diagnosis performance and training consumption.
\end{tcolorbox}

\subsection{Impact Analysis of Modalities and Inter-Task Affinity (RQ4)}
\label{rq:4}
\textbf{Objectives:} We investigate the contribution of each modality, verifying the necessity of integration of multimodal data in failure diagnosis. Furthermore, we analyze the relationship between the RCL and FTI tasks by measuring their inter-task affinity.

\textbf{Experimental design:} To verify the hypothesis that the diagnosis tasks may be biased towards different modalities, we conducted experiments by discarding one or two modalities in \textit{TVDiag}. For example, $w/o$ Metric denotes a variant where the metric modality is discarded, while $w/$ Metric represents a variant where only the metric modality is retained. The omission of a modality is achieved by ablating its features and corresponding modules. In the case of w/o Trace, we excluded the trace alert features and removed trace-related modules in \textit{TVDiag}, such as the trace alert detector and trace graph encoder. Regarding task-oriented learning and cross-modal association, we omitted the trace modality components in Equation \ref{eq:to} and Equation \ref{eq:cm}, simulating the complete absence of traces. Additionally, we assessed the affinity of one modality towards two diagnosis tasks by omitting the other two modalities  simultaneously.

Regarding interpretability, \textit{TVDiag} not only outputs the diagnostic results but also quantifies the modality-level contribution in each inference, guiding operators to prioritize each modality in subsequent analysis. For the FTI task, we calculate the shapley additive explanations (SHAP) value \cite{SHAP} of graph-level features (i.e., $\textbf{F}^\mathcal{M}$, $\textbf{F}^\mathcal{T}$, $\textbf{F}^\mathcal{L}$) to represent the marginal contribution of features to the model output. SHAP is a game-theoretic approach that calculates the importance of each feature by considering all possible combinations of features to explain the output, ensuring that the contributions are consistent and attributed. In each inference, we calculate the SHAP values of all features in $\textbf{F}$ ($\textbf{F}=\textbf{F}^\mathcal{M}\oplus\textbf{F}^\mathcal{T}\oplus\textbf{F}^\mathcal{L}$). In our implementation, we set $\textbf{F}^\mathcal{M},\textbf{F}^\mathcal{T},\textbf{F}^\mathcal{L}\in\mathbb{R}^{32}$ and $\textbf{F}\in\mathbb{R}^{96}$. Therefore, the SHAP value of metric feature $\textbf{F}^\mathcal{M}$ is equal to the mean SHAP value of the first 32 features of $\textbf{F}$. The SHAP value calculation principle for the trace and the log is the same as the metric. Because \textit{TVDiag} scores each instance in the RCL task, we estimate the SHAP value based on the node-level features (i.e., $\{(\textbf{E}_v^\mathcal{M}, \textbf{E}_v^\mathcal{T}, \textbf{E}_v^\mathcal{L})|v\in \mathcal{V}\}$).

Joint learning of two diagnosis tasks holds the promise of improving overall performance. To demonstrate the effectiveness of multi-task learning on failure diagnosis, we adopted the inter-task affinity \cite{fifty2021efficiently} to measure the mutual effect between tasks. More concretely, given two tasks $\mathbf{T}_1$ and $\mathbf{T}_2$, the affinity $Z_{\mathbf{T}_1 \to \mathbf{T}_2}$ can be approximated as the impact of $\mathbf{T}_1$’s gradient on the loss of $\mathbf{T}_2$:
\vspace{-2mm}
\begin{equation}
	Z_{\mathbf{T}_1 \to \mathbf{T}_2} = 1-\frac{\mathcal{L}_{\mathbf{T}_2}\left(f, \theta_{s|\mathbf{T}_1}, \theta_{\mathbf{T}_2}\right)}{\mathcal{L}_{\mathbf{T}_2}\left(f, \theta_{s}, \theta_{\mathbf{T}_2}\right)},
\end{equation}
where $\theta_{s}$ and $\theta_{\mathbf{T}_2}$ denote the shared parameters of the two tasks and task $\mathbf{T}_2$'s specific parameters, respectively. Given the fused feature $f$, we update the shared parameters $\theta_{s|\mathbf{T}_1}$ by leveraging the gradient of task $\mathbf{T}_1$ and the corresponding loss $\mathcal{L}_{\mathbf{T}_2}$. According to \cite{fifty2021efficiently}, the affinity $Z_{\mathbf{T}_1 \to \mathbf{T}_2}$ is positive when the gradient update of task $\mathbf{T}_1$ leads to a lower loss of another task $\mathcal{L}_{\mathbf{T}_2}$. Moreover, a higher absolute value of affinity $Z_{\mathbf{T}_1 \to \mathbf{T}_2}$ indicates a greater impact of task $\mathbf{T}_1$ on task $\mathbf{T}_2$.
We calculated the inter-task affinity at each step within a training epoch and outputted their mean.



\begin{figure}[t]
	\centering{
		\includegraphics[width=\textwidth]{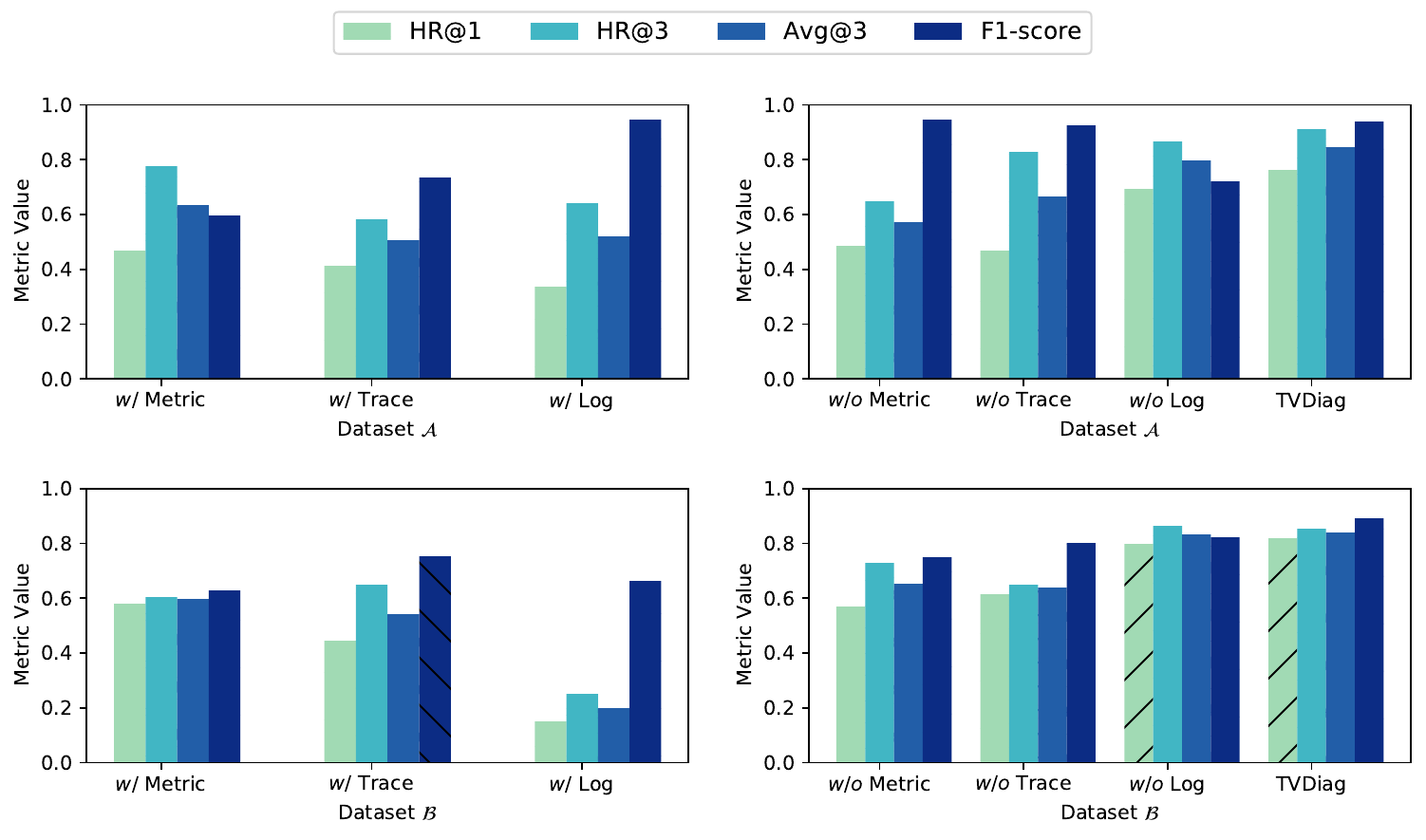} } 
	\caption{Performance comparison of discarding different modalities. $w/ D$ represents a variant of \textit{TVDiag} that only retains modality $D$, while $w/o$ $D$ is the variant which removes modality $D$.}
	\label{fig:modal-impact}
	\vspace{-4mm}
\end{figure}

\begin{figure}[t]
	\centering{
		\subfigure[ Dataset $\mathcal{A}$ ]{
			\includegraphics[width=0.48\textwidth]{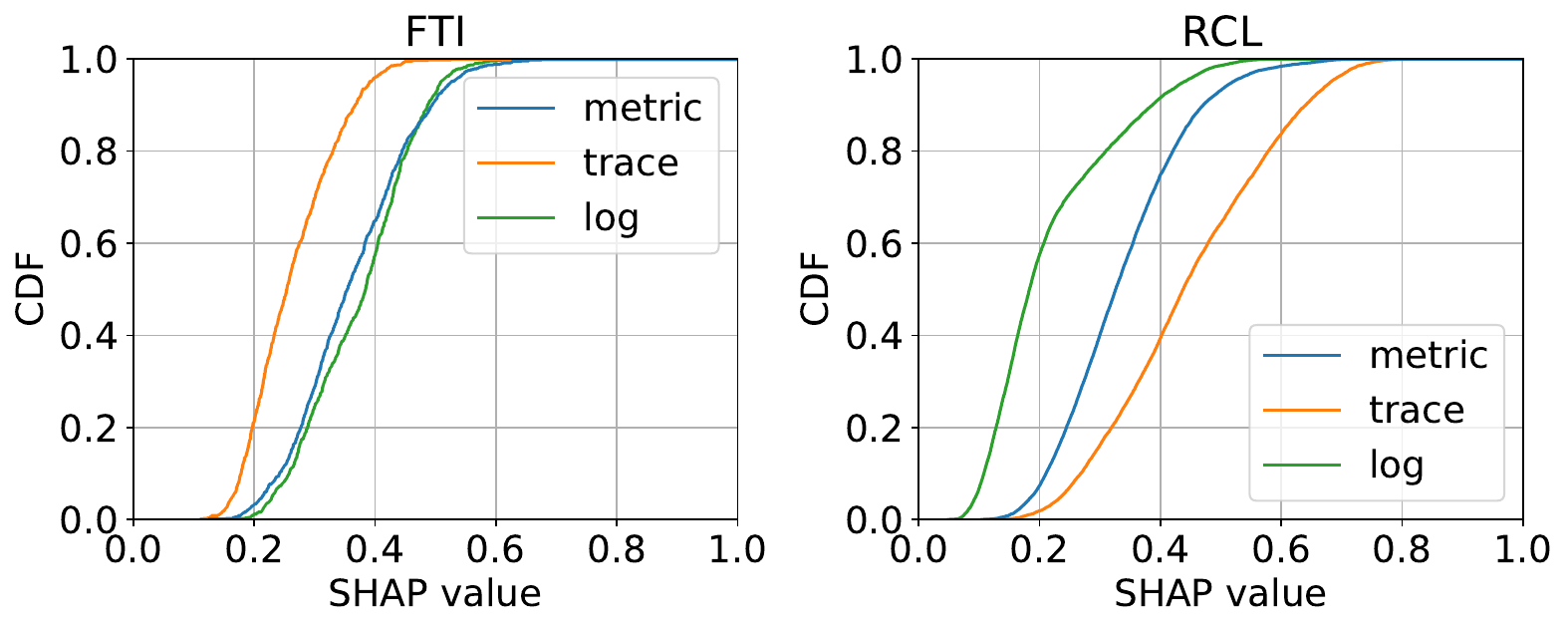} } \hspace{0.5mm}
		\subfigure[ Dataset $\mathcal{B}$ ]{  \includegraphics[width=0.48\textwidth]{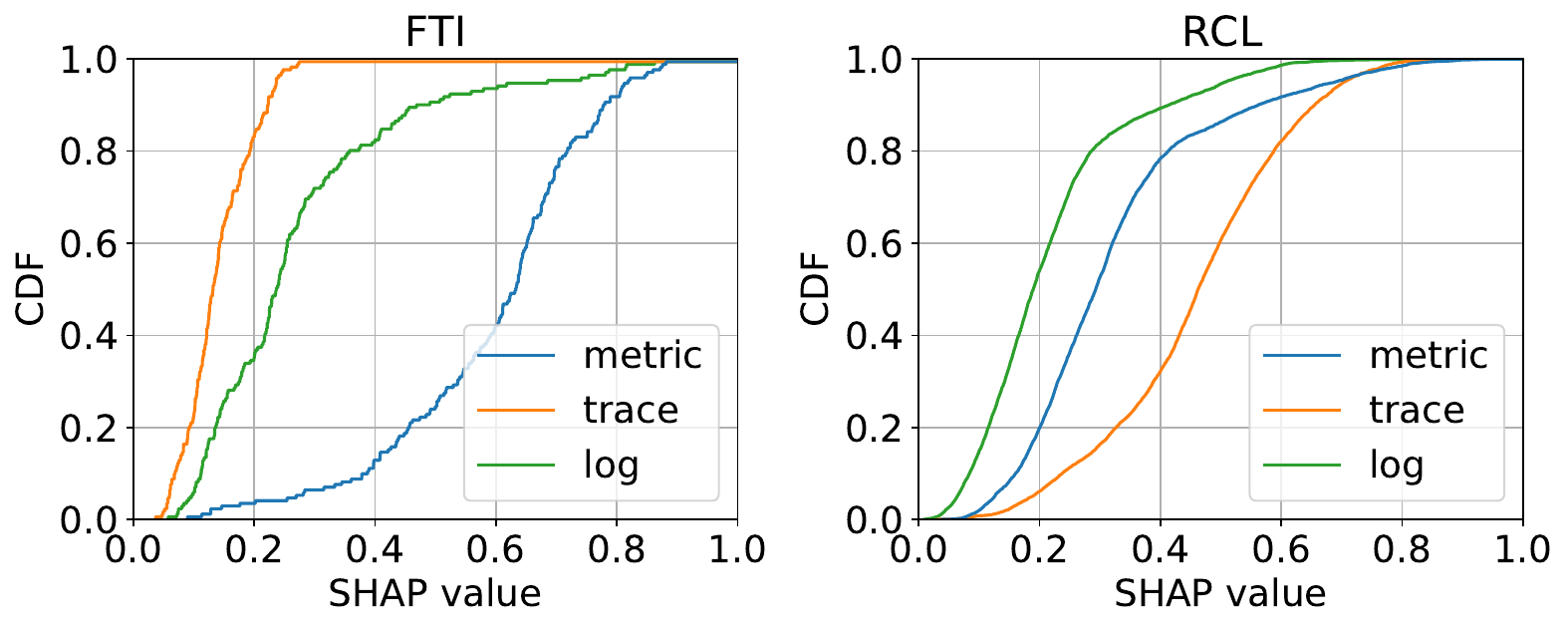}} } 
	\caption{Cumulative distribution of the SHAP value of different modalities.}
	\label{fig:SHAP}
	\vspace{-4mm}
\end{figure}

\textbf{Results:} Fig. \ref{fig:modal-impact} shows the performance comparison of discarding modalities across two datasets. 
For both two datasets, the variants of \textit{TVDiag} based on trace ($w/$ Trace) or metric ($w/$ Metric) achieve over $40\%$ in $HR@1$, while the log-based variant ($w/$ Log) performs modestly in the RCL task but exhibits remarkable effectiveness in the FTI task. This suggests that diagnosis tasks favor specific modalities. On the dataset $\mathcal{B}$, the $w/$ Log performs slightly poorer than other modalities in terms of FTI, as most of the injected failures in dataset $\mathcal{B}$ are resource-related, and the relevant information is recorded by the performance metrics but ignored by the logs. Therefore, FTI does not show a preference for incomplete logs on the dataset $\mathcal{B}$. It can be seen that the comprehensiveness of log information significantly affects the preference of FTI.

We further conducted additional ablation studies by removing one modality at a time to measure the impact of each modality in \textit{TVDiag}.
	Specifically, the lack of metrics ($w/o$ Metric) or traces ($w/o $Trace) causes a significant decline in $HR@1$ ($27.4\%\sim29.4\%$) in dataset $\mathcal{A}$ while having minimal impact on F1-score. This finding validates the importance of metrics and traces in locating the root cause. On the other hand, the absence of logs ($w/o$ Log) degrades the F1-score in identifying failure types (6.9$\%\sim21.6\%$) across two datasets. Moreover, the degradation in $HR@1$ on both datasets is minimal, indicating that the alert information in logs contributes little to the RCL task. The results indicate that the absence of a certain modality has varying impacts on the two diagnosis tasks.


To quantify the modality-level contribution for each failure case, we present the cumulative distribution of SHAP values for two tasks. As shown in Fig. \ref{fig:SHAP}, the trace modality contributes less (SHAP < 0.3) in more than 80\% of failure cases for the FTI task. However, it achieves a SHAP value of 0.5 or higher in 50\% of failure cases for the RCL task, indicating a stronger preference for the trace modality in the RCL task. Similarly, the logs exhibit a significant disparity in SHAP values across the two tasks. For each inference, \textit{TVDiag} provides a quantified expression of the contribution from each modality (i.e., the SHAP value), which helps explain the failure diagnosis results. In this way, we can offer operators a basis for further analyzing the contribution of specific modalities.

Table \ref{tab:task-affinity} presents the inter-task affinity between RCL and FTI (i.e., $Z_{RCL \to FTI}$ and $Z_{FTI \to RCL}$) for two datasets. The positive results suggest that the parameter updates from one task positively contribute to the learning of the other. In dataset $\mathcal{A}$, $Z_{RCL \to FTI}$ and $Z_{FTI \to RCL}$ both are positive, indicating that the two tasks have a complementary relationship during the training process. Although the $Z_{RCL \to FTI}$ in dataset $\mathcal{B}$ is negative, the higher $Z_{FTI \to RCL}$ prompts us to still opt for the fusion of the two tasks. Therefore, we can reasonably speculate that these two diagnosis tasks share certain related features, thereby gaining increased attention in multi-task learning.

\begin{table}[t]
	\caption{Inter-task affinity on two datasets.}
	\vspace{-2mm}
	\label{tab:task-affinity}
	\begin{tabular}{l|cc|cc}
		\toprule
		\cellcolor[HTML]{FFFFFF} & \multicolumn{2}{c}{\cellcolor[HTML]{FFFFFF} $\mathcal{A}$} & \multicolumn{2}{c}{\cellcolor[HTML]{FFFFFF} $\mathcal{B}$} \\ \cmidrule{2-5} 
		\rowcolor[HTML]{FFFFFF} 
		\multirow{-2}{*}{\cellcolor[HTML]{FFFFFF}Dataset} & $Z_{RCL \to FTI}$ & $Z_{FTI \to RCL}$ &  $Z_{RCL \to FTI}$ & $Z_{FTI \to RCL}$\\ \midrule
		Affinity Value & 9.12$\times 10^{-3}$& 8.77$\times 10^{-2}$  & -9.00 $\times 10^{-3}$& 1.08 $\times 10^{-1}$\\ \bottomrule
	\end{tabular}
\end{table}

\begin{tcolorbox}[
	colframe=gray,
	width=\linewidth,boxrule=0.1mm,
	arc=1.0mm,left=0.2mm, auto outer arc,
	breakable]		
	\textbf{Answer to RQ4:} The absence of one or more observability modalities inevitably impairs the diagnostic performance. The experimental results of single-modal methods indicate that diagnostic tasks exhibit a bias towards specific modalities; e.g., the RCL task leans towards metrics and traces, while the FTI task favors logs. Furthermore, the combination of the RCL and FTI tasks manifests a strong inter-task affinity, ensuring the plausibility of joint learning of these two diagnostic tasks.
	
\end{tcolorbox}

\subsection{Efficiency Analysis (RQ5)}
\label{rq:5}
\textbf{Objectives:} We investigate the efficiency of \textit{TVDiag} and existing baselines. Specifically, we calculate their time costs in the offline training phase and the online inference phase.

\textbf{Experimental design:} The efficiency of failure diagnosis comprises two critical phases: offline training and online inference. In the offline training phase, we recorded the total time required for data processing, feature engineering, and model training. Because the model training may incorporate different early stopping mechanisms, we instead calculated the training time per epoch to assess the efficiency of deep-learning models during training. In the online inference phase, we measured the total time for data processing and model inference to evaluate the system's real-time diagnostic capability. We conducted all efficiency experiments on two open-source datasets $\mathcal{A}$ and $\mathcal{B}$.

\begin{table}[t]
	\centering
	\caption{Efficiency comparison of \textit{TVDiag} and baselines. The Offline (s) incorporates the total time cost of offline preparation. The training per epoch (s) records the time taken to complete one forward and backward pass over the entire training dataset. The Online per case (s) indicates the time required to perform a failure diagnosis online.}
	\resizebox{\linewidth}{!}{
		\begin{tabular}{c|ccccccc}
			\toprule
			\multirow{2}[4]{*}{Methods} & \multicolumn{3}{c}{Dataset $\mathcal{A}$} &       & \multicolumn{3}{c}{Dataset $\mathcal{B}$} \\
			\cmidrule{2-4}\cmidrule{6-8}          & Offline (s) & Training per epoch (s) & Online per case (s) &       & Offline (s) & Training per epoch (s) & Online per case (s) \\
			\midrule
			TVDiag & 433.738 & 0.654 & 3.075 &       & 1909.628 & 1.718 & 3.638 \\
			DiagFusion & 388.246 & 0.077 & 3.070  &       & 1563.837 & 0.183 & 3.621 \\
			Eadro & 2484.160 & 0.214 & 15.312 &       & 5420.206 & 0.288 & 13.138 \\
			MicroHECL & 350.012 & -     & 0.147 &       & 21.600  & -     & 0.385 \\
			LogCluster & 52.343 &   -    & 0.335 &       & 666.112 &  -     & 1.602 \\
			TraceRCA & - & -     & 0.702 &       & - & -     & 0.281 \\
			MicroRCA & -     & -     & 2.585 &       & -     & -     & 0.838 \\
			
			\bottomrule
	\end{tabular}}%
	\label{tab:efficiency}%
\end{table}%

\textbf{Results:} Table \ref{tab:efficiency} presents the efficiency comparison of \textit{TVDiag} and several baselines. For both datasets, most methods can diagnose failures online within several seconds, which is efficient enough in practice \cite{li2022actionable}. Although the trained deep-learning model (\textit{TVDiag} and DiagFusion) can quickly infer failures, the data processing of alert extractors, such as \textit{logKey} matching of logs, still incurs a few seconds of time overhead. Similarly, Eadro incorporates the Hawkes process after \textit{logKey} matching, which is time-consuming when the log volume within the failure window is large.
	
Regarding the training time per epoch, \textit{TVDiag} performs modestly because it uses graph augmentation to increase the training data volume. Most diagnostic methods incorporate an offline module to process historical data, such as model training, anomaly detection threshold selection, and knowledge base construction. The offline training time of \textit{TVDiag} is slightly longer than that of DiagFusion due to the separate training of an isolation forest extractor for the trace modality. Note that "-" indicates that the method does not require model training. As explained in Section \ref{sec:failure-diagnosis}, retraining of \textit{TVDiag} is infrequent because it can adapt to dynamic changes in the number of microservice instances. Therefore, the offline training duration does not significantly affect the efficiency of \textit{TVDiag}.

\begin{tcolorbox}[
	colframe=gray,
	width=\linewidth,boxrule=0.1mm,
	arc=1.0mm,left=0.2mm, auto outer arc,
	breakable]		
	\textbf{Answer to RQ5:} \textit{TVDiag} consumes only a few seconds to diagnose failures online, making it suitable for real-world production environments. The offline processing time of \textit{TVDiag} is acceptable, especially considering the infrequent nature of model updates. Furthermore, \textit{TVDiag}'s inherent ability to adapt to dynamic changes in microservice instances reduces the frequency of retraining, further ensuring its efficiency and stability.
	
\end{tcolorbox}

\subsection{Threats to Validity}
The \textit{internal threat to validity} concerns the bias and repeatability of experimental results. The major threat to internal validity lies in the implementation of \textit{TVDiag} and other competing methods. For the RCL task, because Eadro did not publish the preprocessing code, we independently completed the data preprocessing steps introduced in \cite{lee2023eadro} and applied them consistently to the two open datasets. Additionally, we rigorously reproduced the essential modules like log parsing while keeping the main model unchanged.
DiagFusion \cite{zhang2023robust}, MicroRCA \cite{wu2020microrca}, and MicroRank \cite{yu2021microrank} have released their complete code, enabling us to directly utilize them for reproducibility assurance. For the FTI task, LogCluster \cite{lin2016log} also has not released the corresponding source code. To mitigate this threat, we strive to replicate the techniques exactly as described in the original paper, such as log parsing and log clustering. In terms of \textit{TVDiag}, we implemented data augmentation, contrastive loss, and automatic weight adjustment using widely used open-source code to reduce the risk of invalidity. The overall architecture of \textit{TVDiag} was also established based on prevalent frameworks such as PyTorch, scikit-learn, and DGL. To guarantee the reproducibility of our method, we fixed the random seed and preserved the intermediate alert features during experiments. Another internal threat is the potential loss of valuable high-frequency \textit{logKeys}. The characteristics of \textit{logKeys} can vary significantly across different microservice systems, making it challenging to develop an automated tool that can assess the value of \textit{logKeys} universally. Furthermore, manual evaluation is time-consuming and labor-intensive, and it’s not feasible to evaluate all \textit{logKeys} before failure diagnosis. Therefore, we must prioritize the \textit{logKeys} (such as error-level and low-frequency ones) that are more likely to contain failure-related information, while disregarding those that might contain noise or less relevant content. While we do discard some valuable high-frequency \textit{logKeys}, this trade-off is necessary to ensure the universality of our method across various systems. In the future, we plan to investigate how to automatically and accurately identify valuable \textit{logKeys} in different systems.

The \textit{external threat to validity} is primarily related to the generalizability of our proposed method. One major external threat is the limited size of labeled data in the three datasets. In particular, there are only 113 records of failure injection in the original AIOps-22 dataset. To mitigate this threat, we expanded the dataset with a sliding window for each failure record and further performed a data augmentation strategy. In real-world production environments, \textit{TVDiag} may face the cold start problem, i.e., there is a complete lack of labeled historical failure data. We can collect sufficient labeled data by using chaos engineering techniques to simulate failures in the testing environment. This approach can, to some extent, alleviate this issue. Besides, supervised models often suffer from locating failures occurring in new microservice instances due to the inherent elasticity of microservices. The threat of data scarcity of new instances cannot be alleviated even through chaos engineering techniques. To overcome this limitation, we score all known instances instead of fixing the model output to the number of instances, which allows the model to adapt to dynamic changes in the number of instances. In the future, we will explore unsupervised failure diagnosis methods to eliminate the costs associated with manual labeling. Another external threat is the uncertain data quality of logs, given the significant variations in log standards among microservice-based systems. The logs of the GAIA dataset contain rich events at both software and hardware levels, whereas the log information in AIOps-22 is more limited. This variability can directly impact task-oriented learning in \textit{TVDiag}. How to automatically select appropriate diagnosis tasks for guidance in task-oriented learning according to the contents of logs remains a part of our future work. 

Although the three datasets mentioned above may not represent all online microservices systems, we believe \textit{TVDiag} is sufficiently robust to perform well in most scenarios. This is because the input multimodal monitoring data is consistent in terms of data models, even across different microservice-based systems. For example, trace data mainly adhere to the specifications of OpenTracing \cite{OpenTracing} or OpenTelemetry \cite{OpenTelemetry}. Metrics are typically recorded at fixed time intervals in time-series databases \cite{xie2024pbscaler}. Logs are also derived from a small number of templates. These data can be easily transformed into unified alerts with the methods described in Section \ref{sec:AE}.



\section{RELATED WORK}
\label{sec:related-work}
In a microservice-based system, a simple failure in one microservice may prevent the overall system from running normally and seriously affect the user experience \cite{wang2022operation, guo2020graph}. Recently, numerous failure diagnosis methods have been proposed to locate the root cause and identify the failure type. 

\subsection{Single-modal Failure Diagnosis}

Existing single-modal failure diagnosis methods can be categorized into three types according to the dependent modality \cite{zhang2023robust}: metric-based, trace-based, and log-based approaches.

\textbf{Metric-based approaches} \cite{li2022actionable,he2022graph, wu2021microdiag,zhang2021aamr,lin2018microscope,ma2020automap,chen2016causeinfer,wu2020microrca,ma2020self,wu2021identifying,shan2019diagnosis}. Metrics record the running status and performance of a microservice-based system in the form of time series data. Many approaches \cite{meng2020localizing, lin2018microscope, chen2016causeinfer} attempt to construct causal relationships between microservices through metrics and infer the root cause based on simple backtracking. For example, MicroCause \cite{meng2020localizing} builds a causal dependency graph based on collected metrics and performs a random walk to locate the root cause. However, these causal-based methods often construct certain spurious relationships due to the interference of data noise. To overcome this limitation, many methods \cite{li2022actionable,he2022graph,li2022root} use deep learning to exploit the hidden metrics information. For instance, DejaVu \cite{li2022actionable} extracts the temporal features of metrics with recurrent neural networks, aggregates them in a failure dependency graph, and finally scores all components to filter out the root cause and identify the failure type. However, the failure dependency graph requires engineers to build in advance based on their own experience.

\textbf{Trace-based approaches} \cite{li2021practical,yu2021tracerank,li2022microsketch,yu2021microrank, rios2022localizing, mi2013toward,zhou2019latent,guo2020graph}. Traces record the invocation details between microservices, which plays an indispensable role in reproducing the internal correlations of a microservice-based system. Numerous studies focus on extracting the critical path of failures by tracing the correlation between microservices \cite{li2021practical,yu2021microrank,guo2020graph}. For example, MicroRank \cite{yu2021microrank} establishes a microservice correlation graph from traces and analyzes root causes by combining spectrum analysis and the PageRank algorithm. In contrast, MEPFL \cite{zhou2019latent} directly performs supervised classification on traces of failures, achieving good performance as well.

\textbf{Log-based approaches} \cite{pan2021faster,fu2014digging, du2017deeplog,amar2019mining, he2018identifying, rosenberg2020spectrum, bansal2020decaf, li2020swisslog, yuan2019approach}. Logs are a primary source for diagnosing failures and debugging faults in microservice-based systems. Various log-based failure diagnosis techniques have been proposed, leveraging classical machine learning and deep learning approaches. For example, DyCause \cite{pan2021faster} combines API logs from multiple applications based on the crowdsourcing strategy to locate the root cause. However, DyCause only leverages the response time recorded in API logs. DeepLog \cite{du2017deeplog}, on the other hand, parses logs and extracts features to predict the next log, enabling the timely detection of abnormal logs. 

Single-modal methods are limited in their ability to observe only the restricted information provided by a single view, which makes it challenging to diagnose certain failures that are only apparent in other views.

\vspace{-2mm}
\subsection{Multimodal Failure Diagnosis}
In recent years, the fusion of multimodal monitoring data has gained attention in the field of anomaly detection \cite{zhao2021identifying, zhang2022deeptralog, lee2023heterogeneous}. Researchers have also shown interest in integrating the advantages of different modalities to further improve the performance of failure diagnosis \cite{gu2023trinityrcl, he2022perfsig, hou2021diagnosing, nedelkoski2020multi, wang2020root, yu2023nezha,tao2024giving,sun2024art,wang2024mrca}. For instance, Wang et al. \cite{wang2024mrca} identified metric-level root causes, and their novel approach provides an efficient causal analysis method that is terminated using reinforcement learning.
Nezha \cite{yu2023nezha} extracts event patterns from heterogeneous multimodal data, fostering the interpretability of RCL results by comparing normal and abnormal patterns. However, it integrates logs and traces by manually inserting trace IDs into log messages, which involves additional invasive code configurations for microservices. We did not choose Nezha as a comparison method since the two public datasets selected for the experiments do not include this configuration.
In addition, several studies leverage deep learning to mine and fuse the latent information of the three modalities. DiagFusion \cite{zhang2023robust}, for example, fuses all modalities during data preparation by unifying them into events, which are then encoded as features for downstream diagnostic tasks. DiagFusion adopts an early fusion strategy to provide unified information from heterogeneous modalities,  reducing the complexity of data processing. In contrast, Eadro \cite{lee2023eadro} concatenates the representation of each modality during model training. This intermediate fusion can integrate high-dimensional knowledge from all modalities to improve overall performance. Medicine \cite{tao2024giving} encodes each modality separately to address the inherent limitations of single-modal diagnostic approaches, leveraging adaptive optimization to balance the learning progress across modalities. Furthermore, there are studies that use unsupervised methods to achieve fault diagnosis with low human annotation costs. ART \cite{sun2024art} enhances the efficiency of incident management by extracting shared knowledge across these closely related tasks. However, it does not deeply explore the intricate relationships between these tasks and modalities.

These approaches directly fuse the multimodal data, which may hardly capture the underlying associations among various modalities and neglect the potential relationship between diagnosis tasks and modalities.
To address this limitation, we combine the strengths of the above two state-of-the-art approaches by fusing high-dimensional graph-level features of alerts from three modalities. Our \textit{TVDiag} designs a modality-specific feature learning method based on task-oriented learning and constructs cross-modal associations to fully exploit the potential contribution of each modality.

\section{Conclusion and Future Work}
\label{sec:conclusion}

This paper presents \textit{TVDiag}, a multimodal failure diagnosis framework that integrates task-oriented learning and cross-modal association to locate the root causes and identify the type of failures in microservice-based systems. To fully leverage the potential of each modality in corresponding tasks, \textit{TVDiag} employs a novel task-oriented learning method by maximizing the commonality between training samples belonging to the specific modality with the same failure type. Unlike previous multimodal diagnosis methods that ignore the view-invariant information among modalities, \textit{TVDiag} incorporates a cross-modal association method based on contrastive learning to reinforce the view-invariant failure information. Furthermore, \textit{TVDiag} randomly inactivates non-root cause instances to augment the training data, mitigating the issue of insufficient labeled data and incorporating failure scenarios with incomplete observability. Extensive experimental results demonstrate the effectiveness of \textit{TVDiag} in both root cause localization and failure type identification.


In the future, we will explore additional combinations of diagnosis tasks in \textit{TVDiag}, such as anomaly detection and failure prediction. Furthermore, we plan to investigate how to automate task-oriented learning rather than manually designate the relationship between tasks and modalities.

\begin{acks}
	This work is supported by the National Key Research and Development Program of China (No. 2022YFF0902701) and the National Natural Science Foundation of China (No. 62032016).
\end{acks}

\bibliographystyle{ACM-Reference-Format}
\bibliography{TVDiag}


\begin{thebibliography}{80}


\ifx \showCODEN    \undefined \def \showCODEN     #1{\unskip}     \fi
\ifx \showDOI      \undefined \def \showDOI       #1{#1}\fi
\ifx \showISBNx    \undefined \def \showISBNx     #1{\unskip}     \fi
\ifx \showISBNxiii \undefined \def \showISBNxiii  #1{\unskip}     \fi
\ifx \showISSN     \undefined \def \showISSN      #1{\unskip}     \fi
\ifx \showLCCN     \undefined \def \showLCCN      #1{\unskip}     \fi
\ifx \shownote     \undefined \def \shownote      #1{#1}          \fi
\ifx \showarticletitle \undefined \def \showarticletitle #1{#1}   \fi
\ifx \showURL      \undefined \def \showURL       {\relax}        \fi
\providecommand\bibfield[2]{#2}
\providecommand\bibinfo[2]{#2}
\providecommand\natexlab[1]{#1}
\providecommand\showeprint[2][]{arXiv:#2}

\bibitem[AIOps(2024)]%
        {AIOps}
\bibfield{author}{\bibinfo{person}{AIOps}.} \bibinfo{year}{2024}\natexlab{}.
\newblock \bibinfo{title}{[Online]. Available}.
\newblock
\newblock
\newblock
\shownote{\url{https://competition.aiops-challenge.com}}.


\bibitem[Amar and Rigby(2019)]%
        {amar2019mining}
\bibfield{author}{\bibinfo{person}{Anunay Amar} {and} \bibinfo{person}{Peter~C Rigby}.} \bibinfo{year}{2019}\natexlab{}.
\newblock \showarticletitle{Mining historical test logs to predict bugs and localize faults in the test logs}. In \bibinfo{booktitle}{\emph{2019 IEEE/ACM 41st International Conference on Software Engineering (ICSE)}}. IEEE, \bibinfo{pages}{140--151}.
\newblock


\bibitem[Bansal et~al\mbox{.}(2020)]%
        {bansal2020decaf}
\bibfield{author}{\bibinfo{person}{Chetan Bansal}, \bibinfo{person}{Sundararajan Renganathan}, \bibinfo{person}{Ashima Asudani}, \bibinfo{person}{Olivier Midy}, {and} \bibinfo{person}{Mathru Janakiraman}.} \bibinfo{year}{2020}\natexlab{}.
\newblock \showarticletitle{Decaf: Diagnosing and triaging performance issues in large-scale cloud services}. In \bibinfo{booktitle}{\emph{Proceedings of the ACM/IEEE 42nd International Conference on Software Engineering: Software Engineering in Practice}}. \bibinfo{pages}{201--210}.
\newblock


\bibitem[Bigelow(2022)]%
        {Stephen2022What}
\bibfield{author}{\bibinfo{person}{Stephen~J. Bigelow}.} \bibinfo{year}{2022}\natexlab{}.
\newblock \bibinfo{title}{What is observability? A beginner's guide}.
\newblock
\newblock
\newblock
\shownote{\url{https://www.techtarget.com/searchitoperations/definition/observability/}}.


\bibitem[Bojanowski et~al\mbox{.}(2017)]%
        {bojanowski2017enriching}
\bibfield{author}{\bibinfo{person}{Piotr Bojanowski}, \bibinfo{person}{Edouard Grave}, \bibinfo{person}{Armand Joulin}, {and} \bibinfo{person}{Tomas Mikolov}.} \bibinfo{year}{2017}\natexlab{}.
\newblock \showarticletitle{Enriching word vectors with subword information}.
\newblock \bibinfo{journal}{\emph{Transactions of the association for computational linguistics}}  \bibinfo{volume}{5} (\bibinfo{year}{2017}), \bibinfo{pages}{135--146}.
\newblock


\bibitem[chaosmesh(2025)]%
        {chaosmesh}
\bibfield{author}{\bibinfo{person}{chaosmesh}.} \bibinfo{year}{2025}\natexlab{}.
\newblock \bibinfo{title}{[Online]. Available}.
\newblock
\newblock
\newblock
\shownote{\url{https://chaos-mesh.org}}.


\bibitem[Chen et~al\mbox{.}(2022)]%
        {chen2022representation}
\bibfield{author}{\bibinfo{person}{Long Chen}, \bibinfo{person}{Fei Wang}, \bibinfo{person}{Ruijing Yang}, \bibinfo{person}{Fei Xie}, \bibinfo{person}{Wenjing Wang}, \bibinfo{person}{Cai Xu}, \bibinfo{person}{Wei Zhao}, {and} \bibinfo{person}{Ziyu Guan}.} \bibinfo{year}{2022}\natexlab{}.
\newblock \showarticletitle{Representation learning from noisy user-tagged data for sentiment classification}.
\newblock \bibinfo{journal}{\emph{International Journal of Machine Learning and Cybernetics}} \bibinfo{volume}{13}, \bibinfo{number}{12} (\bibinfo{year}{2022}), \bibinfo{pages}{3727--3742}.
\newblock


\bibitem[Chen et~al\mbox{.}(2016)]%
        {chen2016causeinfer}
\bibfield{author}{\bibinfo{person}{Pengfei Chen}, \bibinfo{person}{Yong Qi}, {and} \bibinfo{person}{Di Hou}.} \bibinfo{year}{2016}\natexlab{}.
\newblock \showarticletitle{CauseInfer: Automated end-to-end performance diagnosis with hierarchical causality graph in cloud environment}.
\newblock \bibinfo{journal}{\emph{IEEE transactions on services computing}} \bibinfo{volume}{12}, \bibinfo{number}{2} (\bibinfo{year}{2016}), \bibinfo{pages}{214--230}.
\newblock


\bibitem[Chen et~al\mbox{.}(2020)]%
        {chen2020simple}
\bibfield{author}{\bibinfo{person}{Ting Chen}, \bibinfo{person}{Simon Kornblith}, \bibinfo{person}{Mohammad Norouzi}, {and} \bibinfo{person}{Geoffrey Hinton}.} \bibinfo{year}{2020}\natexlab{}.
\newblock \showarticletitle{A simple framework for contrastive learning of visual representations}. In \bibinfo{booktitle}{\emph{International conference on machine learning}}. PMLR, \bibinfo{pages}{1597--1607}.
\newblock


\bibitem[Du and Li(2018)]%
        {du2018spell}
\bibfield{author}{\bibinfo{person}{Min Du} {and} \bibinfo{person}{Feifei Li}.} \bibinfo{year}{2018}\natexlab{}.
\newblock \showarticletitle{Spell: Online streaming parsing of large unstructured system logs}.
\newblock \bibinfo{journal}{\emph{IEEE Transactions on Knowledge and Data Engineering}} \bibinfo{volume}{31}, \bibinfo{number}{11} (\bibinfo{year}{2018}), \bibinfo{pages}{2213--2227}.
\newblock


\bibitem[Du et~al\mbox{.}(2017)]%
        {du2017deeplog}
\bibfield{author}{\bibinfo{person}{Min Du}, \bibinfo{person}{Feifei Li}, \bibinfo{person}{Guineng Zheng}, {and} \bibinfo{person}{Vivek Srikumar}.} \bibinfo{year}{2017}\natexlab{}.
\newblock \showarticletitle{Deeplog: Anomaly detection and diagnosis from system logs through deep learning}. In \bibinfo{booktitle}{\emph{Proceedings of the 2017 ACM SIGSAC conference on computer and communications security}}. \bibinfo{pages}{1285--1298}.
\newblock


\bibitem[Fifty et~al\mbox{.}(2021)]%
        {fifty2021efficiently}
\bibfield{author}{\bibinfo{person}{Chris Fifty}, \bibinfo{person}{Ehsan Amid}, \bibinfo{person}{Zhe Zhao}, \bibinfo{person}{Tianhe Yu}, \bibinfo{person}{Rohan Anil}, {and} \bibinfo{person}{Chelsea Finn}.} \bibinfo{year}{2021}\natexlab{}.
\newblock \showarticletitle{Efficiently identifying task groupings for multi-task learning}.
\newblock \bibinfo{journal}{\emph{Advances in Neural Information Processing Systems}}  \bibinfo{volume}{34} (\bibinfo{year}{2021}), \bibinfo{pages}{27503--27516}.
\newblock


\bibitem[Foundation(2024)]%
        {OpenTelemetry}
\bibfield{author}{\bibinfo{person}{C.~N.~C. Foundation}.} \bibinfo{year}{2024}\natexlab{}.
\newblock \bibinfo{title}{OpenTelemetry}.
\newblock
\newblock
\newblock
\shownote{\url{https://opentelemetry.io/docs/concepts/sampling/}}.


\bibitem[Fu et~al\mbox{.}(2014)]%
        {fu2014digging}
\bibfield{author}{\bibinfo{person}{Xiaoyu Fu}, \bibinfo{person}{Rui Ren}, \bibinfo{person}{Sally~A McKee}, \bibinfo{person}{Jianfeng Zhan}, {and} \bibinfo{person}{Ninghui Sun}.} \bibinfo{year}{2014}\natexlab{}.
\newblock \showarticletitle{Digging deeper into cluster system logs for failure prediction and root cause diagnosis}. In \bibinfo{booktitle}{\emph{2014 IEEE International Conference on Cluster Computing (CLUSTER)}}. IEEE, \bibinfo{pages}{103--112}.
\newblock


\bibitem[GAIA(2024)]%
        {GAIA}
\bibfield{author}{\bibinfo{person}{GAIA}.} \bibinfo{year}{2024}\natexlab{}.
\newblock \bibinfo{title}{[Online]. Available}.
\newblock
\newblock
\newblock
\shownote{\url{https://github.com/CloudWise-OpenSource/GAIA-DataSet}}.


\bibitem[Google(2024)]%
        {OnlineBoutique}
\bibfield{author}{\bibinfo{person}{Google}.} \bibinfo{year}{2024}\natexlab{}.
\newblock \bibinfo{title}{[Online]. Available}.
\newblock
\newblock
\newblock
\shownote{\url{https://github.com/GoogleCloudPlatform/microservices-demo}}.


\bibitem[Gu et~al\mbox{.}(2023)]%
        {gu2023trinityrcl}
\bibfield{author}{\bibinfo{person}{Shenghui Gu}, \bibinfo{person}{Guoping Rong}, \bibinfo{person}{Tian Ren}, \bibinfo{person}{He Zhang}, \bibinfo{person}{Haifeng Shen}, \bibinfo{person}{Yongda Yu}, \bibinfo{person}{Xian Li}, \bibinfo{person}{Jian Ouyang}, {and} \bibinfo{person}{Chunan Chen}.} \bibinfo{year}{2023}\natexlab{}.
\newblock \showarticletitle{TrinityRCL: Multi-Granular and Code-Level Root Cause Localization Using Multiple Types of Telemetry Data in Microservice Systems}.
\newblock \bibinfo{journal}{\emph{IEEE Transactions on Software Engineering}} (\bibinfo{year}{2023}).
\newblock


\bibitem[Guo et~al\mbox{.}(2020)]%
        {guo2020graph}
\bibfield{author}{\bibinfo{person}{Xiaofeng Guo}, \bibinfo{person}{Xin Peng}, \bibinfo{person}{Hanzhang Wang}, \bibinfo{person}{Wanxue Li}, \bibinfo{person}{Huai Jiang}, \bibinfo{person}{Dan Ding}, \bibinfo{person}{Tao Xie}, {and} \bibinfo{person}{Liangfei Su}.} \bibinfo{year}{2020}\natexlab{}.
\newblock \showarticletitle{Graph-based trace analysis for microservice architecture understanding and problem diagnosis}. In \bibinfo{booktitle}{\emph{Proceedings of the 28th ACM Joint Meeting on European Software Engineering Conference and Symposium on the Foundations of Software Engineering}}. \bibinfo{pages}{1387--1397}.
\newblock


\bibitem[Hamilton et~al\mbox{.}(2017)]%
        {hamilton2017inductive}
\bibfield{author}{\bibinfo{person}{Will Hamilton}, \bibinfo{person}{Zhitao Ying}, {and} \bibinfo{person}{Jure Leskovec}.} \bibinfo{year}{2017}\natexlab{}.
\newblock \showarticletitle{Inductive representation learning on large graphs}.
\newblock \bibinfo{journal}{\emph{Advances in neural information processing systems}}  \bibinfo{volume}{30} (\bibinfo{year}{2017}).
\newblock


\bibitem[He et~al\mbox{.}(2022b)]%
        {he2022perfsig}
\bibfield{author}{\bibinfo{person}{Jingzhu He}, \bibinfo{person}{Yuhang Lin}, \bibinfo{person}{Xiaohui Gu}, \bibinfo{person}{Chin-Chia~Michael Yeh}, {and} \bibinfo{person}{Zhongfang Zhuang}.} \bibinfo{year}{2022}\natexlab{b}.
\newblock \showarticletitle{PerfSig: extracting performance bug signatures via multi-modality causal analysis}. In \bibinfo{booktitle}{\emph{Proceedings of the 44th International Conference on Software Engineering}}. \bibinfo{pages}{1669--1680}.
\newblock


\bibitem[He et~al\mbox{.}(2017)]%
        {he2017drain}
\bibfield{author}{\bibinfo{person}{Pinjia He}, \bibinfo{person}{Jieming Zhu}, \bibinfo{person}{Zibin Zheng}, {and} \bibinfo{person}{Michael~R Lyu}.} \bibinfo{year}{2017}\natexlab{}.
\newblock \showarticletitle{Drain: An online log parsing approach with fixed depth tree}. In \bibinfo{booktitle}{\emph{2017 IEEE international conference on web services (ICWS)}}. IEEE, \bibinfo{pages}{33--40}.
\newblock


\bibitem[He et~al\mbox{.}(2018)]%
        {he2018identifying}
\bibfield{author}{\bibinfo{person}{Shilin He}, \bibinfo{person}{Qingwei Lin}, \bibinfo{person}{Jian-Guang Lou}, \bibinfo{person}{Hongyu Zhang}, \bibinfo{person}{Michael~R Lyu}, {and} \bibinfo{person}{Dongmei Zhang}.} \bibinfo{year}{2018}\natexlab{}.
\newblock \showarticletitle{Identifying impactful service system problems via log analysis}. In \bibinfo{booktitle}{\emph{Proceedings of the 2018 26th ACM Joint Meeting on European Software Engineering Conference and Symposium on the Foundations of Software Engineering}}. \bibinfo{pages}{60--70}.
\newblock


\bibitem[He et~al\mbox{.}(2022a)]%
        {he2022graph}
\bibfield{author}{\bibinfo{person}{Zilong He}, \bibinfo{person}{Pengfei Chen}, \bibinfo{person}{Yu Luo}, \bibinfo{person}{Qiuyu Yan}, \bibinfo{person}{Hongyang Chen}, \bibinfo{person}{Guangba Yu}, {and} \bibinfo{person}{Fangyuan Li}.} \bibinfo{year}{2022}\natexlab{a}.
\newblock \showarticletitle{Graph based Incident Extraction and Diagnosis in Large-Scale Online Systems}. In \bibinfo{booktitle}{\emph{37th IEEE/ACM International Conference on Automated Software Engineering}}. \bibinfo{pages}{1--13}.
\newblock


\bibitem[Hou et~al\mbox{.}(2021)]%
        {hou2021diagnosing}
\bibfield{author}{\bibinfo{person}{Chuanjia Hou}, \bibinfo{person}{Tong Jia}, \bibinfo{person}{Yifan Wu}, \bibinfo{person}{Ying Li}, {and} \bibinfo{person}{Jing Han}.} \bibinfo{year}{2021}\natexlab{}.
\newblock \showarticletitle{Diagnosing Performance Issues in Microservices with Heterogeneous Data Source}. In \bibinfo{booktitle}{\emph{2021 IEEE Intl Conf on Parallel \& Distributed Processing with Applications, Big Data \& Cloud Computing, Sustainable Computing \& Communications, Social Computing \& Networking (ISPA/BDCloud/SocialCom/SustainCom)}}. IEEE, \bibinfo{pages}{493--500}.
\newblock


\bibitem[Khosla et~al\mbox{.}(2020)]%
        {khosla2020supervised}
\bibfield{author}{\bibinfo{person}{Prannay Khosla}, \bibinfo{person}{Piotr Teterwak}, \bibinfo{person}{Chen Wang}, \bibinfo{person}{Aaron Sarna}, \bibinfo{person}{Yonglong Tian}, \bibinfo{person}{Phillip Isola}, \bibinfo{person}{Aaron Maschinot}, \bibinfo{person}{Ce Liu}, {and} \bibinfo{person}{Dilip Krishnan}.} \bibinfo{year}{2020}\natexlab{}.
\newblock \showarticletitle{Supervised contrastive learning}.
\newblock \bibinfo{journal}{\emph{Advances in neural information processing systems}}  \bibinfo{volume}{33} (\bibinfo{year}{2020}), \bibinfo{pages}{18661--18673}.
\newblock


\bibitem[Lee et~al\mbox{.}(2023a)]%
        {lee2023eadro}
\bibfield{author}{\bibinfo{person}{Cheryl Lee}, \bibinfo{person}{Tianyi Yang}, \bibinfo{person}{Zhuangbin Chen}, \bibinfo{person}{Yuxin Su}, {and} \bibinfo{person}{Michael~R Lyu}.} \bibinfo{year}{2023}\natexlab{a}.
\newblock \showarticletitle{Eadro: An end-to-end troubleshooting framework for microservices on multi-source data}. In \bibinfo{booktitle}{\emph{2023 IEEE/ACM 45th International Conference on Software Engineering (ICSE)}}. IEEE, \bibinfo{pages}{1750--1762}.
\newblock


\bibitem[Lee et~al\mbox{.}(2023b)]%
        {lee2023heterogeneous}
\bibfield{author}{\bibinfo{person}{Cheryl Lee}, \bibinfo{person}{Tianyi Yang}, \bibinfo{person}{Zhuangbin Chen}, \bibinfo{person}{Yuxin Su}, \bibinfo{person}{Yongqiang Yang}, {and} \bibinfo{person}{Michael~R Lyu}.} \bibinfo{year}{2023}\natexlab{b}.
\newblock \showarticletitle{Heterogeneous anomaly detection for software systems via semi-supervised cross-modal attention}. In \bibinfo{booktitle}{\emph{2023 IEEE/ACM 45th International Conference on Software Engineering (ICSE)}}. IEEE, \bibinfo{pages}{1724--1736}.
\newblock


\bibitem[Li et~al\mbox{.}(2022a)]%
        {li2022enjoy}
\bibfield{author}{\bibinfo{person}{Bowen Li}, \bibinfo{person}{Xin Peng}, \bibinfo{person}{Qilin Xiang}, \bibinfo{person}{Hanzhang Wang}, \bibinfo{person}{Tao Xie}, \bibinfo{person}{Jun Sun}, {and} \bibinfo{person}{Xuanzhe Liu}.} \bibinfo{year}{2022}\natexlab{a}.
\newblock \showarticletitle{Enjoy your observability: an industrial survey of microservice tracing and analysis}.
\newblock \bibinfo{journal}{\emph{Empirical Software Engineering}}  \bibinfo{volume}{27} (\bibinfo{year}{2022}), \bibinfo{pages}{1--28}.
\newblock


\bibitem[Li et~al\mbox{.}(2020)]%
        {li2020swisslog}
\bibfield{author}{\bibinfo{person}{Xiaoyun Li}, \bibinfo{person}{Pengfei Chen}, \bibinfo{person}{Linxiao Jing}, \bibinfo{person}{Zilong He}, {and} \bibinfo{person}{Guangba Yu}.} \bibinfo{year}{2020}\natexlab{}.
\newblock \showarticletitle{Swisslog: Robust and unified deep learning based log anomaly detection for diverse faults}. In \bibinfo{booktitle}{\emph{2020 IEEE 31st International Symposium on Software Reliability Engineering (ISSRE)}}. IEEE, \bibinfo{pages}{92--103}.
\newblock


\bibitem[Li et~al\mbox{.}(2022c)]%
        {li2022microsketch}
\bibfield{author}{\bibinfo{person}{Yufeng Li}, \bibinfo{person}{Guangba Yu}, \bibinfo{person}{Pengfei Chen}, \bibinfo{person}{Chuanfu Zhang}, {and} \bibinfo{person}{Zibin Zheng}.} \bibinfo{year}{2022}\natexlab{c}.
\newblock \showarticletitle{MicroSketch: Lightweight and Adaptive Sketch Based Performance Issue Detection and Localization in Microservice Systems}. In \bibinfo{booktitle}{\emph{Service-Oriented Computing: 20th International Conference, ICSOC 2022, Seville, Spain, November 29--December 2, 2022, Proceedings}}. Springer, \bibinfo{pages}{219--236}.
\newblock


\bibitem[Li et~al\mbox{.}(2021)]%
        {li2021practical}
\bibfield{author}{\bibinfo{person}{Zeyan Li}, \bibinfo{person}{Junjie Chen}, \bibinfo{person}{Rui Jiao}, \bibinfo{person}{Nengwen Zhao}, \bibinfo{person}{Zhijun Wang}, \bibinfo{person}{Shuwei Zhang}, \bibinfo{person}{Yanjun Wu}, \bibinfo{person}{Long Jiang}, \bibinfo{person}{Leiqin Yan}, \bibinfo{person}{Zikai Wang}, {et~al\mbox{.}}} \bibinfo{year}{2021}\natexlab{}.
\newblock \showarticletitle{Practical root cause localization for microservice systems via trace analysis}. In \bibinfo{booktitle}{\emph{2021 IEEE/ACM 29th International Symposium on Quality of Service (IWQOS)}}. IEEE, \bibinfo{pages}{1--10}.
\newblock


\bibitem[Li et~al\mbox{.}(2022b)]%
        {li2022root}
\bibfield{author}{\bibinfo{person}{Zhongliang Li}, \bibinfo{person}{Yaofeng Tu}, {and} \bibinfo{person}{Zongmin Ma}.} \bibinfo{year}{2022}\natexlab{b}.
\newblock \showarticletitle{Root Cause Analysis of Anomalies Based on Graph Convolutional Neural Network}.
\newblock \bibinfo{journal}{\emph{International Journal of Software Engineering and Knowledge Engineering}} \bibinfo{volume}{32}, \bibinfo{number}{08} (\bibinfo{year}{2022}), \bibinfo{pages}{1155--1177}.
\newblock


\bibitem[Li et~al\mbox{.}(2022d)]%
        {li2022actionable}
\bibfield{author}{\bibinfo{person}{Zeyan Li}, \bibinfo{person}{Nengwen Zhao}, \bibinfo{person}{Mingjie Li}, \bibinfo{person}{Xianglin Lu}, \bibinfo{person}{Lixin Wang}, \bibinfo{person}{Dongdong Chang}, \bibinfo{person}{Xiaohui Nie}, \bibinfo{person}{Li Cao}, \bibinfo{person}{Wenchi Zhang}, \bibinfo{person}{Kaixin Sui}, {et~al\mbox{.}}} \bibinfo{year}{2022}\natexlab{d}.
\newblock \showarticletitle{Actionable and interpretable fault localization for recurring failures in online service systems}. In \bibinfo{booktitle}{\emph{Proceedings of the 30th ACM Joint European Software Engineering Conference and Symposium on the Foundations of Software Engineering}}. \bibinfo{pages}{996--1008}.
\newblock


\bibitem[Liebel and K{\"o}rner(2018)]%
        {liebel2018auxiliary}
\bibfield{author}{\bibinfo{person}{Lukas Liebel} {and} \bibinfo{person}{Marco K{\"o}rner}.} \bibinfo{year}{2018}\natexlab{}.
\newblock \showarticletitle{Auxiliary tasks in multi-task learning}.
\newblock \bibinfo{journal}{\emph{arXiv preprint arXiv:1805.06334}} (\bibinfo{year}{2018}).
\newblock


\bibitem[Lin et~al\mbox{.}(2018)]%
        {lin2018microscope}
\bibfield{author}{\bibinfo{person}{JinJin Lin}, \bibinfo{person}{Pengfei Chen}, {and} \bibinfo{person}{Zibin Zheng}.} \bibinfo{year}{2018}\natexlab{}.
\newblock \showarticletitle{Microscope: Pinpoint performance issues with causal graphs in micro-service environments}. In \bibinfo{booktitle}{\emph{Service-Oriented Computing: 16th International Conference, ICSOC 2018, Hangzhou, China, November 12-15, 2018, Proceedings 16}}. Springer, \bibinfo{pages}{3--20}.
\newblock


\bibitem[Lin et~al\mbox{.}(2016)]%
        {lin2016log}
\bibfield{author}{\bibinfo{person}{Qingwei Lin}, \bibinfo{person}{Hongyu Zhang}, \bibinfo{person}{Jian-Guang Lou}, \bibinfo{person}{Yu Zhang}, {and} \bibinfo{person}{Xuewei Chen}.} \bibinfo{year}{2016}\natexlab{}.
\newblock \showarticletitle{Log clustering based problem identification for online service systems}. In \bibinfo{booktitle}{\emph{Proceedings of the 38th International Conference on Software Engineering Companion}}. \bibinfo{pages}{102--111}.
\newblock


\bibitem[Liu et~al\mbox{.}(2024)]%
        {liu2024integrating}
\bibfield{author}{\bibinfo{person}{Bingqian Liu}, \bibinfo{person}{Duantengchuan Li}, \bibinfo{person}{Jian Wang}, \bibinfo{person}{Zhihao Wang}, \bibinfo{person}{Bing Li}, {and} \bibinfo{person}{Cheng Zeng}.} \bibinfo{year}{2024}\natexlab{}.
\newblock \showarticletitle{Integrating user short-term intentions and long-term preferences in heterogeneous hypergraph networks for sequential recommendation}.
\newblock \bibinfo{journal}{\emph{Information Processing \& Management}} \bibinfo{volume}{61}, \bibinfo{number}{3} (\bibinfo{year}{2024}), \bibinfo{pages}{103680}.
\newblock


\bibitem[Liu et~al\mbox{.}(2021)]%
        {liu2021microhecl}
\bibfield{author}{\bibinfo{person}{Dewei Liu}, \bibinfo{person}{Chuan He}, \bibinfo{person}{Xin Peng}, \bibinfo{person}{Fan Lin}, \bibinfo{person}{Chenxi Zhang}, \bibinfo{person}{Shengfang Gong}, \bibinfo{person}{Ziang Li}, \bibinfo{person}{Jiayu Ou}, {and} \bibinfo{person}{Zheshun Wu}.} \bibinfo{year}{2021}\natexlab{}.
\newblock \showarticletitle{Microhecl: High-efficient root cause localization in large-scale microservice systems}. In \bibinfo{booktitle}{\emph{2021 IEEE/ACM 43rd International Conference on Software Engineering: Software Engineering in Practice (ICSE-SEIP)}}. IEEE, \bibinfo{pages}{338--347}.
\newblock


\bibitem[Liu et~al\mbox{.}(2008)]%
        {liu2008isolation}
\bibfield{author}{\bibinfo{person}{Fei~Tony Liu}, \bibinfo{person}{Kai~Ming Ting}, {and} \bibinfo{person}{Zhi-Hua Zhou}.} \bibinfo{year}{2008}\natexlab{}.
\newblock \showarticletitle{Isolation forest}. In \bibinfo{booktitle}{\emph{2008 eighth ieee international conference on data mining}}. IEEE, \bibinfo{pages}{413--422}.
\newblock


\bibitem[Livens(2023)]%
        {Jay2023What}
\bibfield{author}{\bibinfo{person}{Jay Livens}.} \bibinfo{year}{2023}\natexlab{}.
\newblock \bibinfo{title}{What is observability? Not just logs, metrics and traces}.
\newblock
\newblock
\newblock
\shownote{\url{https://www.dynatrace.com/news/blog/what-is-observability-2/}}.


\bibitem[Lundberg and Lee(2017)]%
        {SHAP}
\bibfield{author}{\bibinfo{person}{Scott~M Lundberg} {and} \bibinfo{person}{Su-In Lee}.} \bibinfo{year}{2017}\natexlab{}.
\newblock \showarticletitle{A Unified Approach to Interpreting Model Predictions}. In \bibinfo{booktitle}{\emph{Advances in Neural Information Processing Systems}}, \bibfield{editor}{\bibinfo{person}{I.~Guyon}, \bibinfo{person}{U.~Von Luxburg}, \bibinfo{person}{S.~Bengio}, \bibinfo{person}{H.~Wallach}, \bibinfo{person}{R.~Fergus}, \bibinfo{person}{S.~Vishwanathan}, {and} \bibinfo{person}{R.~Garnett}} (Eds.), Vol.~\bibinfo{volume}{30}. \bibinfo{publisher}{Curran Associates, Inc.}
\newblock
\urldef\tempurl%
\url{https://proceedings.neurips.cc/paper_files/paper/2017/file/8a20a8621978632d76c43dfd28b67767-Paper.pdf}
\showURL{%
\tempurl}


\bibitem[Ma et~al\mbox{.}(2020a)]%
        {ma2020self}
\bibfield{author}{\bibinfo{person}{Meng Ma}, \bibinfo{person}{Weilan Lin}, \bibinfo{person}{Disheng Pan}, {and} \bibinfo{person}{Ping Wang}.} \bibinfo{year}{2020}\natexlab{a}.
\newblock \showarticletitle{Self-adaptive root cause diagnosis for large-scale microservice architecture}.
\newblock \bibinfo{journal}{\emph{IEEE Transactions on Services Computing}} \bibinfo{volume}{15}, \bibinfo{number}{3} (\bibinfo{year}{2020}), \bibinfo{pages}{1399--1410}.
\newblock


\bibitem[Ma et~al\mbox{.}(2020b)]%
        {ma2020automap}
\bibfield{author}{\bibinfo{person}{Meng Ma}, \bibinfo{person}{Jingmin Xu}, \bibinfo{person}{Yuan Wang}, \bibinfo{person}{Pengfei Chen}, \bibinfo{person}{Zonghua Zhang}, {and} \bibinfo{person}{Ping Wang}.} \bibinfo{year}{2020}\natexlab{b}.
\newblock \showarticletitle{Automap: Diagnose your microservice-based web applications automatically}. In \bibinfo{booktitle}{\emph{Proceedings of The Web Conference 2020}}. \bibinfo{pages}{246--258}.
\newblock


\bibitem[Meng et~al\mbox{.}(2020)]%
        {meng2020localizing}
\bibfield{author}{\bibinfo{person}{Yuan Meng}, \bibinfo{person}{Shenglin Zhang}, \bibinfo{person}{Yongqian Sun}, \bibinfo{person}{Ruru Zhang}, \bibinfo{person}{Zhilong Hu}, \bibinfo{person}{Yiyin Zhang}, \bibinfo{person}{Chenyang Jia}, \bibinfo{person}{Zhaogang Wang}, {and} \bibinfo{person}{Dan Pei}.} \bibinfo{year}{2020}\natexlab{}.
\newblock \showarticletitle{Localizing failure root causes in a microservice through causality inference}. In \bibinfo{booktitle}{\emph{2020 IEEE/ACM 28th International Symposium on Quality of Service (IWQoS)}}. IEEE, \bibinfo{pages}{1--10}.
\newblock


\bibitem[Mi et~al\mbox{.}(2013)]%
        {mi2013toward}
\bibfield{author}{\bibinfo{person}{Haibo Mi}, \bibinfo{person}{Huaimin Wang}, \bibinfo{person}{Yangfan Zhou}, \bibinfo{person}{Michael Rung-Tsong Lyu}, {and} \bibinfo{person}{Hua Cai}.} \bibinfo{year}{2013}\natexlab{}.
\newblock \showarticletitle{Toward fine-grained, unsupervised, scalable performance diagnosis for production cloud computing systems}.
\newblock \bibinfo{journal}{\emph{IEEE Transactions on Parallel and Distributed Systems}} \bibinfo{volume}{24}, \bibinfo{number}{6} (\bibinfo{year}{2013}), \bibinfo{pages}{1245--1255}.
\newblock


\bibitem[Nedelkoski et~al\mbox{.}(2020)]%
        {nedelkoski2020multi}
\bibfield{author}{\bibinfo{person}{Sasho Nedelkoski}, \bibinfo{person}{Jasmin Bogatinovski}, \bibinfo{person}{Ajay~Kumar Mandapati}, \bibinfo{person}{Soeren Becker}, \bibinfo{person}{Jorge Cardoso}, {and} \bibinfo{person}{Odej Kao}.} \bibinfo{year}{2020}\natexlab{}.
\newblock \showarticletitle{Multi-source distributed system data for ai-powered analytics}. In \bibinfo{booktitle}{\emph{Service-Oriented and Cloud Computing: 8th IFIP WG 2.14 European Conference, ESOCC 2020, Heraklion, Crete, Greece, September 28--30, 2020, Proceedings 8}}. Springer, \bibinfo{pages}{161--176}.
\newblock


\bibitem[OpenTracing(2024)]%
        {OpenTracing}
\bibfield{author}{\bibinfo{person}{OpenTracing}.} \bibinfo{year}{2024}\natexlab{}.
\newblock \bibinfo{title}{OpenTracing}.
\newblock
\newblock
\newblock
\shownote{\url{https://opentracing.io/specification}}.


\bibitem[Pan et~al\mbox{.}(2021)]%
        {pan2021faster}
\bibfield{author}{\bibinfo{person}{Yicheng Pan}, \bibinfo{person}{Meng Ma}, \bibinfo{person}{Xinrui Jiang}, {and} \bibinfo{person}{Ping Wang}.} \bibinfo{year}{2021}\natexlab{}.
\newblock \showarticletitle{Faster, deeper, easier: crowdsourcing diagnosis of microservice kernel failure from user space}. In \bibinfo{booktitle}{\emph{Proceedings of the 30th ACM SIGSOFT International Symposium on Software Testing and Analysis}}. \bibinfo{pages}{646--657}.
\newblock


\bibitem[Peng et~al\mbox{.}(2022)]%
        {peng2022trace}
\bibfield{author}{\bibinfo{person}{Xin Peng}, \bibinfo{person}{Chenxi Zhang}, \bibinfo{person}{Zhongyuan Zhao}, \bibinfo{person}{Akasaka Isami}, \bibinfo{person}{Xiaofeng Guo}, {and} \bibinfo{person}{Yunna Cui}.} \bibinfo{year}{2022}\natexlab{}.
\newblock \showarticletitle{Trace analysis based microservice architecture measurement}. In \bibinfo{booktitle}{\emph{Proceedings of the 30th ACM Joint European Software Engineering Conference and Symposium on the Foundations of Software Engineering}}. \bibinfo{pages}{1589--1599}.
\newblock


\bibitem[Pukelsheim(1994)]%
        {pukelsheim1994three}
\bibfield{author}{\bibinfo{person}{Friedrich Pukelsheim}.} \bibinfo{year}{1994}\natexlab{}.
\newblock \showarticletitle{The three sigma rule}.
\newblock \bibinfo{journal}{\emph{The American Statistician}} \bibinfo{volume}{48}, \bibinfo{number}{2} (\bibinfo{year}{1994}), \bibinfo{pages}{88--91}.
\newblock


\bibitem[Qiu et~al\mbox{.}(2020)]%
        {qiu2020firm}
\bibfield{author}{\bibinfo{person}{Haoran Qiu}, \bibinfo{person}{Subho~S Banerjee}, \bibinfo{person}{Saurabh Jha}, \bibinfo{person}{Zbigniew~T Kalbarczyk}, {and} \bibinfo{person}{Ravishankar~K Iyer}.} \bibinfo{year}{2020}\natexlab{}.
\newblock \showarticletitle{FIRM: An intelligent fine-grained resource management framework for slo-oriented microservices}. In \bibinfo{booktitle}{\emph{Proceedings of The 14th USENIX Symposium on Operating Systems Design and Implementation (OSDI ‘20)}}.
\newblock


\bibitem[Rios et~al\mbox{.}(2022)]%
        {rios2022localizing}
\bibfield{author}{\bibinfo{person}{Jesus Rios}, \bibinfo{person}{Saurabh Jha}, {and} \bibinfo{person}{Laura Shwartz}.} \bibinfo{year}{2022}\natexlab{}.
\newblock \showarticletitle{Localizing and Explaining Faults in Microservices Using Distributed Tracing}. In \bibinfo{booktitle}{\emph{2022 IEEE 15th International Conference on Cloud Computing (CLOUD)}}. IEEE, \bibinfo{pages}{489--499}.
\newblock


\bibitem[Rosenberg and Moonen(2020)]%
        {rosenberg2020spectrum}
\bibfield{author}{\bibinfo{person}{Carl~Martin Rosenberg} {and} \bibinfo{person}{Leon Moonen}.} \bibinfo{year}{2020}\natexlab{}.
\newblock \showarticletitle{Spectrum-based log diagnosis}. In \bibinfo{booktitle}{\emph{Proceedings of the 14th ACM/IEEE International Symposium on Empirical Software Engineering and Measurement (ESEM)}}. \bibinfo{pages}{1--12}.
\newblock


\bibitem[Shan et~al\mbox{.}(2019)]%
        {shan2019diagnosis}
\bibfield{author}{\bibinfo{person}{Huasong Shan}, \bibinfo{person}{Yuan Chen}, \bibinfo{person}{Haifeng Liu}, \bibinfo{person}{Yunpeng Zhang}, \bibinfo{person}{Xiao Xiao}, \bibinfo{person}{Xiaofeng He}, \bibinfo{person}{Min Li}, {and} \bibinfo{person}{Wei Ding}.} \bibinfo{year}{2019}\natexlab{}.
\newblock \showarticletitle{?-diagnosis: Unsupervised and real-time diagnosis of small-window long-tail latency in large-scale microservice platforms}. In \bibinfo{booktitle}{\emph{The World Wide Web Conference}}. \bibinfo{pages}{3215--3222}.
\newblock


\bibitem[Sockshop(2025)]%
        {Sockshop}
\bibfield{author}{\bibinfo{person}{Sockshop}.} \bibinfo{year}{2025}\natexlab{}.
\newblock \bibinfo{title}{[Online]. Available}.
\newblock
\newblock
\newblock
\shownote{\url{https://github.com/microservices-demo/microservices-demo}}.


\bibitem[Sui et~al\mbox{.}(2023)]%
        {sui2023logkg}
\bibfield{author}{\bibinfo{person}{Yicheng Sui}, \bibinfo{person}{Yuzhe Zhang}, \bibinfo{person}{Jianjun Sun}, \bibinfo{person}{Ting Xu}, \bibinfo{person}{Shenglin Zhang}, \bibinfo{person}{Zhengdan Li}, \bibinfo{person}{Yongqian Sun}, \bibinfo{person}{Fangrui Guo}, \bibinfo{person}{Junyu Shen}, \bibinfo{person}{Yuzhi Zhang}, {et~al\mbox{.}}} \bibinfo{year}{2023}\natexlab{}.
\newblock \showarticletitle{LogKG: Log Failure Diagnosis through Knowledge Graph}.
\newblock \bibinfo{journal}{\emph{IEEE Transactions on Services Computing}} (\bibinfo{year}{2023}).
\newblock


\bibitem[Sun et~al\mbox{.}(2024)]%
        {sun2024art}
\bibfield{author}{\bibinfo{person}{Yongqian Sun}, \bibinfo{person}{Binpeng Shi}, \bibinfo{person}{Mingyu Mao}, \bibinfo{person}{Minghua Ma}, \bibinfo{person}{Sibo Xia}, \bibinfo{person}{Shenglin Zhang}, {and} \bibinfo{person}{Dan Pei}.} \bibinfo{year}{2024}\natexlab{}.
\newblock \showarticletitle{ART: A Unified Unsupervised Framework for Incident Management in Microservice Systems}. In \bibinfo{booktitle}{\emph{Proceedings of the 39th IEEE/ACM International Conference on Automated Software Engineering}}. \bibinfo{pages}{1183--1194}.
\newblock


\bibitem[Tao et~al\mbox{.}(2024)]%
        {tao2024giving}
\bibfield{author}{\bibinfo{person}{Lei Tao}, \bibinfo{person}{Shenglin Zhang}, \bibinfo{person}{Zedong Jia}, \bibinfo{person}{Jinrui Sun}, \bibinfo{person}{Minghua Ma}, \bibinfo{person}{Zhengdan Li}, \bibinfo{person}{Yongqian Sun}, \bibinfo{person}{Canqun Yang}, \bibinfo{person}{Yuzhi Zhang}, {and} \bibinfo{person}{Dan Pei}.} \bibinfo{year}{2024}\natexlab{}.
\newblock \showarticletitle{Giving Every Modality a Voice in Microservice Failure Diagnosis via Multimodal Adaptive Optimization}. In \bibinfo{booktitle}{\emph{Proceedings of the 39th IEEE/ACM International Conference on Automated Software Engineering}}. \bibinfo{pages}{1107--1119}.
\newblock


\bibitem[TVDiag(2024)]%
        {TVDiag}
\bibfield{author}{\bibinfo{person}{TVDiag}.} \bibinfo{year}{2024}\natexlab{}.
\newblock \bibinfo{title}{[Online]. Available}.
\newblock
\newblock
\newblock
\shownote{\url{https://github.com/WHU-AISE/TVDiag}}.


\bibitem[Wang and Liu(2021)]%
        {wang2021understanding}
\bibfield{author}{\bibinfo{person}{Feng Wang} {and} \bibinfo{person}{Huaping Liu}.} \bibinfo{year}{2021}\natexlab{}.
\newblock \showarticletitle{Understanding the behaviour of contrastive loss}. In \bibinfo{booktitle}{\emph{Proceedings of the IEEE/CVF conference on computer vision and pattern recognition}}. \bibinfo{pages}{2495--2504}.
\newblock


\bibitem[Wang et~al\mbox{.}(2022)]%
        {wang2022operation}
\bibfield{author}{\bibinfo{person}{Lu Wang}, \bibinfo{person}{Yu~Xuan Jiang}, \bibinfo{person}{Zhan Wang}, \bibinfo{person}{Qi~En Huo}, \bibinfo{person}{Jie Dai}, \bibinfo{person}{Sheng~Long Xie}, \bibinfo{person}{Rui Li}, \bibinfo{person}{Ming~Tao Feng}, \bibinfo{person}{Yue~Shen Xu}, {and} \bibinfo{person}{Zhi~Ping Jiang}.} \bibinfo{year}{2022}\natexlab{}.
\newblock \showarticletitle{The operation and maintenance governance of microservices architecture systems: A systematic literature review}.
\newblock \bibinfo{journal}{\emph{Journal of Software: Evolution and Process}} (\bibinfo{year}{2022}), \bibinfo{pages}{e2433}.
\newblock


\bibitem[Wang et~al\mbox{.}(2020)]%
        {wang2020root}
\bibfield{author}{\bibinfo{person}{Lingzhi Wang}, \bibinfo{person}{Nengwen Zhao}, \bibinfo{person}{Junjie Chen}, \bibinfo{person}{Pinnong Li}, \bibinfo{person}{Wenchi Zhang}, {and} \bibinfo{person}{Kaixin Sui}.} \bibinfo{year}{2020}\natexlab{}.
\newblock \showarticletitle{Root-cause metric location for microservice systems via log anomaly detection}. In \bibinfo{booktitle}{\emph{2020 IEEE International Conference on Web Services (ICWS)}}. IEEE, \bibinfo{pages}{142--150}.
\newblock


\bibitem[Wang et~al\mbox{.}(2024)]%
        {wang2024mrca}
\bibfield{author}{\bibinfo{person}{Yidan Wang}, \bibinfo{person}{Zhouruixing Zhu}, \bibinfo{person}{Qiuai Fu}, \bibinfo{person}{Yuchi Ma}, {and} \bibinfo{person}{Pinjia He}.} \bibinfo{year}{2024}\natexlab{}.
\newblock \showarticletitle{MRCA: Metric-level Root Cause Analysis for Microservices via Multi-Modal Data}. In \bibinfo{booktitle}{\emph{Proceedings of the 39th IEEE/ACM International Conference on Automated Software Engineering}}. \bibinfo{pages}{1057--1068}.
\newblock


\bibitem[Wu et~al\mbox{.}(2021b)]%
        {wu2021identifying}
\bibfield{author}{\bibinfo{person}{Canhua Wu}, \bibinfo{person}{Nengwen Zhao}, \bibinfo{person}{Lixin Wang}, \bibinfo{person}{Xiaoqin Yang}, \bibinfo{person}{Shining Li}, \bibinfo{person}{Ming Zhang}, \bibinfo{person}{Xing Jin}, \bibinfo{person}{Xidao Wen}, \bibinfo{person}{Xiaohui Nie}, \bibinfo{person}{Wenchi Zhang}, {et~al\mbox{.}}} \bibinfo{year}{2021}\natexlab{b}.
\newblock \showarticletitle{Identifying root-cause metrics for incident diagnosis in online service systems}. In \bibinfo{booktitle}{\emph{2021 IEEE 32nd International Symposium on Software Reliability Engineering (ISSRE)}}. IEEE, \bibinfo{pages}{91--102}.
\newblock


\bibitem[Wu et~al\mbox{.}(2021a)]%
        {wu2021microdiag}
\bibfield{author}{\bibinfo{person}{Li Wu}, \bibinfo{person}{Johan Tordsson}, \bibinfo{person}{Jasmin Bogatinovski}, \bibinfo{person}{Erik Elmroth}, {and} \bibinfo{person}{Odej Kao}.} \bibinfo{year}{2021}\natexlab{a}.
\newblock \showarticletitle{Microdiag: Fine-grained performance diagnosis for microservice systems}. In \bibinfo{booktitle}{\emph{2021 IEEE/ACM International Workshop on Cloud Intelligence (CloudIntelligence)}}. IEEE, \bibinfo{pages}{31--36}.
\newblock


\bibitem[Wu et~al\mbox{.}(2020)]%
        {wu2020microrca}
\bibfield{author}{\bibinfo{person}{Li Wu}, \bibinfo{person}{Johan Tordsson}, \bibinfo{person}{Erik Elmroth}, {and} \bibinfo{person}{Odej Kao}.} \bibinfo{year}{2020}\natexlab{}.
\newblock \showarticletitle{Microrca: Root cause localization of performance issues in microservices}. In \bibinfo{booktitle}{\emph{NOMS 2020-2020 IEEE/IFIP Network Operations and Management Symposium}}. IEEE, \bibinfo{pages}{1--9}.
\newblock


\bibitem[Xie et~al\mbox{.}(2023)]%
        {xie2023impacttracer}
\bibfield{author}{\bibinfo{person}{Ru Xie}, \bibinfo{person}{Jing Yang}, \bibinfo{person}{Jingying Li}, {and} \bibinfo{person}{Liming Wang}.} \bibinfo{year}{2023}\natexlab{}.
\newblock \showarticletitle{ImpactTracer: Root Cause Localization in Microservices Based on Fault Propagation Modeling}. In \bibinfo{booktitle}{\emph{2023 Design, Automation \& Test in Europe Conference \& Exhibition (DATE)}}. IEEE, \bibinfo{pages}{1--6}.
\newblock


\bibitem[Xie et~al\mbox{.}(2024)]%
        {xie2024pbscaler}
\bibfield{author}{\bibinfo{person}{Shuaiyu Xie}, \bibinfo{person}{Jian Wang}, \bibinfo{person}{Bing Li}, \bibinfo{person}{Zekun Zhang}, \bibinfo{person}{Duantengchuan Li}, {and} \bibinfo{person}{Patrick~CK Hung}.} \bibinfo{year}{2024}\natexlab{}.
\newblock \showarticletitle{PBScaler: A Bottleneck-aware Autoscaling Framework for Microservice-based Applications}.
\newblock \bibinfo{journal}{\emph{IEEE Transactions on Services Computing}} (\bibinfo{year}{2024}).
\newblock


\bibitem[You et~al\mbox{.}(2020)]%
        {you2020graph}
\bibfield{author}{\bibinfo{person}{Yuning You}, \bibinfo{person}{Tianlong Chen}, \bibinfo{person}{Yongduo Sui}, \bibinfo{person}{Ting Chen}, \bibinfo{person}{Zhangyang Wang}, {and} \bibinfo{person}{Yang Shen}.} \bibinfo{year}{2020}\natexlab{}.
\newblock \showarticletitle{Graph contrastive learning with augmentations}.
\newblock \bibinfo{journal}{\emph{Advances in neural information processing systems}}  \bibinfo{volume}{33} (\bibinfo{year}{2020}), \bibinfo{pages}{5812--5823}.
\newblock


\bibitem[Yu et~al\mbox{.}(2021a)]%
        {yu2021microrank}
\bibfield{author}{\bibinfo{person}{Guangba Yu}, \bibinfo{person}{Pengfei Chen}, \bibinfo{person}{Hongyang Chen}, \bibinfo{person}{Zijie Guan}, \bibinfo{person}{Zicheng Huang}, \bibinfo{person}{Linxiao Jing}, \bibinfo{person}{Tianjun Weng}, \bibinfo{person}{Xinmeng Sun}, {and} \bibinfo{person}{Xiaoyun Li}.} \bibinfo{year}{2021}\natexlab{a}.
\newblock \showarticletitle{Microrank: End-to-end latency issue localization with extended spectrum analysis in microservice environments}. In \bibinfo{booktitle}{\emph{Proceedings of the Web Conference 2021}}. \bibinfo{pages}{3087--3098}.
\newblock


\bibitem[Yu et~al\mbox{.}(2023b)]%
        {yu2023logreducer}
\bibfield{author}{\bibinfo{person}{Guangba Yu}, \bibinfo{person}{Pengfei Chen}, \bibinfo{person}{Pairui Li}, \bibinfo{person}{Tianjun Weng}, \bibinfo{person}{Haibing Zheng}, \bibinfo{person}{Yuetang Deng}, {and} \bibinfo{person}{Zibin Zheng}.} \bibinfo{year}{2023}\natexlab{b}.
\newblock \showarticletitle{Logreducer: Identify and reduce log hotspots in kernel on the fly}. In \bibinfo{booktitle}{\emph{2023 IEEE/ACM 45th International Conference on Software Engineering (ICSE)}}. IEEE, \bibinfo{pages}{1763--1775}.
\newblock


\bibitem[Yu et~al\mbox{.}(2023a)]%
        {yu2023nezha}
\bibfield{author}{\bibinfo{person}{Guangba Yu}, \bibinfo{person}{Pengfei Chen}, \bibinfo{person}{Yufeng Li}, \bibinfo{person}{Hongyang Chen}, \bibinfo{person}{Xiaoyun Li}, {and} \bibinfo{person}{Zibin Zheng}.} \bibinfo{year}{2023}\natexlab{a}.
\newblock \showarticletitle{Nezha: Interpretable Fine-Grained Root Causes Analysis for Microservices on Multi-modal Observability Data}. In \bibinfo{booktitle}{\emph{Proceedings of the 31st ACM Joint European Software Engineering Conference and Symposium on the Foundations of Software Engineering}}. \bibinfo{pages}{553--565}.
\newblock


\bibitem[Yu et~al\mbox{.}(2021b)]%
        {yu2021tracerank}
\bibfield{author}{\bibinfo{person}{Guangba Yu}, \bibinfo{person}{Zicheng Huang}, {and} \bibinfo{person}{Pengfei Chen}.} \bibinfo{year}{2021}\natexlab{b}.
\newblock \showarticletitle{TraceRank: Abnormal service localization with dis-aggregated end-to-end tracing data in cloud native systems}.
\newblock \bibinfo{journal}{\emph{Journal of Software: Evolution and Process}} (\bibinfo{year}{2021}), \bibinfo{pages}{e2413}.
\newblock


\bibitem[Yuan et~al\mbox{.}(2019)]%
        {yuan2019approach}
\bibfield{author}{\bibinfo{person}{Yue Yuan}, \bibinfo{person}{Wenchang Shi}, \bibinfo{person}{Bin Liang}, {and} \bibinfo{person}{Bo Qin}.} \bibinfo{year}{2019}\natexlab{}.
\newblock \showarticletitle{An approach to cloud execution failure diagnosis based on exception logs in openstack}. In \bibinfo{booktitle}{\emph{2019 IEEE 12th International Conference on Cloud Computing (CLOUD)}}. IEEE, \bibinfo{pages}{124--131}.
\newblock


\bibitem[Zhang et~al\mbox{.}(2022)]%
        {zhang2022deeptralog}
\bibfield{author}{\bibinfo{person}{Chenxi Zhang}, \bibinfo{person}{Xin Peng}, \bibinfo{person}{Chaofeng Sha}, \bibinfo{person}{Ke Zhang}, \bibinfo{person}{Zhenqing Fu}, \bibinfo{person}{Xiya Wu}, \bibinfo{person}{Qingwei Lin}, {and} \bibinfo{person}{Dongmei Zhang}.} \bibinfo{year}{2022}\natexlab{}.
\newblock \showarticletitle{DeepTraLog: Trace-log combined microservice anomaly detection through graph-based deep learning}. In \bibinfo{booktitle}{\emph{Proceedings of the 44th International Conference on Software Engineering}}. \bibinfo{pages}{623--634}.
\newblock


\bibitem[Zhang et~al\mbox{.}(2023a)]%
        {zhang2023robust}
\bibfield{author}{\bibinfo{person}{Shenglin Zhang}, \bibinfo{person}{Pengxiang Jin}, \bibinfo{person}{Zihan Lin}, \bibinfo{person}{Yongqian Sun}, \bibinfo{person}{Bicheng Zhang}, \bibinfo{person}{Sibo Xia}, \bibinfo{person}{Zhengdan Li}, \bibinfo{person}{Zhenyu Zhong}, \bibinfo{person}{Minghua Ma}, \bibinfo{person}{Wa Jin}, {et~al\mbox{.}}} \bibinfo{year}{2023}\natexlab{a}.
\newblock \showarticletitle{Robust failure diagnosis of microservice system through multimodal data}.
\newblock \bibinfo{journal}{\emph{IEEE Transactions on Services Computing}} \bibinfo{volume}{16}, \bibinfo{number}{6} (\bibinfo{year}{2023}), \bibinfo{pages}{3851--3864}.
\newblock


\bibitem[Zhang et~al\mbox{.}(2023b)]%
        {zhang2023efficient}
\bibfield{author}{\bibinfo{person}{Shenglin Zhang}, \bibinfo{person}{Zhongjie Pan}, \bibinfo{person}{Heng Liu}, \bibinfo{person}{Pengxiang Jin}, \bibinfo{person}{Yongqian Sun}, \bibinfo{person}{Qianyu Ouyang}, \bibinfo{person}{Jiaju Wang}, \bibinfo{person}{Xueying Jia}, \bibinfo{person}{Yuzhi Zhang}, \bibinfo{person}{Hui Yang}, {et~al\mbox{.}}} \bibinfo{year}{2023}\natexlab{b}.
\newblock \showarticletitle{Efficient and Robust Trace Anomaly Detection for Large-Scale Microservice Systems}. In \bibinfo{booktitle}{\emph{2023 IEEE 34th International Symposium on Software Reliability Engineering (ISSRE)}}. IEEE, \bibinfo{pages}{69--79}.
\newblock


\bibitem[Zhang et~al\mbox{.}(2021)]%
        {zhang2021aamr}
\bibfield{author}{\bibinfo{person}{Zekun Zhang}, \bibinfo{person}{Bing Li}, \bibinfo{person}{Jian Wang}, {and} \bibinfo{person}{Yongqiang Liu}.} \bibinfo{year}{2021}\natexlab{}.
\newblock \showarticletitle{AAMR: Automated Anomalous Microservice Ranking in Cloud-Native Environment}.
\newblock  (\bibinfo{year}{2021}).
\newblock


\bibitem[Zhao et~al\mbox{.}(2021)]%
        {zhao2021identifying}
\bibfield{author}{\bibinfo{person}{Nengwen Zhao}, \bibinfo{person}{Junjie Chen}, \bibinfo{person}{Zhaoyang Yu}, \bibinfo{person}{Honglin Wang}, \bibinfo{person}{Jiesong Li}, \bibinfo{person}{Bin Qiu}, \bibinfo{person}{Hongyu Xu}, \bibinfo{person}{Wenchi Zhang}, \bibinfo{person}{Kaixin Sui}, {and} \bibinfo{person}{Dan Pei}.} \bibinfo{year}{2021}\natexlab{}.
\newblock \showarticletitle{Identifying bad software changes via multimodal anomaly detection for online service systems}. In \bibinfo{booktitle}{\emph{Proceedings of the 29th ACM Joint Meeting on European Software Engineering Conference and Symposium on the Foundations of Software Engineering}}. \bibinfo{pages}{527--539}.
\newblock


\bibitem[Zhou et~al\mbox{.}(2019)]%
        {zhou2019latent}
\bibfield{author}{\bibinfo{person}{Xiang Zhou}, \bibinfo{person}{Xin Peng}, \bibinfo{person}{Tao Xie}, \bibinfo{person}{Jun Sun}, \bibinfo{person}{Chao Ji}, \bibinfo{person}{Dewei Liu}, \bibinfo{person}{Qilin Xiang}, {and} \bibinfo{person}{Chuan He}.} \bibinfo{year}{2019}\natexlab{}.
\newblock \showarticletitle{Latent error prediction and fault localization for microservice applications by learning from system trace logs}. In \bibinfo{booktitle}{\emph{Proceedings of the 2019 27th ACM Joint Meeting on European Software Engineering Conference and Symposium on the Foundations of Software Engineering}}. \bibinfo{pages}{683--694}.
\newblock


\end{thebibliography}

\end{document}